\documentclass[
 aps,
 rmp,
 longbibliography
]{revtex4-1}

\makeatletter
\def\@hangfrom@section#1#2#3{\@hangfrom{#1#2}#3}
\def\@hangfroms@section#1#2{#1#2}
\makeatother

\usepackage{graphicx}
\usepackage{braket,bm,amssymb,amsmath,xspace,adjustbox}
\usepackage{array,booktabs,multirow,ulem,longtable}
\usepackage[
draft=false,
bookmarks=true,
bookmarksnumbered=false,
bookmarksopen=false,
bookmarksopenlevel=4,
colorlinks=true,
anchorcolor=black,
citecolor=blue,
filecolor=magenta,
linkcolor=blue,
linkbordercolor={1 0 0},
urlcolor=blue,
pdfborder={0 0 1},
pdftitle={},
pdfauthor={},
pdfsubject={},
pdfkeywords={},
pdfpagemode=UseThumbs,
]{hyperref}
\usepackage{orcidlink}

\newcommand{\bk}{{\bm{k}}}
\newcommand{\bp}{{\bm{r}}}
\newcommand{\Pa}{$\mathcal{P}$\xspace}
\newcommand{\T}{$\mathcal{T}$\xspace}
\newcommand{\PT}{$\mathcal{PT}$\xspace}
\newcommand{\cm}{$\checkmark$}

\newcommand{\matfig}[2]{#1 & \adjustbox{valign=c}{\includegraphics[height=2cm]{#2}} }

\newcolumntype{L}{>{$}l<{$}} 
\newcolumntype{C}{>{$}c<{$}} 

\begin{document}
\title{Magnetic parity violation and Parity-time-reversal-symmetric magnets}

\author{Hikaru Watanabe \orcidlink{0000-0001-7329-9638}}
\address{Research Center for Advanced Science and Technology, University of Tokyo, {Meguro-ku}, Tokyo 153-8904, Japan}
\email{hikaru-watanabe@g.ecc.u-tokyo.ac.jp}

\author{Youichi Yanase}
\address{Department of Physics, Graduate School of Science, Kyoto University, Kyoto 606-8502, Japan}
\email{yanase@scphys.kyoto-u.ac.jp}

\begin{abstract}

Parity-time-reversal symmetry (\PT{} symmetry), a symmetry for the combined operations of space inversion (\Pa{}) and time reversal (\T{}), is a fundamental concept of physics and characterizes the functionality of materials as well as \Pa{} and \T{} symmetries.   
In particular, the \PT{}-symmetric systems can be found in the centrosymmetric crystals undergoing the parity-violating magnetic order which we call the odd-parity magnetic multipole order.
While this spontaneous order leaves \PT{} symmetry intact, the simultaneous violation of \Pa{} and \T{} symmetries gives rise to various emergent responses that are qualitatively different from those allowed by the nonmagnetic \Pa{}-symmetry breaking or by the ferromagnetic order. 
In this review, we introduce candidates hosting the intriguing spontaneous order and overview the characteristic physical responses.
Various off-diagonal and/or nonreciprocal responses are identified, which are closely related to the unusual electronic structures such as hidden spin-momentum locking and asymmetric band dispersion.

\end{abstract}

\maketitle

\tableofcontents

\section{Introduction}

Over the past few decades, the exploration of physical responses arising from symmetry breaking has been extensively conducted in various research fields.
One notable example is the response induced by the parity-symmetry (\Pa{}-symmetry) breaking.
Materials lacking space inversion centers, such as ferroelectric materials and zinc-blende type semiconductors, are attracting a lot of interest in light of the interconversion between different degrees of freedom and the nonlinear effects applied to frequency conversion.
The physical properties often involve the bulk electronic structure and are of paramount interest in condensed matter physics.

A lot of studies have been devoted to nonmagnetic systems in terms of the \Pa{} violation, and previous studies have explored a wide range of materials such as nonmagnetic semiconductors with significant spin-orbit coupling by spectroscopic and transport measurements~\cite{Ishizaka2011-dz,Murakawa2013-zc,Krempasky2016-to}.
We also find works on the \Pa{}-violating effect in magnetic materials named \textit{magnetic parity violation}, that is, the violation of \Pa{} and time-reversal (\T{}) symmetries.
The magnetic parity violation occurs trivially due to the noncentrosymmetric crystal structure and external magnetic field by which the \Pa{} and \T{} symmetries are respectively broken.
On the other hand, the simultaneous violation of the two symmetries can happen in a series of antiferromagnets.
The antiferromagnetic materials, called magnetoelectric systems, have been of much interest in the field of multiferroic science because the magnetic-electric interconversion originates from the magnetic parity violation~\cite{Fiebig2005-hj,Spaldin2008-zj}.
Owing to the controllability ensured by the magnetic-electric coupling, the antiferromagnetic state can be nicely computed by external fields~\cite{Van_Aken2007-zf,Zimmermann2014-pv}.
Since the antiferromagnetic domains can be monitored via electric and optical signals~\cite{Kosub2015-hp,Kosub2017-ly,Kocsis2018-ka}, there may exist applications such as memory devices invulnerable to magnetic noises. 

Much attention has been drawn to the magnetoelectric material which is mostly insulating to keep the electric polarization.
Few studies, however, have been performed on the interplay between the magnetic parity violation and itinerant properties such as electric conductivity.
Recent rapid developments paved the way for spintronic physics based on the antiferromagnets, so-called antiferromagnetic spintronics~\cite{Jungwirth2016-gj,Baltz2018-pu}.
The promising candidates for the growing field include not only magnetoelectric insulators but also various kinds of antiferromagnetic metals such as those offering the anomalous Hall effect~\cite{Smejkal2022-rk} and those switchable by the electric current~\cite{Wadley2016-lo}.
The large variety of candidates stems from the fact that the symmetry breaking by the antiferromagnetic order varies depending on the structural degree of freedom coupled to the magnetic order.
Furthermore, since the energy scale relevant to the antiferromagnetic order may be much higher than that of the ferromagnets, the antiferromagnetic spintronics possesses advantageous properties (\textit{e.g.}, large transition temperature and faster magnetic dynamics) when compared to the ferromagnetic spintronics.
These growing interests motivate us to take a deeper look into the role of the magnetic parity violation in metals.
To this end, we overview the physics originating from the magnetic parity violation mainly in terms of itinerant properties.

The organization is the following.
Firstly, we present a brief symmetry analysis of the magnetic parity violation with a comparative study with that of the electric parity violation, the \Pa{}-symmetry violation preserving the \T{} symmetry (Sec.~\ref{Sec_magnetic_parity_violation_symmetry}).
Many candidate materials manifesting the magnetic parity violation show a special property in their crystal structure, namely locally noncentrosymmetric crystal symmetry.
The characteristics arising from the structure are seemingly inaccessible without microscopic measurements (Sec.~\ref{SecSub_locally_noncentrosymmetric}), whereas it is unveiled in a macroscopic manner because of the antiferroic ordering (Sec.~\ref{SecSub_multipolar_classification}).
Such an intimate coupling between the structural degree of freedom and antiferroic ordering cultivates the basic understanding of the itinerant property of magnetically-parity-violating materials.  
In Sec.~\ref{Sec_emergent_response}, we present some representative examples of the emergent responses induced by the magnetic parity violation.
Similarly to discussions in Sec.~\ref{SecSub_multipolar_classification}, there are contrasting roles of the magnetic and electric parity violations in a broad range of physical responses such as electric-elastic coupling (Sec.~\ref{SecSub_MPE}) and nonreciprocal transport and optical responses (Sec.~\ref{Sec_nonreciprocal_electric_phenomena}).
Based on the arguments presented in these sections, one can figure out that the \PT{} symmetry (symmetry of the combined operation of \Pa{} and \T{}) holds an essential role in disentangling the emergent responses induced by the parity violation. 
Moreover, we overview recent studies working on the control of the magnetic parity violation in Sec.~\ref{Sec_control}.
In light of the field of antiferromagnetic spintronics, a lot of efforts have been made to control and utilize the parity-violating magnets.
Similarly to a series of antiferromagnets, the magnetic parity violation may exist in superconductors.
We introduce candidate superconductors hosting the magnetic parity violation and the physical property resulting from the interplay between superconductivity and magnetic parity violation (Sec.~\ref{Sec_SC_and_MPV}). 
Finally, we summarize the review and give some outlooks in Sec.~\ref{Sec_summary}.

\section{Magnetic Parity Violation and Space-Time Symmetry}
\label{Sec_magnetic_parity_violation_symmetry}

The parity violation can be classified by whether it is accompanied by the \T{}-symmetry breaking.
We call the parity violations \textit{magnetic parity violation} and \textit{electric parity violation}, when the symmetry breaking accompanies \T{} violation or not, respectively.
The electric parity violation may be typical and is found in mundane acentric crystals and those undergoing structural transitions such as the ferroelectric order.
The magnetic parity violation trivially occurs by applying the external magnetic field to systems manifesting the electric parity violation. 
On the other hand, the magnetic parity violation can occur not to be admixed with the electric parity violation.
In this section, we explain the symmetry of each parity violation and discuss candidates for magnetically-parity-violating materials (Sec.~\ref{SecSub_P_T_symmetry_parity_violation}).
The candidate materials consist of a class of antiferromagnets called \PT{}-symmetric magnets.
A lot of \PT{}-symmetric magnets show a common feature in their crystal structure, that is, locally-noncentrosymmetric structure.
We briefly discuss the structural feature and associated physical properties such as hidden magnetic degrees of freedom (Sec.~\ref{SecSub_locally_noncentrosymmetric}).

\subsection{Parity-time-reversal symmetry and parity violations}
\label{SecSub_P_T_symmetry_parity_violation}

Two types of parity violation are clearly distinguished by the space-time symmetry.
One may take account of \Pa{}- and \T{}-parities to characterize the space-time symmetry.
We, however, stress that the combination of \Pa{} and \T{} operations plays a fundamental role as well as \Pa{} and \T{} symmetries.
To this end, we raise examples of symmetry breaking arising from the antiferromagnetic order.

                \begin{figure}[htbp]
                \centering
                \includegraphics[width=0.95\linewidth,clip]{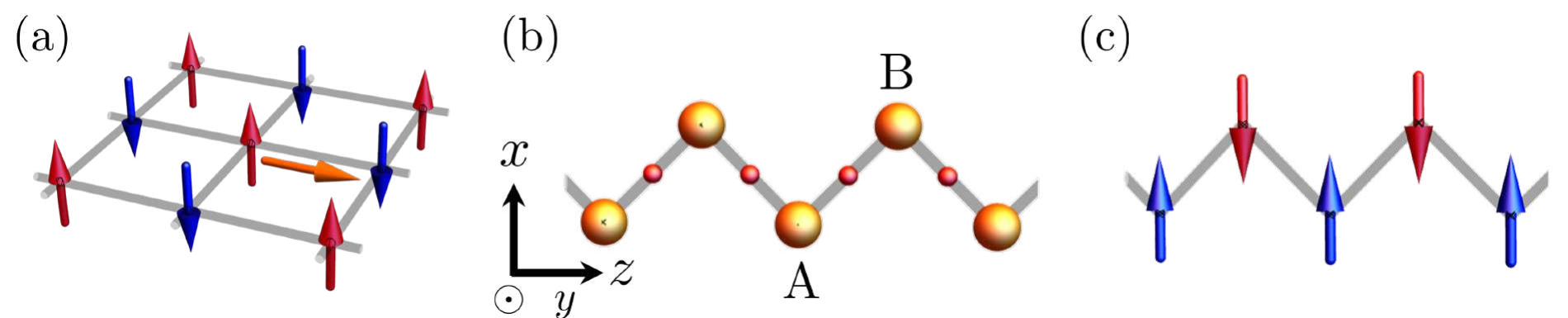}
                \caption{
                        (a) Checkerboard-type antiferromagnetic order in the square lattice. The red- and blue-colored magnetic moments are interchanged under the combined operation of the time-reversal operation and translation operation $\bm{\tau} = (a,0)$ depicted by the orange arrow.
                        (b) Zigzag chain comprised of two sites (orange-colored spheres). The small red-colored spheres are inversion centers with which the parity operation interchanges the two sites.
                        (c) Antiferromagnetic order with the zero propagation vector in the zigzag chain. The red- and blue-colored magnetic moments are placed at different crystallographic sites of the zigzag chain. 
                        }
                \label{Fig_1_AFM_square_zigzag}
                \end{figure}

The antiferromagnetic order is defined by the antiferroic alignment of magnetic moments in crystals.
Owing to the \T{}-odd nature of magnetic moments, the antiferromagnetic order is accompanied by the \T{}-symmetry violation.
The symmetry related to the \T{} operation, however, may be retained; \textit{e.g.}, for the $\bm{q} = (\pi,\pi)$ antiferromagnetic order in the square lattice, there exists \T{} symmetry coupled to the lattice translation $\bm{\tau} = (a,0),(0,a)$ [Fig.~\ref{Fig_1_AFM_square_zigzag}(a)].
The preserved symmetry is equivalent to the \T{} symmetry in terms of the bulk physical properties because the microscopic lattice translation $\bm{\tau}$ does not lead to any symmetry constraint.
Such an effective \T{} symmetry makes it difficult to detect the antiferromagnetic state without microscopic spectroscopy such as neutron diffraction.

The antiferromagnetic order gives rise to various symmetry breaking due to the crystal structure in contrast to the case of checkerboard antiferromagnetic order depicted in Fig.~\ref{Fig_1_AFM_square_zigzag}(a).
Here, we consider the zigzag chain to corroborate the space-time symmetry arising from the magnetic parity violation.
The zigzag chain is comprised of two sites (A and B sites) inside the unit cell~\cite{Yanase2014-mw} [Fig.~\ref{Fig_1_AFM_square_zigzag}(b)].
The system holds the \Pa{} symmetry due to interchanging the sublattices, while each site is not an inversion center.
Let us introduce the antiferromagnetic order with the zero propagation vector ($\bm{q}=\bm{0}$) [Fig.~\ref{Fig_1_AFM_square_zigzag}(c)].
The ordered state is described by using the magnetic moments localized at sites as
                \begin{equation}
                \left( \bm{m}_\text{A},\bm{m}_\text{B} \right) = ( +\hat{x},-\hat{x}).
                \label{zigzag_AFMorder}
                \end{equation}
Intriguingly, the antiferromagnetic order breaks both of the \Pa{} and \T{} symmetries in the macroscopic scale.
It is readily checked by applying the operations to the magnetic configuration given in Eq.~\eqref{zigzag_AFMorder}.
The \Pa{} operation interchanging the two sites does not flip the localized magnetic moments due to the axial symmetry of magnetic moments.
Then, the magnetic moments at each site are transformed as
                \begin{equation}
                        \left( \bm{m}_\text{A},\bm{m}_\text{B} \right)  = ( +\hat{x},-\hat{x}) \xrightarrow{\mathcal{P}} ( -\hat{x},+\hat{x}).
                        \label{zigzag_AFM_p_operation}
                \end{equation}
Similarly, the \T{} operation reverses the magnetic moments with keeping structural degrees of freedom invariant as
                \begin{equation}
                        \left( \bm{m}_\text{A},\bm{m}_\text{B} \right)  = ( +\hat{x},-\hat{x}) \xrightarrow{\mathcal{T}} ( -\hat{x},+\hat{x}). 
                        \label{zigzag_AFM_t_operation}
                \end{equation}
As a result, the antiferromagnetic state shows the odd parity under the \Pa{} and \T{} operation, which is a magnetic parity violation.
The symmetry breaking is not compensated by any symmetry operation such as a lattice translation in sharp contrast to the antiferromagnetic order shown in Fig.~\ref{Fig_1_AFM_square_zigzag}(a).

One can notice the symmetry unique to the antiferromagnetic zigzag chain by considering Eqs.~\eqref{zigzag_AFM_p_operation} and \eqref{zigzag_AFM_t_operation}.
The antiferromagnetic state returns to the original configuration under the combined operation as
                \begin{equation}
                ( +\hat{x},-\hat{x}) \xrightarrow{\mathcal{P}} ( -\hat{x},+\hat{x}) \xrightarrow{\mathcal{T}} ( +\hat{x},-\hat{x}),
                \end{equation}
which indicates the \PT{} symmetry.
In the light of the \PT{}-even nature, the parity-violating magnets may be called \textit{\PT{}-symmetric magnets}.
The \PT{}-symmetric magnets are comprised of intriguing materials such as magnetoelectric materials~\cite{Fiebig2005-hj} and electrically-switchable antiferromagnets~\cite{Wadley2016-lo}.

The intact \PT{} symmetry is a fundamental property distinguishing the \PT{}-symmetric magnets from systems showing the electric parity violation.
For an example of electric parity violation, the macroscopic electric polarization $\bm{P}$ is flipped under the \PT{} operation while it is invariant under the \T{} operation (\PT{}-odd, \T{}-even).
The electric polarization $\bm{P}$ is therefore absent in the \PT{}-symmetric systems.
As a result, parity-violating systems are divided into three classes; the \T{}-symmetric case with only the electric parity violation, \PT{}-symmetric case with only the magnetic parity violation, and the otherwise manifesting both electric and magnetic parity violations.
The third class contains an important series of the multiferroic magnets such as $R$MnO$_3$ ($R$: rare-earth element)~\cite{Kimura2003-ma,Tokura2014-ix}, which is not in the scope of this review.
As a result, we have obtained the classification of parity violations based on the parity under \Pa{}, \T{}, and \PT{} operations.

\begin{figure}[htbp]
                \centering
                \includegraphics[width=0.80\linewidth,clip]{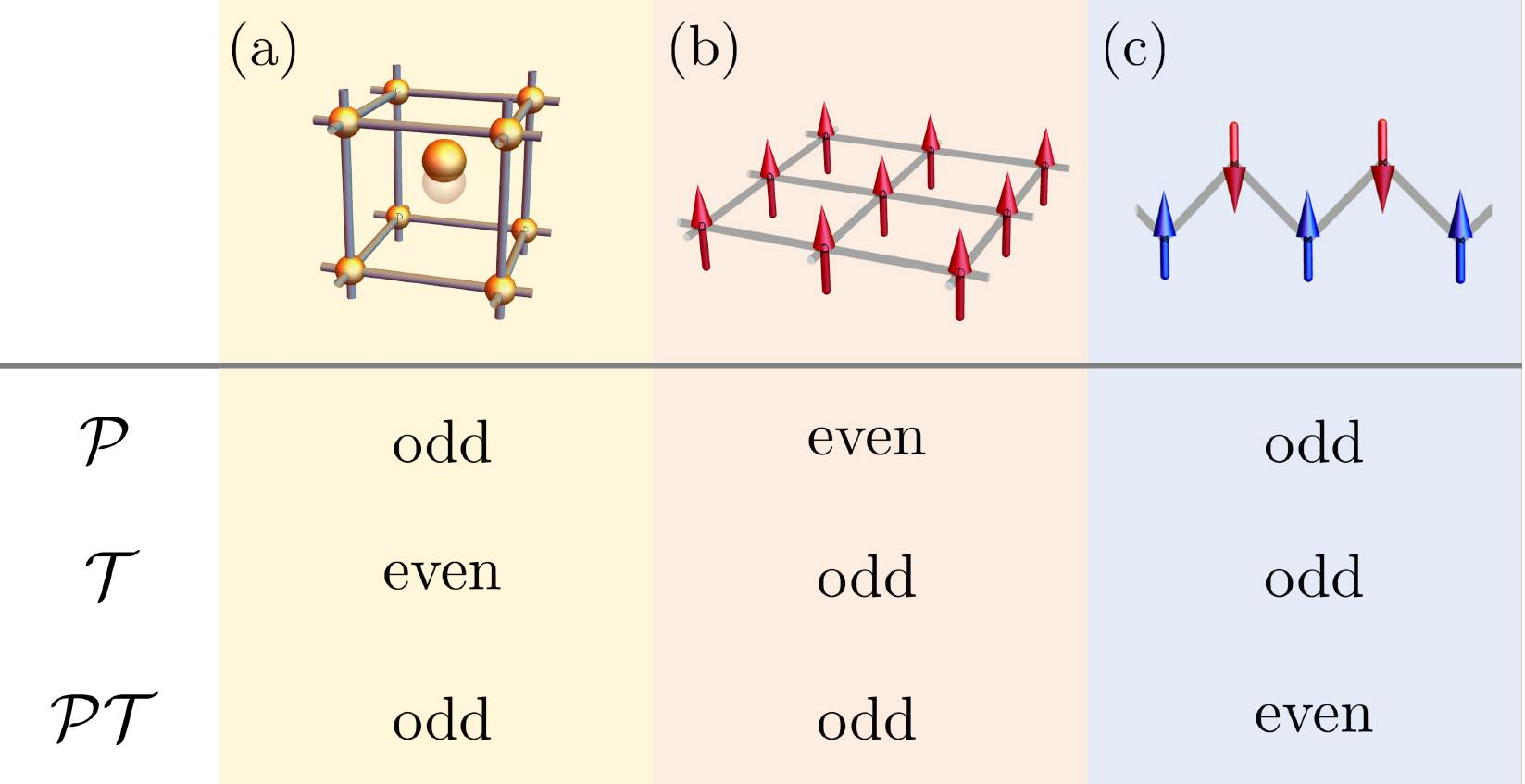}
                \caption{
                        Space-time symmetry of (a) ferroelectric material, (b) ferromagnetic material, and (c) \PT{}-symmetric magnets.
                        The classification is based on the parity under the \Pa{}, \T{}, and \PT{} operations.
                        The electric and magnetic parity violations are classified into the class (a) and (c), respectively.
                }
                \label{Fig_2_space-time-classification}
                \end{figure}

Adding the case of the ferromagnets, we tabulate the space-time symmetry of ordered states in Fig.~\ref{Fig_2_space-time-classification}.
Owing to different space-time symmetry, the order is not admixed with each other unless the preserved symmetry is lost.

Keys to the magnetic parity violation are antiferromagnetism and the crystal structure showing the local parity violation at atomic sites as depicted in Fig.~\ref{Fig_1_AFM_square_zigzag}(b).
The \PT{}-symmetric magnets can be found in a broad range of materials as tabulated in Refs.~\cite{schmid1973magnetoelectric,siratori1992magnetoelectric,Gallego2016-yz,Watanabe2018-cu}.
The structural feature, called \textit{locally-noncentrosymmetric property}, is ascribed to various structural degrees of freedom other than atomic sites; layers, clusters of atoms, chains, and so on (see Sec.~\ref{SecSub_locally_noncentrosymmetric}).
Furthermore, the magnetic parity violation can be realized without the locally-noncentrosymmetric structure by unconventional order such as the loop-current order~\cite{Zhao2015-hu,Seyler2020-yq,Murayama2021-zg,Watanabe2021-yv} and exotic superconductivity~\cite{Wang2017-co,Kanasugi2022-pb,Kitamura2023-ep}.
Interestingly, the magnetic parity violation can be implemented by micro-fabrications as demonstrated in Refs.~\cite{Lehmann2019-xp,Lehmann2020-hi}.

\subsection{Hidden Magnetic Degrees of Freedom in Crystals}
\label{SecSub_locally_noncentrosymmetric}

\begin{figure}[htbp]
                \centering
                \includegraphics[width=0.95\linewidth,clip]{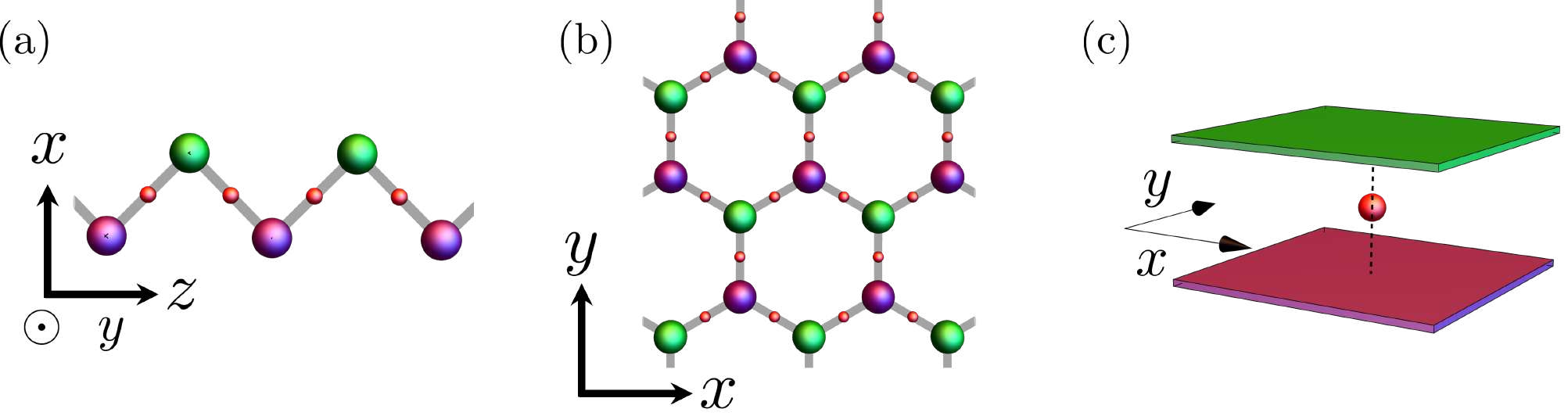}
                \caption{
                        Locally-noncentrosymmetric systems.
                        The objects in green and purple are transformed into each other by the parity operation.
                        Inversion centers are depicted in red.
                        (a) zigzag chain (b) honeycomb net (c) bilayer.
                        Other locally-noncentrosymmetric configurations can be found in Ref.~\cite{Fischer2023-at}. 
                        }
                \label{Fig_locally-noncentrosymmetric}
                \end{figure}

The locally-noncentrosymmetric property can be found in various sectors in crystal structure~\cite{Fischer2023-at}.
The series of locally-noncentrosymmetric crystals consists of the subsector degree of freedom such as atomic site, cluster, chain, and so on (Fig.~\ref{Fig_locally-noncentrosymmetric}).
We here introduce a key ingredient to understand the itinerant property unique to locally-noncentrosymmetric crystals, that is hidden magnetic degrees of freedom such as spin and Berry curvature.

Firstly, we explain spin-charge coupling hidden by the inversion partners in crystals.
The concept has been introduced to predict the \T{}-symmetric counterpart of Chern insulator proposed in Ref.~\cite{Haldane1988-dm}, that is quantum spin Hall insulator~\cite{Kane2005-lv,Fu2007-mo}. 
Independently, the concept has been proposed in the contexts of superconductivity~\cite{Yanase2010-or,Fischer2011-qi}, spintronic applications~\cite{Zelezny2014-qr}, and the first-principles study of spin-momentum coupling~\cite{Zhang2014-jf}.
A prototypical example is the bilayer two-dimensional electron gas which may be found in double quantum wells and layered materials such as the cuprate~\cite{Liu2013-xb} [Fig.~\ref{Fig_locally-noncentrosymmetric}(c)].
Although the inversion center is present between the layers, the \Pa{} symmetry does not hold if each layer is taken to be the origin.
The locally-noncentrosymmetric symmetry indicates that each layer is under unidirectional potential arising from the other layer.
Owing to the global \Pa{} parity, the polarity of the unidirectional field should be opposite between the two layers.

The structural property is built into the peculiar spin-orbit interaction as follows.
The two-dimensional electron gas manifests the so-called Rashba spin-orbit coupling in the presence of the polar field as written by the one-body Hamiltonian
                \begin{equation}
                h_\text{R} = \alpha_\text{R} \left( \bk \times \bm{\sigma} \right)_z = \alpha_\text{R} \left(  k_x \sigma_y - k_y \sigma_y \right),
                \label{Rashba_spin_orbit_coupling}
                \end{equation}
where the polar field is along the $z$ direction, and the electron's momentum $\bk$ and spin $\bm{\sigma}$.
The Rashba spin-orbit coupling gives rise to the vortex-like spin configuration in the momentum space as confirmed in spin-resolved spectroscopy of bulk materials such as BiTeI~\cite{Ishizaka2011-dz,Landolt2012-ju}.
In the case of the bilayer system, the Rashba spin-orbit coupling has the opposite signs as
                \begin{equation}
                \alpha_\text{R}  \left( \bk \times \bm{\sigma} \right)_z,
                \end{equation}
for the upper layer and
                \begin{equation}
                - \alpha_\text{R}  \left( \bk \times \bm{\sigma} \right)_z,
                \end{equation}
for the lower layer.
It indicates that spins are closely coupled to the layer degree of freedom as well as the momentum and that the spin-momentum locking is completely compensated.

\begin{figure}[htbp]
                \centering
                \includegraphics[width=0.80\linewidth,clip]{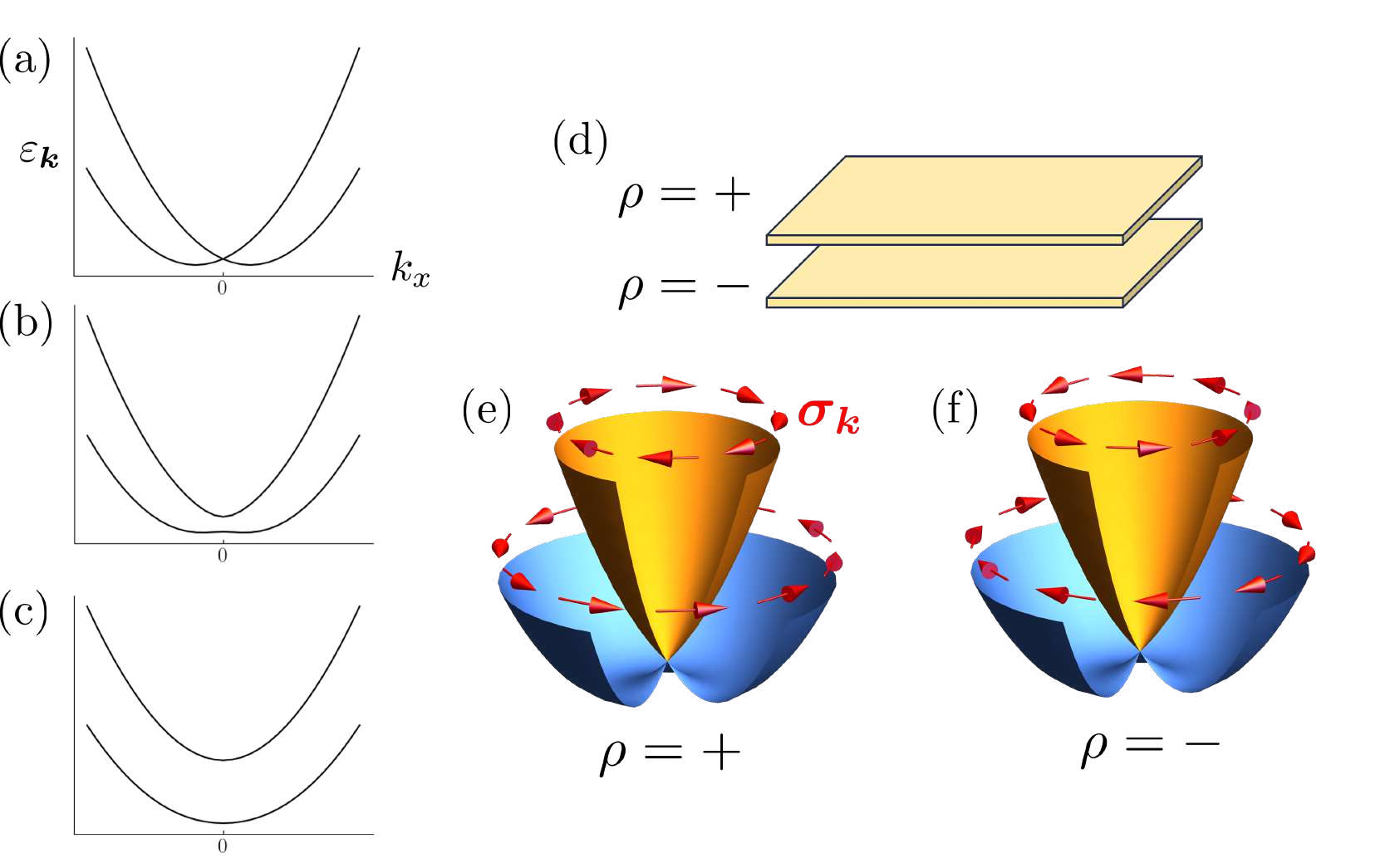}
                \caption{
                        (a,b,c) Energy spectrum of the bilayer two-dimensional electron gas obtained from Eq.~\eqref{Hamiltonian_bilayer}.
                        The spectrum is illustrated along the $k_y =0$ line.
                        Each plot is given by the tunneling parameter as (a) $t=0$, (b) $t=\alpha_\text{R}/2$, and (c) $t \gg \alpha_\text{R}$.
                        Note that each dispersion is doubly degenerate due to the spinful \PT{} symmetry.
                        (d) Bilayer system where the layers are labeled by $\rho = \pm$.
                        (e,f) Energy spectrum for $t=0$. The expectation value of the spin evaluated at each momentum $\bm{\sigma}_\bk$ is depicted by the red-colored arrows for (e) the upper layer and (f) the lower layer.
                        The spin-momentum locking is opposite between the layers.   
                 }
                \label{Fig_4_bilayer_electronic_structure}
                \end{figure}

To elucidate the electronic structure, let us write down a Hamiltonian of the bilayer system in the field-quantization representation as
                \begin{equation}
                H = \sum_{\bk} 
                        \left( {c_{\bk +}}^\dagger, {c_{\bk -}}^\dagger \right)
                        \begin{pmatrix}
                                h_{\bk + } & t\\
                                t^\ast & h_{\bk- }
                        \end{pmatrix}
                        \begin{pmatrix}
                                c_{\bk +}\\ 
                                c_{\bk -} 
                        \end{pmatrix}.
                        \label{Hamiltonian_bilayer}
                \end{equation}
The creation (annihilation) operators ${c_{\bk \rho_z}}^\dagger$ ($c_{\bk, \rho_z}$) are for the upper ($\rho_z=+$) and lower ($\rho_z=-$) layers and implicitly include the spin degree of freedom written by the Pauli matrices $\bm{\sigma}$.
A constituent Hamiltonian reads as
                \begin{equation}
                h_{\bk \rho_z} = \frac{\bk^2}{2m} + \rho_z \, \alpha_\text{R}  \left(  k_x \sigma_y - k_y \sigma_y \right),
                \label{bilayer_bloch_diagonal}
                \end{equation}
for the diagonal components and the tunneling parameter $t$ for the off-diagonal components.
When the constant tunneling parameter $t$ varies, the energy spectrum of Hamiltonian Eq.~\eqref{Hamiltonian_bilayer} changes as schematically depicted in Fig.~\ref{Fig_4_bilayer_electronic_structure}(a,b,c).
It is noteworthy that in the case of no tunneling ($t=0$) one can observe the energy spectrum similar to that with the Rashba spin-orbit coupling nevertheless each spectrum shows double degeneracy due to the \PT{} symmetry [Fig.~\ref{Fig_4_bilayer_electronic_structure}(a)].
This is because the opposite spin-momentum coupling in two layers leads to the vanishing spin polarization at each momentum as
                \begin{equation}
                \sum_\alpha \Braket{\phi_{\bk p \alpha} |\bm{\sigma} | \phi_{\bk p \alpha}} = \bm{0},
                \end{equation}
where $\ket{\phi_{\bk p \alpha}}$ is the eigenstate for energy $\varepsilon_{\bk p}$ of Eq.~\eqref{Hamiltonian_bilayer} denoted by the momentum $\bk$, level index $p$, and Kramers degree of freedom $\alpha$.
On the other hand, there exists the spin-momentum locking in a staggered manner as given by
                \begin{equation}
                \sum_\alpha \Braket{\phi_{\bk p \alpha} | \rho_z \, \bm{\sigma} | \phi_{\bk p \alpha}} \neq  \bm{0}.
                \end{equation}
The obtained subsector-dependent spin-momentum coupling is called hidden spin polarization~\cite{Zhang2014-jf}.
If the hopping $t$ surpasses the spin-orbit coupling quantified by $\alpha_\text{R}$ [Fig.~\ref{Fig_4_bilayer_electronic_structure} (b,c)], the entanglement of wavefunction localized at each layer weakens the hidden spin-polarization~\cite{Maruyama2012-pu}.

The concept of hidden spin polarization is generalized to other locally-noncentrosymmetric crystals~\cite{Kane2005-lv,Fu2007-mo,Yanase2014-mw,Zelezny2014-qr}.
For instance, given that local inversion asymmetry comes from the sublattice degree of freedom as in the zigzag chain and the honeycomb net [Fig.~\ref{Fig_locally-noncentrosymmetric}(b)], the hidden spin polarization is given by the sublattice-dependent spin-orbit coupling 
                \begin{equation}
                h_\text{SOC} = \sum_{\bk} \bm{g}_{\bk} \cdot \bm{\sigma}\,  \rho_z.
                \label{sublattice_dependent_soc}
                \end{equation}
The basis for Pauli matrices $\ket{\rho_z =\pm }$ is spanned by the two sublattice degrees of freedom, and thus $\rho_z$ indicates that the antisymmetric spin-orbit coupling ($\bm{g}_{-\bk} = - \bm{g}_{\bk}$) appears in a staggered manner for electrons localized at each sublattice.
The expression for the vector $\bm{g}_\bk$ is determined by the local site symmetry of the sublattice~\cite{Fischer2023-at,Guan2022-yr} such as
                \begin{equation}
                \bm{g}_\bk = \alpha \sin{k_z} \hat{x},
                \label{zigzag_chain_asoc_gvector}
                \end{equation}
with the coupling constant $\alpha $ given for the next-nearest neighbor hopping path of the zigzag chain [Fig.~\ref{Fig_locally-noncentrosymmetric}(a)], and
                \begin{equation}
                \bm{g}_\bk = \alpha \sin{\frac{k_y}{2}} \left( \cos{\frac{k_y}{2}} - \cos{ \frac{\sqrt{3}}{2}k_x }    \right) \hat{z},
                \label{honeycomb_asoc_gvector}
                \end{equation}
for that of the honeycomb lattice~\cite{Kane2005-lv,Hayami2014-wo} [Fig.~\ref{Fig_locally-noncentrosymmetric}(b)].
The coupling is similarly obtained in more complex cases such as locally-noncentrosymmetric crystals consisting of $n(>2)$ sublattice like Cr$_2$O$_3$~\cite{Sumita2017-cr,Daido2019-iq,Hayami2014-wo, Niu2017}.

After the discovery of the hidden spin polarization, many studies have been devoted to utilizing and maximizing this peculiar charge-spin coupling mostly for the case of local asymmetry of atomic sites.
For instance, since the hidden spin-orbit coupling gives a significant modification of the spin susceptibility, the superconducting state acquires more robustness to the paramagnetic pair breaking~\cite{Maruyama2012-pu} as reviewed in Refs.~\cite{Sigrist2014-gu,Fischer2023-at}.
In addition to the enriched property of nonmagnetic superconductors, the hidden spin-momentum coupling leads to various physical phenomena if coupled to the antiferroic order (see the following sections).

The hidden spin-charge coupling has been quantitatively evaluated for various materials.
The model studies clarified important ingredients for the large sublattice-dependent spin-orbit coupling, that is the atomic spin-orbit coupling and odd-parity hopping allowed by the local parity violation~\cite{Fischer2011-qi,Yanase2014-mw,Hayami2014-wo,Hayami2014-kz}.
It follows that the hidden spin polarization is significant for the bands consisting of orbitals at heavy atoms~\cite{Liu2013-xb,Yao2017-ip,Goh2012-hx}.
On one hand, as implied in Fig.~\ref{Fig_4_bilayer_electronic_structure}(a-c), the inter-sublattice hopping may smear out the hidden spin-charge coupling. 
In this regard, large hidden spin polarization may occur due to negligible inter-sublattice hopping that can be realized in the layered materials such as 2H-stacking transition metal dichalcogenides (\textit{e.g.}, WSe$_2$)~\cite{Gong2013-eg,Riley2014-fm,Jones2014-zq,Gehlmann2016-op,Bertoni2016-yg,Razzoli2017-do,Devarakonda2020-sk,Tu2020-gb} and layered superconductor~\cite{Liu2013-xb,Liu2015-tz,Nakamura2017-eg,Wu2017-jg,Gotlieb2018-mp}.
The layered materials are promising platforms for manipulating the spin-orbit coupling due to the high controllability offered by gating fields and the epitaxial-growth method.~\cite{Gong2013-eg,Goh2012-hx,Shimozawa_2016}.
Interestingly, nonsymmorphic space group symmetry may realize the segregation between sublattice degrees of freedom in the high-symmetry subspace of the Brillouin zone~\cite{Young2015-qd} and therefore leads to enhanced hidden spin polarization.
The hidden spin polarization protected by the nonsymmorphic symmetry has been pointed out by theories~\cite{Yanase_UPt3Weyl,Slawinska2016-qk,Yuan2019-vk} and demonstrated in experiments~\cite{Santos-Cottin2016-ix,Zhang2021-sr}.
In the two-sublattice Hamiltonian Eq.~\eqref{Hamiltonian_bilayer}, the inter-sublattice decoupling is represented by vanishing hopping $t$ at high-symmetry momentum.
Such decoupling can also be protected by symmorphic symmetries, depending on the symmetry of the local orbitals~\cite{Akashi2017-xw,Nakamura2017-eg}.

\begin{figure}[htbp]
                \centering
                \includegraphics[width=0.95\linewidth,clip]{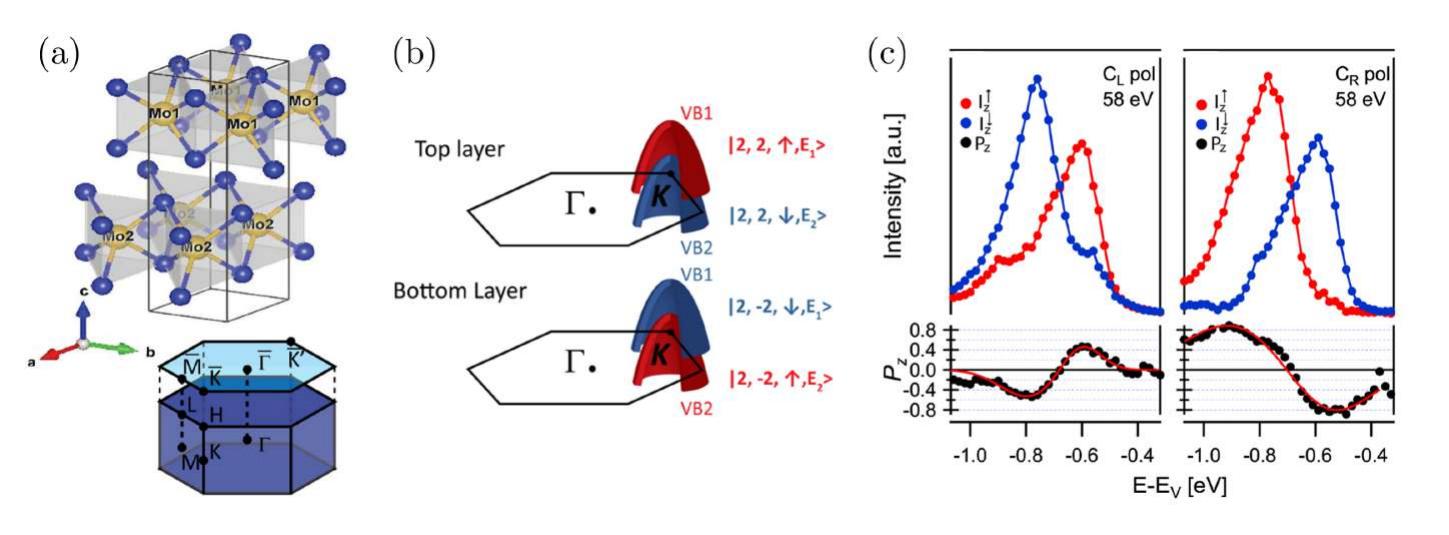}
                \caption{
                        Spin-resolved ARPES study of the bulk MoS$_2$.
                        (a) Crystal structure (upper panel) and its bulk and surface Brillouin zones (lower panel).
                        (b) Layer-resolved electronic structure of $2H$-stacked MoS$_2$.
                        Owing to the locally-noncentrosymmetric structure, the layer, momentum, spin, and valley degrees of freedom are intimately coupled to each other; \textit{e.g.} at the $K$ valley, valence bands manifest significant spin splitting in a staggered manner between the top and bottom layers.
                        (c) Energy distribution curves from the spin-resolved ARPES measurement at the $\bar{K}$ point with the left-handed ($C_L$) / right-handed ($C_R$) circularly-polarized lights.
                        The sizable and staggered out-of-plane spin polarization of two valence bands is consistent with the Zeeman-Ising type spin-orbit coupling of Eq.~\eqref{honeycomb_asoc_gvector} and coupled to the circularly-polarized light via the valley-selective excitation. 
                        Panels are taken from \cite{Razzoli2017-do} (\copyright American Physical Society). 
                        }
                \label{Fig_hidden-spin-polarization}
                \end{figure}

Since the electronic bands of layered materials are described in a spin- and layer-resolved manner, the spin-resolved ARPES study may allow us to measure the hidden spin polarization if there is no complete compensation in the spin polarization between the photoelectrons emitted by the scattering at each layer~\cite{Zhang2014-jf,Riley2014-fm}.
For example, the intimate coupling between the spin, valley, and layer degrees of freedom in van der Waals materials has been reported by Ref.~\cite{Razzoli2017-do} (Fig.~\ref{Fig_hidden-spin-polarization}).
Such an interplay between various degrees of freedom has been investigated by photoluminescence measurements as well~\cite{Wu2013-vn,Zhu2014-qr,Jones2014-zq}.

In the preceding discussions, we considered the electrical activity of spins hidden by the locally noncentrosymmetric structure.
It is noteworthy that other \T{}-odd quantities are coupled to the momentum in a similar manner to spins~\cite{Lin2020-gy}.
For instance, the celebrated Kane-Mele model shows the hidden Berry curvature offering the quantum spin Hall effect~\cite{Kane2005-lv}. 
The hidden magnetic properties have been explored in terms of the orbital magnetic moment and Berry curvature~\cite{Go2018-ke,Cho2018-dj,Beaulieu2020-es}.

\subsection{Multipolar degree of freedom of visible antiferromagnets}
\label{SecSub_multipolar_classification}

The locally-noncentrosymmetric symmetry can be incorporated into the model studies by using the hidden spin-orbit coupling introduced in Sec.~\ref{SecSub_locally_noncentrosymmetric}. 
This peculiar spin-charge coupling seemingly may not dramatically affect the physical property due to the fact that the symmetry-breaking effect is compensated between subsectors.
On the other hand, once the antiferroic ordering such as anti-ferroelectricity and antiferromagnetism occurs, staggered order parameters may give rise to macroscopic symmetry breaking.
As a result, the emergent responses arise from the coupling between the hidden spin-charge coupling and `antiferroic' order.
It is convenient for clarifying the macroscopic physical properties to introduce the multipolar degrees of freedom~\cite{Watanabe2018-cu,Hayami2018-bh,Winkler2023-ju, Watanabe2017-qk}.

                \begin{figure}[htbp]
                \centering
                \includegraphics[width=0.85\linewidth,clip]{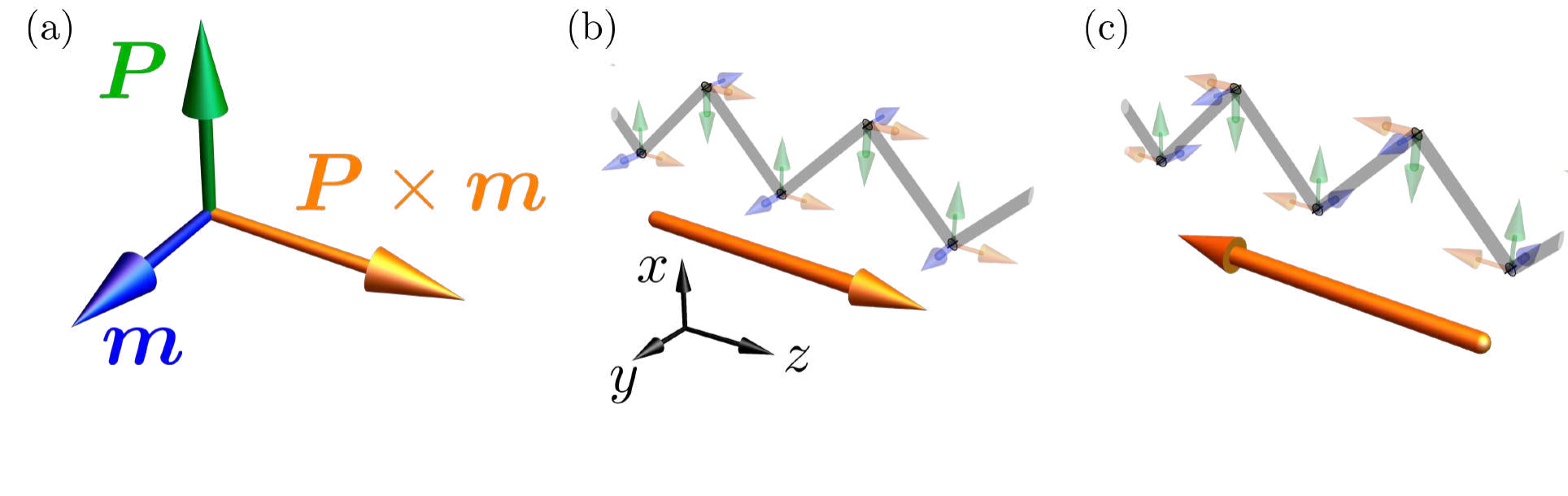}
                \caption{
                        (a) Atomic toroidal moment $\bm{P} \times \bm{m}$ defined by the local polar field $\bm{P}$ and local magnetic moment $\bm{m}$.
                        (b,c) Uniformly aligned toroidal moments in the antiferromagnetic zigzag chain. The atomic toroidal moments are in the (b) $+z$ direction and (c) $-z$ direction depending on the antiferromagnetic pattern.  
                        }
                \label{Fig_uniform_toroidal}
                \end{figure}

Let us again consider the antiferromagnetic state of the zigzag chain where magnetic moments are ordered along the $y$ direction as an example.
Note that the N\'eel vector is taken to be along the $y$ axis different from that depicted in Fig.~\ref{Fig_1_AFM_square_zigzag}(c).
As implied by the form of spin-orbit coupling in Eq.~\eqref{zigzag_chain_asoc_gvector}, each sublattice is under the polar field along the $x$ direction in a staggered manner.
By taking the cross product of the local polar fields $\bm{P}$ and magnetic moments $\bm{m}$, the products are the same between the sublattices A and B  
                \begin{equation}
                        \bm{P}_\text{A} \times \bm{m}_\text{A} = \bm{P}_\text{B} \times \bm{m}_\text{B} \parallel \hat{z},
                \end{equation}
which are termed with an atomic toroidal moment [Fig.~\ref{Fig_uniform_toroidal}(a)].
Thus, the antiferromagnetic state is translated into the ferroic arrangement of atomic toroidal moments, which is in agreement with the zero propagation vector indicating the uniformly manifesting physical quantity.
As depicted in Fig.~\ref{Fig_uniform_toroidal}(b,c), the atomic toroidal moments uniformly align along the $z$ direction, and its direction is opposite between the two antiferromagnetic states. 
Given that the toroidal moment is the polar vector with the \T{}-odd and \PT{}-even parities, one can see that the obtained ferro-toroidal state shows the magnetic parity violation.
Similarly, if the zigzag chain undergoes a $\bm{q}=\bm{0}$ antiferroic ordering of nonmagnetic even-parity objects such as commensurate charge density wave and orbital order, the order results in the ferroic alignment of odd-parity and \T{}-preserving quantities such as electric polarization.
It indicates that the ordered state shows the electric parity violation~\cite{Hitomi2014-is,Hitomi2016-vy}.

These examples imply that the antiferroic order in the locally noncentrosymmetric crystals is a promising playground for the spontaneous parity violation in solids and that the ordered state may be classified in terms of the electric or magnetic anisotropy showing the \Pa{}-odd parity, that is the odd-parity electric or magnetic multipolar field~\cite{Kusunose2008-lh,Kuramoto2009-sn}.
Similar arguments are obtained for the even-parity magnetic multipolar order without uniform magnetization such as collinear magnetic order in a rutile-type magnet, MnF$_2$, and noncollinear magnetic order in Mn$_3$\textit{X} ($X =$ Ga, Ge, Sn, Ir, and Pt)~\cite{Nakatsuji2015-iq,Nayak2016-uo,Smejkal2020-yt}.
In those cases, the seemingly antiferroic order is characterized by $\bm{q} = \bm{0}$ and classified into an even-parity multipolar phase, whereas the local parity violation is not required because the ferroic even-parity multipolar order does not break the \Pa{} symmetry. 
It is noteworthy that the importance of the magnetic octupole moment has been confirmed in Mn$_3$Sn and Mn$_3$Ge~\cite{Suzuki2017-ps,Nomoto2020-dr,Kimata2021-ju,Higo2022-ig,Go2022-xw,Yoon2023-ht}.

Physical consequences of the odd-parity multipolar order can be derived by group-theoretical tools once the symmetry of the phase is identified.
Let us raise some examples of odd-parity multipoles and associated physical properties from the viewpoint of the representation analysis below. 
Notably, the odd-parity multipolar symmetry can emerge not only due to the antiferroic order in the locally-noncentrosymmetric systems but also by other exotic quantum phases such as the loop-current order~\cite{Zhao2015-hu,Murayama2021-zg}.
We emphasize that the classification result obtained by group theory can be applied to any system showing electric and magnetic parity violations.

The uniformly emerging order parameter characterizing the (second-order) phase transition is classified by irreducible representations of a given crystallographic point group.
For instance, when the ordered phase belongs to the $B_{1u}$ representation of the tetragonal point group $4/mmm$ ($D_{4h}$), odd-parity multipole moments appear in the ferroic manner.
When the $B_{1u}$-type phase transition is attributed to multipolar degrees of freedom, candidates for the order parameter are the electric octupole moment
                \begin{equation}
                Q_{31}^+ = xyz,
                \label{B1u_electric_multipole}
                \end{equation}
for the electric parity violation and the magnetic quadrupole moment 
                \begin{equation}
                M_{22}^+ = x m_x - y m_y,
                \label{B1u_magnetic_multipole}
                \end{equation}
for the magnetic parity violation~\cite{Watanabe2017-qk}.
These multipole moments indicate the lowest order of electric or magnetic anisotropy in real space.
Since these multipole moments show the opposite parity under the \T{} or \PT{} operation, the corresponding irreducible representation should be labeled by the \T{}-parity such as $B_{1u}^+$ for $Q_{31}^+$ and $B_{1u}^-$ for $M_{22}^+$; \textit{i.e.,} the odd-parity irreducible representation $\Gamma_u$ is \T{}-even and \PT{}-odd for $\Gamma_{u}^+$, or \T{}-odd and \PT{}-even for $\Gamma_u^-$ (subscript `u' denotes the odd \Pa{} parity).~\footnote{
        We implicitly assume that the para phase is \T{}-symmetric and \Pa{}-symmetric to characterize the multipolar order by the definite parity of each operation.
        Generalized representation analysis is similarly obtained by making use of the magnetic point group~\cite{Erb2020-jb}.
}

As in the classification of order parameters in terms of the real-space basis [Eqs.~\eqref{B1u_electric_multipole} and \eqref{B1u_magnetic_multipole}], the relevant basis functions in the momentum space ($\bk$) are identified in the group-theoretical manner~\cite{Sigrist1991-bd}.
Referring to the classification presented in Refs.~\cite{Watanabe2017-qk,Watanabe2018-cu,Hayami2018-bh}, we can construct a basis function formed by the momentum $\bk$ and magnetic moment $\bm{m}$ as
                \begin{equation}
                k_x m_x - k_y m_y,
                \label{B1u_electric_momentum_basis}
                \end{equation}
for the $B_{1u}^+$ representaion and 
                \begin{equation}
                k_x k_y k_z,
                \label{B1u_magnetic_momentum_basis}
                \end{equation}
for the $B_{1u}^-$ representation.
One may notice that the momentum-space basis for the electric parity violation of Eq.~\eqref{B1u_electric_momentum_basis} is obtained by replacing $\bp$ with $\bk$ in the real-space basis for the magnetic parity violation [Eq.~\eqref{B1u_magnetic_multipole}] and \textit{vice versa}.
The cross-correlation is a consequence of the intact symmetry, that is \T{} or \PT{} symmetry.

                \begin{figure}[htbp]
                \centering
                \includegraphics[width=0.75\linewidth,clip]{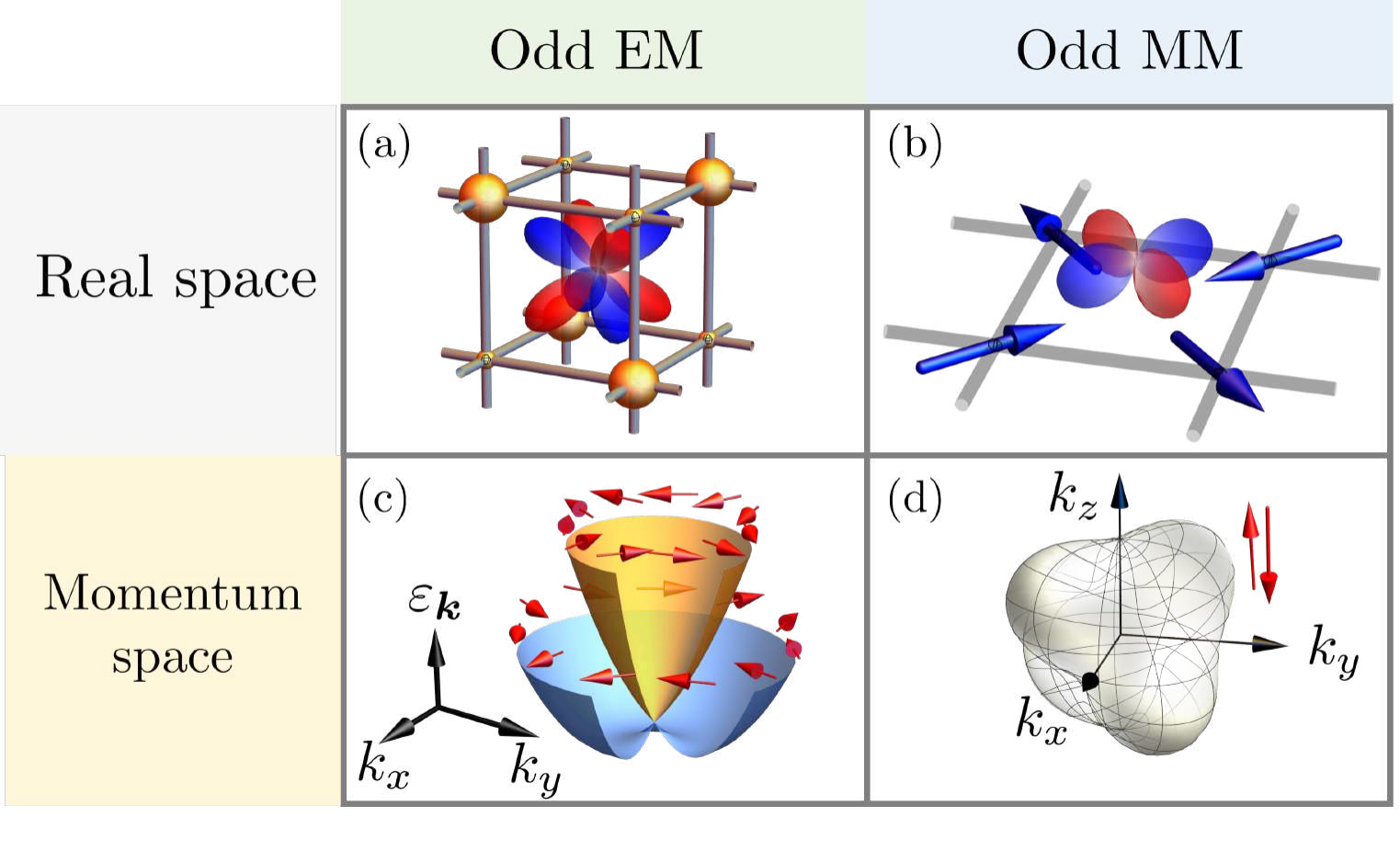}
                \caption{
                        The real- and momentum-space structure of the odd-parity multipolar states.
                        Corresponding to the odd-parity multipolar state depicted in (a,b), the energy spectrum shows the characteristic structure shown in (c,d). 
                        (a) electric octupolar state for an odd-parity electric multipolar (odd EM) state.
                        (b) magnetic quadrupolar state for an odd-parity magnetic multipolar (odd MM) state.
                        (c) Rashba-like spin splitting of the electronic band structure induced by the electric octupolar order. The split bands show the spin-momentum locking in the opposite way (red-colored arrows denote spins coupled to each momentum).
                        (d) Fermi surface undergoing asymmetric and octupolar modulation due to the magnetic quadrupolar order.
                        Note that the Fermi surface is doubly degenerate due to the \PT{} symmetry and that there is no spin splitting as depicted by the spins (red-colored vectors) antiparallel to each other.  
                        }
                \label{Fig_real-momentum-structure-oddparitymultipole}
                \end{figure}

The correspondence between the real- and momentum-space basis functions conveniently gives clues to understanding the itinerant property of the parity-violating materials such as changes in the electronic structure and emergent responses.
Leaving discussions of the physical responses to the following sections, let us consider the band structure undergoing the $B_{1u}$-type odd-parity multipolar order for an example.
In case of the electric parity violation, the odd-parity electric multipolar anisotropy of Eq.~\eqref{B1u_electric_multipole} corresponds to spontaneous emergence of the spin-momentum locking of Eq.~\eqref{B1u_electric_momentum_basis} whose form is similar to that of the Rashba spin-orbit coupling of Eq.~\eqref{Rashba_spin_orbit_coupling}.
Contrastingly, in the case of the magnetic parity violation, the magnetic multipolar anisotropy of Eq.~\eqref{B1u_magnetic_multipole} gives rise to the spin-independent modulation of electronic bands of Eq.~\eqref{B1u_magnetic_momentum_basis}.
The asymmetric energy spectrum has been observed in experiments of a \PT{}-symmetric magnet~\cite{Fedchenko2022-pi,Lytvynenko2023-vf}.
These characteristic changes in the band structure are strictly forbidden if the \T{} or \PT{} symmetry is present; the asymmetric modulation such as Eq.~\eqref{B1u_magnetic_momentum_basis} is forbidden due to the \T{} symmetry, while the spin-momentum locking as in Eq.~\eqref{B1u_electric_momentum_basis} is absent in total if the \PT{} symmetry is retained.
The modified band structures arising from electric and magnetic parity violation are summarized in Fig.~\ref{Fig_real-momentum-structure-oddparitymultipole}.

The microscopic grounds for the multipole moments in solids remain to be completed, though the macroscopic aspects have been addressed from the viewpoint of point group symmetry.
The origin of the multipole moments may be attributed to the atomic orbitals well-localized at rare-earth atoms in some f-electron systems~\cite{Kuramoto2009-sn}.
The quantitative estimates of the multipole moments have been explored by first-principles calculations~\cite{Cricchio2010-fq,Bultmark2009-zo,Suzuki2018-me}; \textit{e.g.}, the close relationship between the magneto-electric coupling and the odd-parity magnetic multipole moments has been pointed out~\cite{Spaldin2013-tt,Thole2016-zk,Thole2020-yq}.

Furthermore, recent theoretical studies revisited the connection between the observable physical property and multipolar degrees of freedom in the context of generalized free energy. 
For instance, the thermodynamic magnetic quadrupole moments are defined to be conjugate to the gradient of magnetic fields, and they are directly related to the magnetoelectric property~\cite{Gao2018-xt,Shitade2018-vh,Bhowal2022-ay}.
It is noteworthy that the thermodynamic multipole moments contribute to other cross-correlated responses such as the nonlinear thermoelectric response~\cite{Gao2018-gl}, temperature-gradient-induced magnetization (gravito-magnetoelectric effect)~\cite{Shitade2019-xn,Shinada2023orb_gravitoME}, and spin accumulations~\cite{Shitade2022-zb}.
In the same spirit, the electric quadrupole moments are defined and estimated in the language of thermodynamics and quantum geometry~\cite{Daido2020-EQ,kitamura2021-EQ}.
It is desirable to perform further explorations for higher-order multipolar degrees of freedom in solids~\cite{Tahir2023-ti}.

Finally, we comment on material realizations of the \PT{}-symmetric (odd-parity magnetic multipolar) magnets.
Although the uniform odd-parity magnetic multipolar fields show up in the presence of special crystal and antiferromagnetic structures, one can notice that they exist in a broad range of materials as we tabulate candidates in Appendix~\ref{SecApp-magnetic_materials}.
Historically, enormous efforts have been made to explore materials undergoing the \PT{}-symmetric magnetic order, with special attention to the oxides motivated by the interests in the magnetoelectric coupling in magnetic insulators such as Cr$_2$O$_3$ and LiCoPO$_4$~\cite{Fiebig2005-hj}.
On the other hand, as mentioned in the introductory part, recent studies shed light on magnetic metals as well with the developments in the field of antiferromagnetic spintronics~\cite{Jungwirth2016-gj,Baltz2018-pu}.
The candidates cover a diverse class of materials such as ferro-pnictide (\textit{e.g.}, BaMn$_2$As$_2$, EuMnBi$_2$, CuMnAs, YbMnBi$_2$)~\cite{Borisenko2019-ke,Wadley2015-ra,Sakai2022-dq,Tanida2022-fm} and rare-earth-based magnetic conductors~\cite{Saito2018-hr,Arakawa2023-nx,Ota2022-bh}.
It is expected that the magnetic parity violation may show intriguing interplays with the physical properties unique to the metals like topological electrons nearby the Fermi energy~\cite{Tang2016-au,Smejkal2018-ks}, giant magnetoresistance~\cite{Aoyama2022-gg,Sun2021-fe,Ogasawara2021-yb}, and what is discussed in the following sections (\textit{e.g.}, magnetopiezoelectric effect and nonreciprocal electrical transport induced by the magnetic parity violation).

\section{Emergent Responses of \PT{}-symmetric magnets}
\label{Sec_emergent_response}

\subsection{Electric-Magnetic Classification of Response Function}
\label{SecSub_EM_decomposition}

The contrasting space-time symmetry of the odd-parity electric/magnetic multipolar materials is reflected in the response as well as the electronic structure described in Sec.~\ref{SecSub_multipolar_classification}.
In this section, taking the linear response function, we introduce the \T{}-odd/\T{}-even classification of physical responses and argue that the \T{} and \PT{} symmetries play complementary roles.

The response formula is generally given by
        \begin{equation}
        X_i  = \chi_{ij}^{XY} F_j^Y,
        \label{linear_response_function}
        \end{equation}
where the physical quantity $\bm{X}$ responds to the external field $\bm{F}^Y$ conjugate to a physical quantity $\bm{Y}$ by the perturbed Hamiltonian $H_\text{ex} = \bm{Y} \cdot \bm{F}^Y $.
Based on the Kubo formula, we derive the ac response function $\chi_{ij}^{XY} (\omega)$ in the Lehmann representation
    \begin{equation}
        \chi_{ij}^{XY} (\omega) =  \sum_{a,b} \frac{\rho_a - \rho_b}{\omega +i \eta + \epsilon_a - \epsilon_b} X_{ab}^i Y_{ba}^j,
    \end{equation}
with the summation over the many-body eigenstates $(a,b)$ for the unperturbed Hamiltonian.
We introduced the eigenenergy $\epsilon_a$, Boltzmann factor $\rho_a = \exp{(-\epsilon_a /T)} /Z$ ($Z$: partition function), and infinitesimal positive parameter $\eta =+0$.
Physical quantities are given by the matrix element for the eigenstates 
$X_{ab}^i = \Braket{a | X_i | b}$.
The response function is classified by the parity $\tau_g$ under a symmetry operation $g$ such as \Pa{}, \T{}, and \PT{} operations.
In the case of magnetization response to the electric field [$(\bm{X},\bm{Y}) = (\bm{M},\bm{E})$] known as the magnetoelectric effect, the response function $\hat{\chi}^{ME}$ has odd- ($\tau_{I} = -1 $), odd- ($\tau_{\theta} = -1 $), and even-parity ($\tau_{I\theta} = +1 $) under the \Pa{}, \T{}, and \PT{} operations, respectively.
Note that the parity satisfies the relation $\tau_{I\theta} = \tau_I \cdot \tau_{\theta}$.

The response function is further divided into the symmetric and antisymmetric parts under the permutation of $X_i$ and $Y_j$ as
                \begin{equation}
                        \chi_{ij}^{XY} = \chi_{ij}^{XY;\text{s}} + \chi_{ij}^{XY;\text{a}}.
                \end{equation}
The components are respectively given by
        \begin{equation}
            \chi_{ij}^{XY;\text{a}} (\omega) = \sum_{a,b} \left( \rho_a - \rho_b \right)\frac{\omega +i \eta}{\left( \omega +i \eta \right)^2  - \left( \epsilon_a - \epsilon_b \right)^2}  X_{ab}^i Y_{ba}^j,
            \label{antisymmetric_response}
        \end{equation}
and
        \begin{equation}
            \chi_{ij}^{XY;\text{s}}(\omega) = \sum_{a,b} \left( \rho_a - \rho_b \right) \frac{\epsilon_b - \epsilon_a}{\left( \omega +i \eta \right)^2  - \left( \epsilon_a - \epsilon_b \right)^2}  X_{ab}^i Y_{ba}^j.
            \label{symmetric_response}
        \end{equation}
We may gain an intuitive understanding of the symmetric and antisymmetric terms by considering the dc limit ($\omega \to 0$).
By replacing $\eta$ with the phenomenological scattering effect $\gamma >0$, each contribution is given as 
        \begin{equation}
            \hat{\chi}^{XY;\text{a}} \sim \frac{ \gamma}{\gamma^2 + \left( \epsilon_a - \epsilon_b \right)^2},~~~\hat{\chi}^{XY;\text{s}} \sim \frac{ \epsilon_a - \epsilon_b}{\gamma^2 + \left( \epsilon_a - \epsilon_b \right)^2}.
            \label{dc_symmetric_antisymmetric_with_scattering}
        \end{equation}
The antisymmetric contribution explicitly depends on the sign of $\gamma$ and is proportional to $\gamma^{-1}$ for the equi-energy transitions ($\epsilon_a = \epsilon_b$).
As a result, the response is in an intimate relation with the dissipative phenomenon accompanied by energy absorption.
On the other hand, the symmetric term does depend on $|\gamma|$ and remains finite even in the limit of $\gamma \to +0$.
This property indicates that the symmetric term is generically irrelevant to the scattering process and may occur without dissipation. 
In the light of these contrasting aspects with respect to dissipation, the antisymmetric and symmetric terms are also called dissipative and dissipation-less responses, respectively~\cite{Freimuth2014-zz,Zelezny2017-ov,Watanabe2017-qk}.

Following the symmetry arguments in Ref.~\cite{Watanabe2024-ju}, we finally obtain the \T{}-parity decomposition of the antisymmetric and symmetric responses in Eqs.~\eqref{antisymmetric_response} and \eqref{symmetric_response}.
We here introduce the \textit{\T{}-even} and \textit{\T{}-odd} contributions which appear with and without \T{}-symmetry, respectively.
As displayed in Table~\ref{Table_EM_decomposition_response_function}, the symmetric and antisymmetric parts of a given response function are classified by the \T{}-parity $\tau_\theta$; for a response with $\tau_\theta =+1$, only the symmetric part is allowed in the \T{}-symmetric system, while the antisymmetric contribution is admixed by the \T{}-symmetry breaking.
On the other hand, for the case of $\tau_\theta = -1$, the response in the \T{}-symmetric system is solely attributed to the antisymmetric part, but it contains both antisymmetric and symmetric contributions when the \T{} symmetry is lost.
For instance, the dc electric conductivity tensor ($J_a = \sigma_{ab} E_b$) formally shows the odd-parity under the \T{} operation, and thus $\tau_\theta = -1$.
Therefore, only the antisymmetric part is finite in the \T{}-symmetric system, consistent with the fact that the allowed longitudinal response is dissipative as formulated in the Boltzmann's semiclassical theory of transport phenomena. 
The symmetric part is relevant to the \T{}-symmetry breaking.
This argument agrees with the fact that the electric conductivity tensor hosts the symmetric part such as the Hall conductivity which can generically be free from dissipation.~\footnote{The symmetric and antisymmetric terms of the electric conductivity do not indicate the tensor symmetry of $\sigma_{ab}$; for instance, the antisymmetric term we introduced is not the Hall response ($\sigma_{ab} = -\sigma_{ba}$) but the dissipative and longitudinal conductivity ($\sigma_{ab} = \sigma_{ba}$).
This is because we relate the electric conductivity with the inverse response written by $P_a = \chi_{ab} A_b$ ($\bm{P}$ is the electric polarization, $\bm{A}$ is the vector potential).
The antisymmetric term therefore satisfies $\sigma_{ab} = - \chi_{ba}$.
The well-known Onsager relation is reproduced by using the relations $\bm{J} = -i\omega \bm{P}$ and $\bm{E} = i\omega \bm{A}$.
Similarly, one can derive the symmetric-antisymmetric decomposition by using the linear response formula in the canonical-correlation representation, by which the Onsager reciprocity is obtained more explicitly~\cite{Watanabe2024-ju}. 
}

Similarly, the response function with $\tau_\theta = +1$ is decomposed into the \T{}-even and \T{}-odd contributions.
An example is the nonmagnetic and magnetic spin Hall effects which are of high interest in recent studies on spintronic effects.
The spin-polarized current response to the electric fields is given by the formula
        \begin{equation}
        J_a^{s_c} = \sigma_{ab}^c E_b,
        \end{equation}
where $\bm{J}^{s_c}$ denotes the current with the spin polarization along the $c$ direction.
Specifically, the spin Hall conductivity is defined by the off-diagonal elements $ \sigma_{ab}^c + \sigma_{ba}^c \neq 0  $.
The \T{}-even component can make contributions through the symmetric part of the tensor [Eq.~\eqref{symmetric_response}], consistent with the dissipationless nature of the spin-Hall effect, as demonstrated in intensive studies of the spin Hall effect of nonmagnetic semiconductors~\cite{Sinova2015-uq}.
On the other hand, the \T{}-symmetry violation gives rise to the antisymmetric counterpart [Eq.~\eqref{antisymmetric_response}] of spin Hall response called magnetic spin Hall effect~\cite{Kimata2019-xe}.
Since its dissipative aspect is closely linked with the metallic conductivity, the magnetic spin Hall effect is characteristic of magnetic (\T{}-odd) metals~\cite{Zelezny2017-tf,Mook2020-ae}.

        \begin{table}[htbp]
        \caption{
                \T{}-parity classification of the linear response tensor.
                For a given response with the time-reversal parity $\tau_\theta = \pm 1$, its antisymmetric and symmetric parts are attributed to the \T{}-even and \T{}-odd contributions.
                }
        \label{Table_EM_decomposition_response_function}
        \centering
        \renewcommand{\arraystretch}{2.0}
        \begin{tabular}{ccc}
         \hline
         &  Symmetric $\hat{\chi}^\text{s}$ & Antisymmetric $\hat{\chi}^\text{a}$\\
         \hline
         $\tau_\theta: +1$ & \T{} even & \T{} odd\\
         $\tau_\theta: -1$ & \T{} odd & \T{} even\\
         \hline
        \end{tabular}
        \end{table}

It is noteworthy that either \T{}-even or \T{}-odd contribution may appear without being concomitant with the other due to the additional symmetry constraint.
For example, the magnetoelectric effect is \Pa{}-odd ($\tau_I = -1$) and can occur in parity-violating materials.
The electric parity violation allows the system to have the antisymmetric part of $\hat{\chi}^{ME;\text{A}}$ because of the preserved \T{} symmetry.
This response is accompanied by dissipation and is called the inverse magnetogalvanic effect (Edelstein effect).
On the other hand, those with the pure magnetic parity violation manifest only the symmetric counterpart not being admixed with the antisymmetric contribution, because the magnetoelectric effect is incompatible with the space-time symmetry of magnetic parity violation due to the \PT{}-odd nature.
Consequently, the \T{}- and \PT{}-symmetries play complementary roles in \Pa{}-odd physical responses due to the \Pa{}-symmetry constraint.
The symmetry argument is applicable to the nonlinear responses as well as the linear response, as we see later in Sec.~\ref{Sec_nonreciprocal_electric_phenomena}.

\subsection{Magnetic Counterpart of Piezoelectric Effect}
\label{SecSub_MPE}

The piezoelectric effect is the interconversion between the stress and electric polarization allowed in noncentrosymmetric materials.
The effect has been widely used for applications such as the transducer.
The response formula reads as
        \begin{equation}
        P_a = \bar{e}_{abc} s_{bc},~~~ \varepsilon_{ab} = e_{abc} E_c,
        \label{piezoelectric_repsonse}
        \end{equation}
with electric polarization $\bm{P}$, stress $\hat{s}$, and strain $\hat{\varepsilon}$.
Typically, the piezoelectric-active materials are limited to insulators such as ferroelectric ceramics to hold electric polarization.
By taking $\bm{X} = \bm{P}$ and $\bm{F}^{\bm{Y}}= \hat{s}$ in Eq.~\eqref{linear_response_function}, the time-reversal parity $\tau_{\theta} = +1$ indicates that the \T{}-even part is the symmetric part (see Table~\ref{Table_EM_decomposition_response_function}) and may be attributed to the equilibrium property in the DC limit, being consistent with the piezoelectric effect.
On the other hand, the \T{}-parity classification in Table~\ref{Table_EM_decomposition_response_function} also makes us aware of the possibility of the \T{}-odd and antisymmetric contribution to the electric-elastic coupling.
Notably, owing to the \Pa{}-odd property of the response ($\tau_I =-1$), the \T{}-odd contribution is allowed without being admixed with the conventional piezoelectric effect in the \PT{}-symmetric systems.

Let us take the response function given by $\bm{X} = \hat{\varepsilon}$ and $\bm{F}^{Y} = \bm{E}$ 
and corroborate the DC limit.
Owing to the \PT{}-symmetry constraint, the surviving term is purely the antisymmetric part as
        \begin{equation}
        e^\text{m}_{abc} \equiv \left( \hat{\chi}^{\varepsilon P; \text{a}} \right)_{abc} = \sum_{p,q}  \left( \rho_p - \rho_q \right)\frac{-i \gamma}{\gamma^2  + \left( \epsilon_p - \epsilon_q \right)^2}  \varepsilon_{pq}^{ab} P_{qp}^c.
        \label{MPE_Lehmann_representation}
        \end{equation}
We introduced the phenomenological scattering effect parametrized by $\gamma$ similarly to Eq.~\eqref{dc_symmetric_antisymmetric_with_scattering}.
In the clean limit ($\gamma \to +0$), the response is dominated by the equi-energy transition process ($\epsilon_a = \epsilon_b $) giving rise to the response as much as $O(\gamma^{-1})$.

In the independent particle approximation, the formula in the clean limit is recast as~\cite{Watanabe2017-qk}
                \begin{equation}
                e^\text{m}_{abc} = - \frac{e}{\gamma}  \int \frac{d\bk}{(2\pi)^d} \sum_p \varepsilon_{pp}^{ab} v_{pp}^c \frac{\partial f (\varepsilon)}{\partial \varepsilon}\bigg|_{\varepsilon = \varepsilon_{\bk p}},
                \label{MPE_band_electron}
                \end{equation}
where $e>0$ is the elementary charge of electrons and the summation is over the momentum ($\bk$) and the band indices ($p$).
We introduced the velocity operator $\bm{v}$ for electrons and the Fermi-Dirac distribution function $f (\varepsilon)$.
The strain and velocity operators are evaluated by the Bloch states $\ket{\psi_{\bk p}}$ having the eigenvalue $\varepsilon_{\bk p}$ for the Hamiltonian.
The Fermi-surface factor $\partial_\varepsilon f (\varepsilon)$ indicates that metallic conductivity is required for this piezoelectric-like response in sharp contrast to the conventional piezoelectric response allowed even in insulators. 

This unconventional type of ``piezoelectricity'' is termed as \textit{magnetopiezoelectric effect} and was predicted by theories with a semiclassical theory including the quantum-geometrical effect~\cite{Varjas2016-sw} and with full-quantum treatment based on the linear response theory~\cite{Watanabe2017-qk}.
When the stimulus $E_z$ is replaced with the electric current $J_z$, the response formula is rewritten by
                \begin{equation}
                \varepsilon_{ab} = \kappa_{abc} J_c,
                \label{current_induced_strain}
                \end{equation}
where the response function is obtained as $\kappa_{abc} = \varepsilon_{abc}^\text{m} / \sigma_{cc}$ by using the formula for longitudinal conductivity $J_a = \sigma_{aa} E_a$.
The obtained magnetopiezoelectric response function $\kappa_{abc}$ is not sensitive to the phenomenological scattering parameter.
This is because the Drude-type conductivity $\sigma_{aa} = \text{O} (\gamma^{-1})$ leads to the scattering-rate dependence of $\kappa_{abc} = e_{abc}^\text{m} / \sigma_{cc} = \text{O} (\gamma^{0})$ with Eq.~\eqref{MPE_Lehmann_representation}.
As a result, the response function $\hat{\kappa}$ does depend on the generic material properties as in the case of inverse magneto-galvanic effect~\cite{Levitov1985-oy,Edelstein1990-jr}.
It indicates the fact that the electric current plays an essential role in this magnetic counterpart of the piezoelectric response rather than the electric field.
It follows that the magnetopiezoelectric response occurs under the electric current flow and is inevitably accompanied by the Joule heating.
The energy loss may be unfavorable for future applications based on the magnetic metals.
The undesirable heating effect may be alleviated by utilizing the superconducting property (see Sec.~\ref{Sec_SC_and_MPV}).
The metallic property could also be an advantage of the magnetopiezoelectric effect for applications as we discuss at the end of this subsection.
We summarize the contrasting properties of the known piezoelectric effect and the magnetopiezoelectric effect in Table~\ref{table_comparison_piezoelectric}.
For the switchability of the \PT{}-symmetric magnetic order and the magnetopiezoelectric effect, one can refer to the discussions in Sec.~\ref{Sec_control}.

        \begin{table}[htbp]
        \centering
        \caption{
        Comparison of the conventional piezoelectric effect and the magnetopiezoelectric effect.
                }
        \label{table_comparison_piezoelectric}
        \renewcommand{\arraystretch}{1.3}
        \begin{tabular}{lcc}
         \hline
        &Piezoelectric effect ($\varepsilon_{ab} = e_{abc} E_c$) ~~~ &Magnetopiezoelectric effect ($\varepsilon_{ab} = e_{abc}^\text{m} E_c$) \\ 
         \hline
        \T{} parity&Even&Odd\\
        Stimulus&Electric field $\bm{E}$&Electric current $\bm{J}$\\
        System&Insulator \& Metal&Metal\\
        Relaxation time dependence& $\hat{e} \propto \tau^{2n}$&$\hat{e}^\text{m} \propto \tau^{2n+1}$\\
        Joule heating&No &Yes\\
        Switching&$\bm{E}$ w/ or w/o $\hat{s}$&$\bm{J}$ w/ or w/o $\hat{s}$\\
         \hline
        \end{tabular}
        \end{table}

The magnetopiezoelectric effect can also be observed in nonmagnetic materials with the electric parity violation by applying the external magnetic field.
In this case, not only the \T{} symmetry but also the \PT{} symmetry is broken, and the elastic-electric coupling is obtained as the combination of the conventional piezoelectric response and the magnetopiezoelectric response. 
The admixture may prevent us from specifying the magnetopiezoelectric response~\cite{Varjas2016-sw}.
On the other hand, there is no such concern in the case of the \PT{}-symmetric magnets because the \PT{} symmetry exactly forbids the conventional piezoelectric response.

\begin{figure}[htbp]
                \centering
                \includegraphics[width=0.95\linewidth,clip]{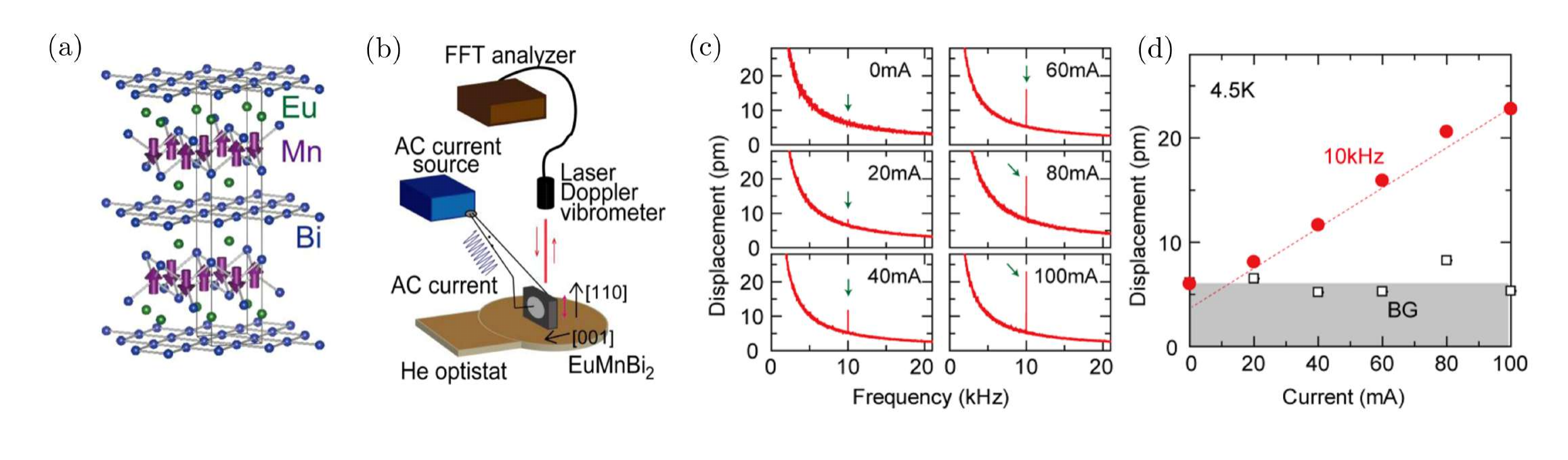}
                \caption{
                        Experimental results of the magnetopiezoelectric response in EuMnBi$_2$.
                        (a) The crystal and magnetic structures of EuMnBi$_2$.
                        (b) Measurement set-up of the magnetopiezoelectric effect. The laser Doppler vibrometer optically monitors the displacement responding to the AC current.
                        (c) Frequency profile of the displacement signals with varying intensity of AC electric current. The frequency of the electric current is 10~kHz. As the electric current increases, the current-induced signals marked by green arrows monotonically grow.
                        (d) The dependence of current-induced signals on the applied electric current. The signal is clearly distinguished from the background (BG) and is proportional to the electric current.
                        The figures are taken from Ref.~\cite{Shiomi2020-ux}.
                }
                \label{Fig_mpe_experiment}
                \end{figure}

In the following, we explain the microscopic grounds for the magnetopiezoelectric effect by taking a \PT{}-symmetric magnetic 
metal, hole-doped BaMn$_2$\textit{Pn}$_2$ (\textit{Pn}=As, Sb, Bi)~\cite{Watanabe2017-qk}.
BaMn$_2$As$_2$, for instance, undergoes the G-type antiferromagnetic order at high N\'eel temperature $T_{\text{N}} \sim 600\, \mathrm{K}$ which breaks the \Pa{} symmetry.
The \Pa{} violation originates from the coupling between the antiferromagnetic order and the locally noncentrosymmetric environment of Mn sites [Fig.~\ref{Fig_mpe_experiment}(a)].
Considering the site symmetry at Mn atoms, one can derive the antisymmetric spin-orbit coupling as 
                \begin{equation}
                \bm{g}_\bk = \left( \alpha_1 \sin{k_y}, \alpha_1 \sin{k_x}, \alpha_3 \sin{\frac{k_x}{2}}\sin{\frac{k_y}{2}}\sin{\frac{k_z}{2}}  \right),
                \label{spin-orbit-coupling_BaMn2As2}
                \end{equation}
where $\alpha_1,\alpha_3$ denote the coupling constants of the sublattice-dependent anti-symmetric spin-orbit coupling.
The locally noncentrosymmetric Mn atoms show hidden spin polarization described by Eq.~\eqref{sublattice_dependent_soc} with Eq.~\eqref{spin-orbit-coupling_BaMn2As2}, and it couples to the antiferromagnetic order.

The ordered state manifests the magnetic parity violation.
Specifically, the magnetic point group symmetry is
                \begin{equation}
                \mathbf{G} = 4'/m'm'm = \bar{4}2m \cup I\theta \cdot \bar{4}2m,
                \label{BaMn2As2_magnetic_point_group}
                \end{equation}
consisting of the unitary symmetry with the point group $\bar{4}2m$ and of the anti-unitary symmetry including the \PT{} symmetry ($I\theta$).
Although the parent compound BaMn$_2$As$_2$ shows insulating behavior with a small gap $\sim 10 $ (meV), the metallic conductivity is acquired by doping hole carriers or by applying high pressure.
Thus, the material is a promising candidate for \PT{}-symmetric magnetic metals by which we can demonstrate the interplay between the metallic conductivity and magnetic parity violation.

By imposing the unitary symmetry $\bar{4}2m$ in Eq.~\eqref{BaMn2As2_magnetic_point_group}, the allowed components of the response tensor Eq.~\eqref{MPE_Lehmann_representation} are 
        \begin{equation}
            e^\text{m}_{xyz},~~ e^\text{m}_{zxy} = e^\text{m}_{yzx}.
        \end{equation}
We again stress that the preserved \PT{} symmetry forbids the conventional (inverse) piezoelectric effect of Eq.~\eqref{piezoelectric_repsonse}.
Taking the independent-particle approximation, the microscopic origin can be inferred from the formula Eq.~\eqref{MPE_band_electron}.
To grasp the intuitive picture of the magnetopiezoelectric effect in BaMn$_2$As$_2$, let us focus on the electronic structure unique to it.
The metallic conductivity is ascribed to the hole pocket placed at $\Gamma$ point, which is expected to realize the interplay between the magnetic parity violation and itinerant property.
According to the magnetic point group symmetry of Eq.~\eqref{BaMn2As2_magnetic_point_group}, the active odd-parity magnetic multipoles are such as the magnetic quadrupole [Eq.~\eqref{B1u_magnetic_multipole}] and magnetic hexadecapole moment~\cite{Watanabe2018-cu}.
The group-theoretical argument introduced in Sec.~\ref{SecSub_multipolar_classification} allows us to identify the anti-symmetric modulation in the electronic structure given by the tetrahedral modulation of the energy spectrum $\delta \varepsilon_{\bk p} \sim k_x k_y k_z$ [Eq.~\eqref{B1u_magnetic_momentum_basis}].

The obtained asymmetric electronic band structure plays an essential role in the magnetopiezoelectric effect.
When applying the electric field to metals, the Fermi surface is shifted along the applied direction as obtained in the semiclassical theory of transport as
                \begin{equation}
                \delta f (\varepsilon_{\bk p}) \sim \gamma^{-1} \bm{E} \cdot \bm{v}_{pp} \partial_\varepsilon f (\varepsilon_{\bk p})= \gamma^{-1} E_k \cdot \partial_\bk \varepsilon_{\bk p} \partial_\varepsilon f (\varepsilon_{\bk p}).
                \label{boltzmann_fermi_surface_shift}
                \end{equation}
The field-induced shift ($\delta f$) is coupled to the asymmetry in the band dispersion ($\delta \varepsilon_{\bk p}$) and may induce additional anisotropy in the Fermi surface.
In the case of BaMn$_2$As$_2$, the coupling between the tetrahedral modulation and the field-induced shift along the $z$-axis results in the nematic anisotropy in the $k_x $-$ k_y$ plane.
The resultant nematic anisotropy is absent in equilibrium according to the magnetic symmetry of Eq.~\eqref{BaMn2As2_magnetic_point_group}.

By Letting $\varepsilon_{xy}^{(\bk)}$ be $k_xk_y$-type nematic anisotropy in the Fermi surface, the response formula reads as
        \begin{equation}
        \varepsilon_{xy}^{(\bk)} = e_{xyz}^\text{m} (\bk) E_z,
        \end{equation}
where the electric-elastic coupling is defined with the electronic strain $\varepsilon_{xy}^{(\bk)} \sim k_x k_y$.
The strain is induced by the dissipative electrical stimulus leading to the shift in Eq.~\eqref{boltzmann_fermi_surface_shift}.
Then, the response formula may be better to be rewritten by using the electric current $\bm{J}$ as
        \begin{equation}
        \varepsilon_{xy}^{(\bk)} = \overline{e}_{xyz}^\text{m} (\bk) J_z,
        \end{equation}
agreeing with Eq.~\eqref{current_induced_strain}.
The current-induced electronic strain $\varepsilon_{xy}^{(\bk)}$ may subsequently induce the lattice strain $\varepsilon_{xy}$ through the electron-phonon coupling.

The theoretical prediction of magnetopiezoelectric responses has been confirmed by a series of experiments~\cite{Shiomi2019-co,Shiomi2019-eo,Shiomi2020-ux}.
The experiments with the \PT{}-symmetric magnets such as EuMnBi$_2$ and CaMn$_2$Bi$_2$ are consistent with theories.
It has been verified that the elastic response is proportional to the applied current density and follows the aforementioned symmetry analysis~\cite{Shiomi2020-ux} (Fig.~\ref{Fig_mpe_experiment}).
While extensive research has been devoted to maximizing the magnitude of the conventional piezoelectric effect, the magnetopiezoelectric effect has not been fully explored for its material dependence and potential applications.
For instance, the magnetopiezoelectric effect varies by the magnetic order and may apply to future elastic devices due to its compatibility with the functionality unique to metals (\textit{e.g.}, giant magnetoresistance~\cite{Aoyama2022-gg,Huynh2019-gy,Ogasawara2021-yb,Sun2021-fe}) and nontrivial temperature dependence coming from that of magnetic moments.
Notably, the switchable property of \PT{}-symmetric magnetic order (see Sec.~\ref{Sec_control}) may be favorable for that purpose.    
More explorations by experiments and quantitative estimates by the first-principles calculations are highly desirable.


\subsection{Nonreciprocal Property of Electronic Transport and Optical Response}
\label{Sec_nonreciprocal_electric_phenomena}

The \T{}-parity classification introduced in Sec.~\ref{SecSub_EM_decomposition} is generalized to the nonlinear responses, while the parameter dependence of each contribution gets more complicated than the linear response functions.
For an example of the nonlinear response, let us consider the electric current response to double electric fields
                \begin{equation}
                J_a (\omega) = \int \frac{ d\omega_1 d \omega_2}{2\pi} \delta (\omega - \omega_1 -\omega_2) ~\sigma_{a;bc} (\omega_1,\omega_2) E_b (\omega_1) E_c (\omega_2).
                \label{nonreciprocal_current_generation_formula}
                \end{equation} 
The generated current is not flipped under the reversal of the electric field.
Thus, the response can be termed as the nonreciprocal current generation and is characteristic of parity-violating materials~\cite{Tokura2018-ht}.
The formula covers various phenomena attracting a lot of interest for a long time such as the nonreciprocal conductivity ($\omega_1 = \omega_2 = 0$), second-harmonic generation ($\omega_1 = \omega_2$), photocurrent generation ($\omega_1 = - \omega_2$), and parametric generation ($\left| \omega_1 \right| \neq \left| \omega_2 \right| $).
In the following, we mainly discuss the nonreciprocal conductivity and photocurrent generation from the viewpoint of the \PT{} symmetry.~\footnote{
        The photocurrent may respond to the irradiating light in materials without the parity violation such as due to the drag effect~\cite{Plank2016-zw}.
        In this review, we consider only the nonreciprocal photocurrent response arising from the parity violation.
        }

\subsubsection{Nonreciprocal conductivity}
\label{SecSubSub_nonreciprocal_conductivity}

A typical example of the DC nonreciprocal current generation is the diode effect of the p-n semiconductor junction where the parity violation comes from the artificial and mesoscale electric field in the depletion layer.
Bulk materials similarly show unidirectional behaviors in their conductivity because of the microscopic symmetry breaking.
The (bulk) nonreciprocal conductivity is further divided into that with the external magnetic field and that without it, called field-induced and field-free nonreciprocal conductivity, respectively.

Firstly, we briefly overview each type of nonreciprocal conductivity realized in materials manifesting the electric parity violation.
There exist extensive works on the field-induced nonreciprocal conductivity realized by the electric parity violation and external magnetic field.
For the longitudinal component ($\sigma_{a;aa}$) in Eq.~\eqref{nonreciprocal_current_generation_formula}, the response is quantified by the nonlinear resistivity
                \begin{equation}
                R = R_0 \left( 1 + \Gamma J \right),
                \label{nonlinear_resistivity}
                \end{equation}
up to the correction linear in the electric current.
The linear resistance $R_0$ usually surpasses the nonreciprocal correction, and thus the nonreciprocal correction $\Gamma$ is approximately obtained as
                \begin{equation}
                \Gamma \propto \frac{\sigma_{a;aa}}{\sigma_{aa}^3}.
                \end{equation}
Then, the nonreciprocal conductivity $\sigma_{a;aa}$ quantifies the nonreciprocal electric transport. 
Under a weak magnetic field, the nonreciprocal resistivity is approximately written as
                \begin{equation}
                \Gamma =  \Gamma_0 + \gamma H + O(H^2),
                \end{equation}
where a field-induced contribution $\gamma$ is determined by the effect of electric parity violation such as the strength of antisymmetric spin-orbit coupling of Eq.~\eqref{Rashba_spin_orbit_coupling}.
Although the nonreciprocal correction is typically smaller than the linear response, it can be identified by using the AC measurement with small but finite frequency~\cite{Ideue2017-py}.

From the viewpoint of symmetry, the field-induced nonreciprocal longitudinal transport can be classified into two classes; one comes from the magnetoelectric anisotropy and the other from the trigonal anisotropy~\cite{Szaller2013-dn}.
In the case of the magnetoelectric anisotropy, the unidirectionality is parallel to the \T{}-odd polar vector resulting from the coupling of electric parity violation and external magnetic fields. 
In the seminal work of Ref.~\cite{Rikken2001-aj}, the \T{}-odd polar vector, given by $\bm{E} \times \bm{H} \neq \bm{0}$, is built onto the two-dimensional electron gas in a semiconductor device with the perpendicular external electric and magnetic fields.
The \T{}-odd polar vector similarly appears under the magnetic field along the whirling spiral in a chiral system~\cite{Krstic2002-vq,Rikken2005-sp,Yokouchi2017-iy,Aoki2019-qc,Akaike2021-lb,Jiang2020-dr}, namely magnetochiral anisotropy, since the chirality couples axial vectors to polar vectors with their \T{}-parity kept.
On the other hand, the trigonal anisotropy does not require the polar asymmetry of systems as demonstrated in MoS$_2$ under the out-of-plane magnetic field~\cite{Wakatsuki2017-ft}.

The nonreciprocal conductivity manifests enhancement by the strong spin-orbit coupling and sizable \T{}-breaking effect.
In spin-orbit-coupled materials such as a topological insulator~\cite{Yasuda2016-wc,Yasuda2017-ng}, Weyl semimetal~\cite{Morimoto2016-ll,Wang2022-ip}, and polar semiconductor consisting of heavy elements~\cite{Ideue2017-py,Li2021-nu}.
Similarly, the field-induced nonreciprocal conductivity significantly increases by large spin splitting via the coupling to the ferromagnetic order~\cite{Zelezny2021-wm,Yoshimi2022-ng}.

The field-free nonreciprocal conductivity of nonmagnetic systems is also of much interest, though the studies have been devoted to the transverse response, that is, nonlinear Hall effect $\sigma_{a;bb}$ ($a \neq b$).
Notably, the nonlinear Hall response occurs without being admixed with the linear-response signal since the preserved \T{} symmetry forbids the linear Hall response.
The known mechanism for the nonlinear Hall response stems from the intrinsic or extrinsic origin.
The intrinsic mechanism originates from, so-called, the Berry curvature dipole~\cite{Deyo2009-dv,Moore2010-xi,Sodemann2015-bf}, where the Hall effect occurs due to the imbalance in the Berry curvature at $\pm \bk$ stimulated by the electric current.
The extrinsic effect denotes the mechanism essentially beyond the independent-particle approximation such as the electron-disorder scattering and electron-electron interaction.
In the case of the disorder effect, the skew and side-jump scatterings contribute to the nonreciprocal conductivity~\cite{Du2019-sh,Xiao2019-lt,Nandy2019-xo,Isobe2020-ax,Du2021-kt} similarly to the anomalous Hall and spin Hall responses~\cite{Nagaosa2010-dn,Sinova2015-uq}.
The intrinsic and extrinsic contributions can be comparable to each other~\cite{Du2021-xu}; \textit{e.g.,} a spin-orbit-coupled semiconductor WTe$_2$ shows the large nonlinear Hall effect which may be from the comparable contributions from the intrinsic and extrinsic effects~\cite{Ma2019-xf,Kang2019-bj}. 
Note that one can exclude the Berry curvature dipole effect by taking the highly-symmetric noncentrosymmetric materials with nongyrotropic point group symmetry since gyrotropic symmetry is required for the Berry curvature dipole~\cite{Toshio2020-cz,Isobe2020-ax,He2021-xh,Dzsaber2021-to}.

Next, let us consider the nonreciprocal conductivity of the \PT{}-symmetric magnets, that is the response purely induced by the magnetic parity violation.
The nonreciprocal conductivity of the \PT{}-symmetric materials received delayed attention despite intensive interest in that of the \T{}-symmetric materials, possibly because parity-violation-induced phenomena have been rarely explored in the antiferromagnetic conductors.
Circumventing this situation, recent experimental work has identified nonreciprocal transport of the \PT{}-symmetric magnets~\cite{Godinho2018-fn,Ota2022-bh,Gao2023-fz,Wang2023-fi}.
The nonreciprocal transport of \PT{}-symmetric magnets is expected to be huge due to the remarkable parity violation.
The energy scale of the parity violation can be as much as the Hund's coupling, which is much larger than the external Zeeman field.
It is worth mentioning that the nonreciprocal nature is in intimate relation with the domain state of \PT{}-symmetric magnetic order and that it can be applied to the antiferromagnetic spintronics (see also Sec.~\ref{Sec_control}).

Let us look into theoretical backgrounds for the nonreciprocal transport mainly with respect to the band-electron system.
A basic understanding of the phenomena is obtained by examining the intrinsic mechanism identified by a simple theoretical set-up where the perturbation calculation is performed under the independent-particle approximation.
The scattering effect is phenomenologically taken into account by the relaxation-time approximation in the semiclassical transport theory~\cite{Ideue2017-py} and by introducing the damping term into the von-Neumann equation of the quantum transport theory~\cite{Ventura2017-db,Matsyshyn2019-wi}.
Including the mechanisms originating from the electric and magnetic parity violations, there exist three intrinsic effects in the clean limit; the nonlinear Drude, Berry curvature dipole, and intrinsic Hall effects.
The formula for each mechanism is given as 
                \begin{align}
                \sigma_{a;bc}^\text{D} 
                        &= - \frac{e^3 }{\gamma^2} \int \frac{d\bk}{\left( 2\pi \right)^d} \sum_p   v_{pp}^a \frac{\partial^2 f (\varepsilon_{\bk p})}{\partial k_b\partial k_c } ,\\
                        &= - \frac{e^3 }{\gamma^2}  \int  \frac{d\bk}{\left( 2\pi \right)^d} \sum_p   v_{pp}^a v_{pp}^b v_{pp}^c \frac{\partial^2 f (\varepsilon)}{\partial^2 \varepsilon}|_{\varepsilon = \varepsilon_{\bk p}} , 
                        \label{nonlinear_drude_term}
                \end{align}
for the nonlinear Drude effect~\cite{Ideue2017-py}, 
                \begin{align}
                \sigma_{a:bc}^\text{BCD} 
                        &= -\frac{e^3}{2\gamma} \int \frac{d\bk}{\left( 2\pi \right)^d} \sum_{p} \epsilon_{abd}  \frac{\partial f(\epsilon_{\bk p})}{\partial k_c}   \Omega^d_p  + \left[ b \leftrightarrow c \right],\\
                        &= \frac{e^3}{2\gamma} \int \frac{d\bk}{\left( 2\pi \right)^d} \sum_{p} \epsilon_{abd}   f(\epsilon_{\bk p}) \frac{\partial \Omega^d_p}{\partial k_c}  + \left[ b \leftrightarrow c \right],\\
                        &= \frac{e^3}{2\gamma} \epsilon_{abd}   \mathcal{D}^{\,cd} + \left[ b \leftrightarrow c \right],
                        \label{BCD_term}
                \end{align}
for the Berry-curvature-dipole effect~\cite{Deyo2009-dv,Sodemann2015-bf}, and
                \begin{equation}
                \sigma_{a;bc}^\text{int} = -e^3 \int \frac{d \bk}{(2 \pi)^d} \sum_{p \neq q} \frac{1}{\epsilon_{\bk p} - \epsilon_{\bk q}} \left[  g_{pq}^{ab}  \frac{\partial f (\varepsilon_{\bk p})}{\partial k_c } + g_{pq}^{ac} \frac{\partial f (\varepsilon_{\bk p})}{\partial k_b } -2  g_{pq}^{bc} \frac{\partial f (\varepsilon_{\bk p})}{\partial k_a } \right],
                \label{intrinsic_hall_effect}
                \end{equation}
for the intrinsic Hall effect~\cite{Gao2014-sq,Watanabe2020-oe,Watanabe2021thesis}.
All the formulas comprise the Fermi-surface effect $\partial_\varepsilon f (\varepsilon)$, and hence the above intrinsic mechanisms are allowed in conductors and prohibited in insulators.
The derivations have been presented in the density-matrix formalism~\cite{Matsyshyn2019-wi,Watanabe2020-oe,Yatsushiro2022-ui,Kaplan2022-vc} and diagrammatic approach~\cite{Joao2020-mk,Michishita2022-fh,Oiwa2022-he}.

Since the intraband matrix element of the velocity operator is given as $v_{pp}^a = \partial_{k_a} \varepsilon_{\bk a}$ in the band-electron system, only the band energy $\varepsilon_{\bk p}$ is relevant to the nonlinear Drude effect.
This is in agreement with the fact that the nonlinear Drude term can be captured by Boltzmann's transport theory.
On the other hand, the multiband property plays an essential role in the latter two effects as they include the Berry curvature 
                \begin{equation}
                \Omega_{p}^{a} = \epsilon_{abc} \frac{\partial \xi_{pp}^c}{\partial k_b},
                \end{equation} 
and the band-resolved quantum metric~\cite{Gao2020-eh}
                \begin{equation}
                g_{pq}^{ab} = \text{Re}\left[  \xi_{pq}^a \xi_{qp}^b \right],
                \end{equation}
with the Berry connection $\xi_{pq}^a = i \Braket{u_{\bk p} | \partial_{k_a} u_{\bk q}}$ ($\ket{u_{\bk q}}$ is the periodic part of the Bloch state).
We have also introduced the Berry curvature dipole
                \begin{equation}
                \mathcal{D}^{\, ab } = \int \frac{d\bm{k}}{\left( 2\pi \right)^d} \sum_p  f(\epsilon_{\bm{k}p}) \partial_{k_a} \Omega^b_p,
                \end{equation}
indicating the dipolar distribution of the Berry curvature along the Fermi surface.
The multiband nature is captured by the full-quantum theory and by the semiclassical transport theory taking account of the quantum-geometric corrections; \textit{e.g.}, following the semiclassical transport theory, the Berry curvature dipole effect comes from the anomalous-velocity correction ($ \bm{v}_\text{ano} \sim \bm{\Omega} \times \bm{E}$)~\cite{Xiao2010-tk}.
Although the intrinsic Hall effect originates from the anomalous velocity as well, the Berry curvature comes from the $\bm{E}$-induced virtual transition, namely positional-shift effect~\cite{Gao2014-sq}. 

The Berry curvature dipole effect depends on the scattering rate as $O(\gamma)$.
The sensitivity to scatterings is intuitively figured out by the imbalance of Berry curvature dipole driven by the electric current~\cite{Toshio2020-cz}.
Similarly to the current-induced correction to the distribution function in Eq.~\eqref{boltzmann_fermi_surface_shift}, the compensation between opposite Berry curvature at $\pm \bk$ is removed under the ohmic electric current.
The resultant total Berry curvature is finite in the steady state and allows for the deflection of electrons.
The obtained Hall response originating from the current-induced Berry curvature is consistent with the symmetry of the nonlinear Hall effect.
On the other hand, the intrinsic Hall effect does not show prominent dependence on the scattering rate as much as $O(\gamma^0)$, because of the uncompensated Berry curvature as well as the positional shift stem from the interband mixing.

The mechanism of nonreciprocal transport is classified by the preserved anti-unitary symmetry (\T{} and \PT{} symmetries) as summarized in Table~\ref{Table_relaxation_time_dependence_2nd_conductivity} based on its dependence on the phenomenological relaxation time defined by $\tau = \gamma^{-1}$.
As in the case of linear response function (see Sec.~\ref{SecSub_EM_decomposition}), the anti-unitary symmetry determines the $\tau$ dependence; \textit{i.e.}, $O(\tau^{2n+1})$ for the \T{}-symmetric case and $O(\tau^{2n})$ for the \PT{}-symmetric case.
When the system manifests both electric and magnetic parity violations (without \T{} or \PT{} symmetry), all of the effects contribute~\cite{Shao2020-kb}. 
Note that the nonlinear Drude effect gives rise to the longitudinal nonreciprocity by taking an appropriate geometry of measurement, while the Berry curvature and intrinsic Hall effects lead to only the Hall response.
As a result of the adopted approximations, the \PT{}-symmetric magnetic metals offer both longitudinal and transverse nonreciprocal conductivity, while the \T{}-symmetric conductors do show only the Hall response if without the electron correlation effect or more rigorous treatment of disorder scattering~\cite{Morimoto2018-dq,Du2019-sh,Isobe2020-ax}.

        \begin{table}[htbp]
        \caption{Classification of intrinsic nonreciprocal conductivity in the clean limit in terms of the relaxation time $\tau = \gamma^{-1}$ and preserved anti-unitary symmetry~\cite{Watanabe2020-oe}. ``N/A'' denotes that contribution is forbidden by the symmetry.}
        \label{Table_relaxation_time_dependence_2nd_conductivity}
        \centering
        \renewcommand{\arraystretch}{2.0}
                \begin{tabular}{LCCC}
                \hline
                &\text{Nonlinear Drude}	&\text{Berry curvature dipole}	&\text{Intrinsic Hall}	\\ 
                \hline
                \text{\T{}}	&\text{N/A}&O(\tau)&O(\tau^{-1})\\
                \text{\PT{}}&O(\tau^2) &\text{N/A} &O(\tau^{0})\\
                \hline
                \end{tabular}
        \end{table}

The \PT{}-symmetric magnetic metals show the nonlinear Drude and intrinsic Hall effects which may dominate the nonreciprocal conductivity with the scarce and moderate disorder concentration, respectively.
When the Fermi energy is placed in the vicinity of the Dirac dispersion, the quantum-geometrical effect is so significant as to overwhelm the nonlinear Drude effect~\cite{Wang2021-jm,Liu2021-nd} as implied in even-layer MnBi$_2$Te$_4$~\cite{Gao2023-fz,Wang2023-fi}.
For an example of the nonreciprocal conductivity of \PT{}-symmetric magnets, Ref.~\cite{Watanabe2020-oe} reported that the nonlinear Drude effect of the carrier-doped BaMn$_2$As$_2$ is estimated to be
                \begin{equation}
                \sigma_{z;xy} = \frac{e^3 \alpha_3 n \tau^2}{4} \text{sgn}\left( m_\text{AF} \right) + O(\tau^0),
                \end{equation}
with the small density of electron $n$ in the clean limit ($\tau \to + \infty$).
The result shows that the nonreciprocal conductivity is informative for investigating the \PT{}-symmetric magnetic order.
Firstly, the response is related to the magnitude and sign of the hidden spin-momentum coupling constant $\alpha_3$ in Eq.~\eqref{spin-orbit-coupling_BaMn2As2} which may be hard to directly measure since it is originally hidden by the sublattice degree of freedom.
Secondly, the response directly indicates the AFM-domain state by $\text{sgn}\left( m_\text{AF} \right)$ ($m_\text{AF}$: sign of N\'eel vector) and will be a promising tool for the electronics based on \PT{}-symmetric magnets.

The nonreciprocal conductivity shows the richer property when one goes beyond the independent-particle approximation such as by taking account of the electron-correlation and disorder-scattering effects~\cite{Morimoto2018-dq,Isobe2020-ax,Michishita2022-fh,Kofuji2021-ol,Kappl2023-ym}.
For instance, the scattering event occurs in the presence of impurities~\cite{Du2019-sh,Xiao2019-lt,Nandy2019-xo,Ma2022-hw}, spin degree of freedom~\cite{Yasuda2017-ng,Ishizuka2020-qo}, and magnetic-multipolar object~\cite{Isobe2022-qa,Liu2022-se}. 
Recent studies further clarified that superconductivity leads to the giant nonreciprocal conductivity by the superconducting fluctuation and the vortex motion.
Although we do not discuss it in detail, interested readers can refer to Refs.~\cite{Tokura2018-ht,Nagaosa2024-fr}.

The disorder effect is particularly of importance for the conduction phenomena as intensively investigated in the studies of the anomalous Hall effect~\cite{Nagaosa2010-dn}, where the so-called skew-scattering effect due to disorders can surpass the intrinsic mechanism determined by the Berry curvature~\cite{Onoda2008-wq}.
For the nonreciprocal transport, the field-free nonlinear Hall effect may be attributed to comparable contributions of the intrinsic mechanism [Eq.~\eqref{BCD_term}] and of the skew scattering effect~\cite{Kang2019-bj}.
It is therefore important to take a brief look at the disorder effect on the nonreciprocal conductivity.

The disorder effect beyond the relaxation-time approximation has been corroborated in recent theories.
The known mechanisms such as the side-jump and skew-scattering effects can be incorporated by the semiclassical theory and by a full quantum-mechanical theory~\cite{Du2021-xu}.
Following a semiclassical formulation with the Boltzmann kinetic equation, disorder scattering affects the collision integral and correction to the energy spectrum.
In the clean limit, the extrinsic mechanism (skew-scattering term $\sigma_{a;bc}^\text{sk}$ and side-jump term $\sigma_{a;bc}^\text{sj}$) contributes to the nonreciprocal conductivity with the relaxation time dependence
                \begin{equation}
                \sigma_{a;bc}^\text{sk} = O(\tau^3 \cdot  \tau_\text{sk}^{-1}),~ \sigma_{a;bc}^\text{sj}= O(\tau),
                \label{extrinsic_nonreciprocal_conductivity_without_spinflipping}
                \end{equation}
where $\tau \propto (n_\text{imp} V_0^2)^{-1}$ and $\tau_\text{sk} \propto (n_\text{imp} V_0^3)^{-1}$ denote the characteristic time for relaxation due to the symmetric and antisymmetric scattering, respectively~\cite{Du2019-sh,Isobe2020-ax}.
Specifically, the antisymmetric scattering event is given by the difference in the scattering rates between $\bk \to \bk'$ and $\bk' \to \bk$, the latter of which can be substituted by the scattering process $-\bk \to - \bk'$ in the presence of the \T{} symmetry.

\begin{figure}[htbp]
                \centering
                \includegraphics[width=0.95\linewidth,clip]{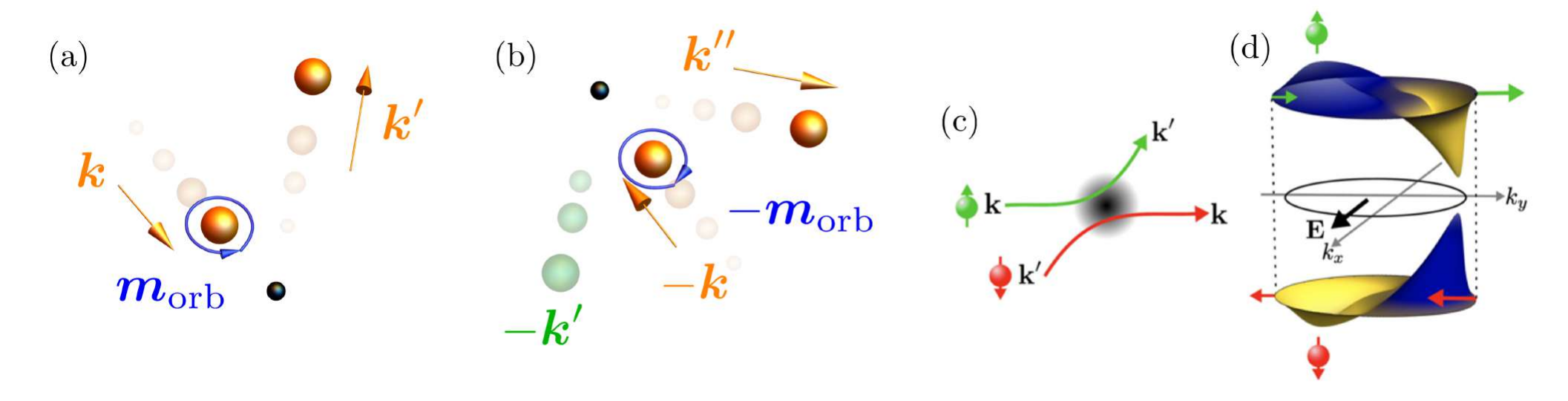}
                \caption{
                        Illustration of skew-scattering mechanisms in (a,b) \T{}- and (c,d) \PT{}-symmetric systems.
                        (a) Spinning wave packet (orange-colored) is scattered by the impurity (black-colored), resulting in the momentum transfer as $\bk \to \bk'$.
                        The self-rotation leading to deflection arises from the orbital angular momentum $\bm{m}_\text{orb} (\bk)$ due to the electric parity violation.  
                        (b) The impurity-scattering event for the oppositely-incident wave packet.
                        Owing to the opposite spinning [$\bm{m}_\text{orb} (-\bk) = -\bm{m}_\text{orb} (\bk)$], the deflected direction does not show the reciprocity, that is, the wave packet is deflected into $\bk''$, not into $-\bk'$.
                        (c) Spin-dependent scattering of electrons.
                        (d) Nonlinear Hall response induced by the anomalous skew scatterings in \PT{}-symmetric systems.
                        The spin-dependent skew scattering gives the correction to the electron's distribution function (colored in blue and yellow).
                        Because of the \PT{}-symmetric magnetic order, the skew scattering mostly occurs for up and down electrons in the sublattice A ($\rho =+$) and B ($\rho =-$), respectively.
                        The Berry curvature roughly appears upwards and downwards in the sublattices A and B due to the coupling between the antiferromagnetic exchange splitting and hidden Berry curvature.
                        The resultant electric-field-induced anomalous velocity shows the opposite sign between sublattice (green- and red-colored arrows).  
                        The staggered signs of anomalous velocity and deviation in the electrons' distribution are canceled to produce a finite Hall current.
                        The panels (c,d) are taken from Ref.~\cite{Ma2022-hw} (\copyright~American Physical Society).
                        }
                \label{Fig_skew}
                \end{figure}

We illustrate the origin of the asymmetric scattering rate in the light of wave-packet dynamics~\cite{Isobe2020-ax}.
In the presence of the electric parity violation, the wave-packet comprised of the Bloch states manifests spinning behavior due to the orbital angular momentum $\bm{m}_\text{orb} (\bk)$~\cite{Chang1996-hi,Xiao2010-tk}, which is highly related to the Berry curvature $\bm{\Omega} (\bk)$~\cite{Sinitsyn2008-dg}.
Let us consider that the spinning wave packet with the momentum $\bk$ gets deflected into the orbit with $\bk' $ by the impurity scattering like the Magnus effect [Fig.~\ref{Fig_skew}(a)].
If the incident momentum flipped, the wave packet with the momentum $-\bk$ shows self-rotation in the opposite way as ensured by the \T{} symmetry as $\bm{m}_\text{orb} (\bk) = -\bm{m}_\text{orb} (-\bk)$ and thereby shows nonreciprocal scattering [Fig.~\ref{Fig_skew}(a,b)].

The \PT{}-symmetric magnets do not allow for such a nonreciprocal-deflection process because of the zero Berry curvature at every momentum [$\bm{\Omega} (\bk) = \bm{0}$].
The absence of extrinsic contributions follows from the fact that the scattering matrix is reciprocal in the presence of the \PT{} symmetry unless the spin-flip process is taken into account.
As a result, the nonreciprocal conductivity of the \PT{}-symmetric magnets is free from the extrinsic mechanism of Eq.~\eqref{extrinsic_nonreciprocal_conductivity_without_spinflipping} in contrast to that of the nonmagnetic materials.
Note that the argument can be applied to systems with the effective \PT{} symmetry; \textit{e.g.}, the isotropic two-dimensional Dirac electron retains the antiunitary symmetry $\theta 2_\perp$, the combination of \T{} operation and out-of-plane two-fold rotation.
The symmetry satisfying $\left( \theta 2_\perp \right)^2 = +1$ makes the scattering process reciprocal in the two-dimensional plane.
It implies that an ideal Dirac electron on the topological-insulator surface shows the nonreciprocal conductivity tolerant of the disorder scattering.

Equation~\eqref{extrinsic_nonreciprocal_conductivity_without_spinflipping} claims that extrinsic effects are comparable to or dominating the intrinsic effects when they are not forbidden by the \PT{} symmetry, since the leading contribution in the clean limit is the skew-scattering mechanism in the order of $\sim \tau^3 / \tau_\text{sk}$.
The contributions in Eq.~\eqref{extrinsic_nonreciprocal_conductivity_without_spinflipping} comes from scattering events without spin flip, whereas the spin-flip process gives rise to another extrinsic mechanism characteristic of \PT{}-symmetric magnets~\cite{Ma2022-hw} [Fig.~\ref{Fig_skew}(c)].
For instance, there exists anomalous skew-scattering effect $\sigma_{a;bc}^\text{Ask}$, which depends on the relaxation time as
                \begin{equation}
                \sigma_{a;bc}^\text{Ask} = O(\tau^2 \cdot \tau_\text{sk}^{-1}),
                \end{equation} 
being comparable to the nonlinear Drude effect of Eq.~\eqref{nonlinear_drude_term} with a moderate antisymmetric scattering rate $\tau_\text{sk}^{-1}$.
The nonvanishing skew-scattering effect is closely related to the hidden Berry curvature of \PT{}-symmetric magnets.
For the \PT{}-symmetric magnets comprised of A/B sublattice, the strong exchange splitting may result in electric transport carried by spin-up electrons on the sublattice A and the spin-down electrons on the sublattice B.
The two kinds of carriers are connected by the \PT{} symmetry and undergo the opposite Berry curvature even at the same momentum $\left[ \bm{\Omega}_\text{A} (\bk) = -\bm{\Omega}_\text{B} (\bk)  \right]$, by which the total Berry curvature is completely compensated at each momentum.
The hidden Berry curvature gives rise to the sublattice-dependent anomalous velocity under the electric field as
                \begin{equation}
                \bm{v}'_\text{ano} (\rho_z) \sim \rho_z \bm{\Omega}(\bk) \times \bm{E}, 
                \end{equation}
where $\rho_z = +1$ $(-1)$ for the sublattice A (B) [Fig.~\ref{Fig_skew}(d)].
The staggered anomalous velocity may offer a nonvanishing nonlinear Hall response, since electrons at each site may experience different spin-dependent antisymmetric scattering events which are correlated with the sublattice [Fig.~\ref{Fig_skew}(c)]. 
The emergence of the hidden Berry curvature stems from the \PT{}-symmetric magnetic order by which the spin degeneracy is lifted at each sublattice in a staggered manner.
If disorder concentration is not negligible, another mechanism for the nonreciprocal conductivity also plays an important role.
For instance, the side-jump effect similarly gives contributions differently from Eq.~\eqref{extrinsic_nonreciprocal_conductivity_without_spinflipping}~\cite{Ma2022-hw,Atencia2023-be} and may be dominant in the presence of moderate disorder scattering.

Finally, the space-time classification of Table~\ref{Table_relaxation_time_dependence_2nd_conductivity} remains meaningful even when taking into account the extrinsic mechanism~\cite{Watanabe2020-oe,Ma2022-hw};
in other words, the (conventional) skew-scattering and side-jump effects of Eq.~\eqref{extrinsic_nonreciprocal_conductivity_without_spinflipping} vanishes by the \PT{} symmetry, while the anomalous skew-scattering effect survives in the \PT{}-symmetric system due to its \T{}-odd and \PT{}-even nature.
The classification is similarly extended to cover the self-energy effect as corroborated in the Green's-function fashion~\cite{Michishita2022-fh}.

\subsubsection{Photocurrent generation}
\label{SecSubSub_photocurrent}

The photocurrent generation (photogalvanic effect, photovoltaic effect) is a response extensively applied to our daily lives such as solar panels and photodetector.
The response can occur due to microscopic parity violation as the nonreciprocal conductivity does, while it has been implemented by mesoscale parity violation such as the internal electric fields of the semiconductor-based p-n junction and ferroelectric materials~\cite{Fridkin1978-vw,Sturman1992-bf}.
The mechanism derived from microscopic symmetry breaking is called bulk photocurrent generation.
In contrast to typical photo-electric rectifiers, bulk photocurrent generation is allowed not only in polar materials but also in noncentrosymmetric but nonpolar materials including well-known zinc-blende-type semiconductors such as GaAs.
The bulk photocurrent response has been attracting enormous interest from theoretical and experimental investigators due to an active discussion of applications to conversion efficiency~\cite{Nagaosa2017-np,Spanier2016-fe,Liu2020-mw,Pusch2023-iu}.

The (bulk) photocurrent responses have been microscopically investigated in studies of the band-electron systems~\cite{Sturman1992-bf,von-Baltz1981-yf,Kristoffel1985-wz,Sipe2000-sb}.
For instance, it was shown that the electric parity violation gives rise to various mechanisms for the photocurrent response.
The mechanism is in close relation to the quantum geometry of electronic structure similar to the mechanism of nonreciprocal conductivity (see Sec.~\ref{SecSubSub_nonreciprocal_conductivity})~\cite{Moore2010-xi,de-Juan2020-ev}.
In the light of topological material science, a lot of theoretical and experimental works have been devoted to the photocurrent responses in topological materials~\cite{Ma2021-xq,Orenstein2021-hu,Ma2022-pm,Morimoto2023-mg}.
For example, the significant photocurrent generation is attributed to the electron-hole creations around Weyl nodes of TaAs~\cite{Ma2017-wr} and RhSi~\cite{Rees2020-xk,de-Juan2017-ph}.
These results indicate that the photocurrent response is a possible probe of the quantum geometry in solids in addition to its potential for energy harvesting devices and other engineering applications.
Furthermore, the photocurrent measurement being sensitive to the symmetry of materials has been applied to various quantum materials to examine exotic quantum phases such as those of cuprate superconductor and excitonic insulators~\cite{Xu2020-ye,Lim2020-vi}.
Diagnosis based on the photocurrent response may be advantageous compared to the widely-used other nonlinear optical probe such as second-harmonic generation where the interference with the reference signals should be prepared~\cite{Ma2022-pm}.

Recent works have further addressed the microscopic mechanism of the photocurrent responses induced by the magnetic parity violation after the symmetry analysis~\cite{Men-shenin2000-kp}.
We first introduce the photocurrent mechanism induced by the electric and magnetic parity violations in the context of the band-electron picture.
Similarly to the nonreciprocal transport response, the electric and magnetic parity violations play contrasting roles in the photocurrent generation.
Then, several remarks will be made about modifications coming from the disorder scattering and electron correlation.

\begin{figure}[htbp]
                \centering
                \includegraphics[width=0.70\linewidth,clip]{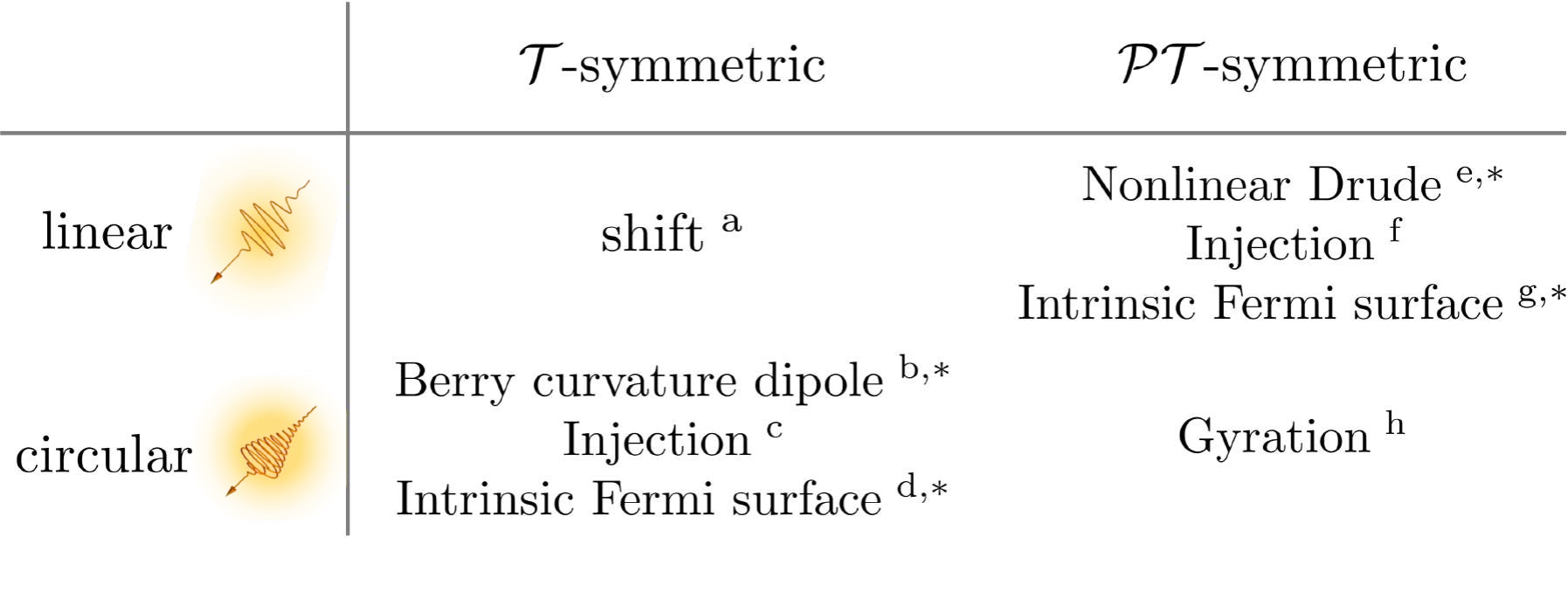}
                \caption{
                        Classification of the intrinsic photocurrent mechanism 
                        in terms of the space-time symmetry (allowed in the \T{}-symmetric or \PT{}-symmetric systems) and the polarization state of the incident light (linearly-polarized or circularly-polarized light).
                        The mechanisms with the superscript ``$\ast$'' denote those allowed in metals.
                        Each mechanism has been identified in a~\cite{von-Baltz1981-yf}, b~\cite{Moore2010-xi}, c~\cite{Sipe2000-sb}, d~\cite{de-Juan2020-ev}, e~\cite{Holder2020-tv}, f~\cite{Zhang2019-uy}, g~\cite{Watanabe2021-bt}, h~\cite{Watanabe2021-bt,Ahn2020-ec}.           
                        }
                \label{Fig_photocurrent-classification}
                \end{figure}

The mechanism of photocurrent generation has been thoroughly investigated in the framework of the independent-particle approximation.
Such intrinsic mechanism stems from the carrier dynamics such as due to the Fermi-surface (FS) and particle-hole creation (PH) effects, which give the photocurrent conductivity
                \begin{equation}
                \sigma_{a:bc} = \sigma_{a:bc}^\text{FS} + \sigma_{a:bc}^\text{PH},
                \label{total_photocurrent_response_normal}
                \end{equation}
where the frequency dependence of $\sigma_{a:bc}(\omega,-\omega)$ in Eq.~\eqref{nonreciprocal_current_generation_formula}
is implicit. 
Assuming the absence of scatterers, the known Fermi-surface effects are given by
                \begin{equation}
                \sigma_{a:bc}^\text{FS} = \sigma_{a:bc}^\text{D} + \sigma_{a:bc}^\text{BCD} + \sigma_{a:bc}^\text{iFS},
                \label{photocurrent_fermi_surface_total}
                \end{equation}
that is nonlinear Drude, Berry curvature dipole, intrinsic Fermi-surface effects, respectively~\cite{Holder2020-tv,Moore2010-xi,de-Juan2020-ev}.
The mechanism based on particle-hole excitations gives contributions that are similarly decomposed into 
                \begin{equation}
                \sigma_{a:bc}^\text{PH} = \sigma_{a:bc}^\text{e-inj} + \sigma_{a:bc}^\text{m-inj} + \sigma_{a:bc}^\text{shift} + \sigma_{a:bc}^\text{gyro},
                \end{equation}
called electric-injection, magnetic-injection, shift, and gyration-current mechanisms~\cite{Sipe2000-sb,Zhang2019-uy,von-Baltz1981-yf,Watanabe2021-bt,Ahn2020-ec,Okumura2021-iw}.
These contributions to the photocurrent conductivity are classified in terms of their definite parity under the \T{} and \PT{} operations as tabulated in Fig.~\ref{Fig_photocurrent-classification}.
In contrast to the DC response in Sec.~\ref{SecSubSub_nonreciprocal_conductivity}, the space-time classification of photocurrent response is related to the degrees of freedom of irradiating light such as frequency ($\omega$) and polarization state (linearly-polarized light and circularly-polarized light).~\footnote{
        More accurately, the photocurrent responds to linearly polarized, circularly polarized, and unpolarized light.
        The mechanism for the unpolarized-light-induced photocurrent response is the same as that for the linearly-polarized light, while it contributes even under the circularly-polarized light since the unpolarized-light contribution originates from $|\bm{E}|^2$.
        Then, the circular contribution is obtained by taking the difference in the responses to the light with opposite circular polarizations.
}

Let us write down the explicit formulas for each contribution.
The Fermi-surface effects are formulated as
                \begin{align}
                \sigma_{a;bc}^\text{D} &= - \frac{e^3}{2 \omega^2}\int \frac{d\bk}{(2\pi)^d} \sum_p v_{pp}^a \frac{\partial^2 f (\varepsilon_{\bk p})}{\partial k_b \partial k_c},
                \label{nonlinear_drude_photocurrent}\\ 
                \sigma_{a;bc}^\text{BCD} &= -\frac{i e^3}{2 \omega}\left(  \epsilon_{abd} \mathcal{D}_{cd} - \epsilon_{acd} \mathcal{D}_{bd}  \right),
                \label{bcd_photocurrent}\\ 
                \sigma_{a;bc}^\text{iFS} &= - \frac{e^3}{2} \int \frac{d\bk}{(2\pi)^d} \sum_{p \neq q}  \xi_{pq}^b \xi_{qp}^c \frac{1}{\omega + \varepsilon_{\bk q}- \varepsilon_{\bk p}} \partial_{k_a} \left\{ f (\varepsilon_{\bk p}) -f (\varepsilon_{\bk q}) \right\},
                \label{intrinsic_fermi_surface_photocurrent}
                \end{align}
all of which consist of the Fermi-surface effect $\partial_{\varepsilon} f (\varepsilon)$ and vanish without any gapless quasiparticle excitations.
In particular, the nonlinear Drude and Berry curvature dipole effects resemble the contributions to the nonreciprocal conductivity of Eqs.~\eqref{nonlinear_drude_term},~\eqref{BCD_term} and are therefore allowed in the presence of the magnetic and electric parity violations, respectively (see also Table~\ref{Table_relaxation_time_dependence_2nd_conductivity}).
Under the nearly-static electric field, the formulas for dc and ac nonreciprocal current generation are given in a unified manner by properly taking into account the scattering effect~\cite{Du2019-sh}.

The intrinsic Fermi-surface effect is further divided into 
                \begin{equation}
                \sigma_{a;bc}^\text{iFS} =  \sigma_{a;bc}^\text{e-iFS}  +\sigma_{a;bc}^\text{m-iFS},
                \end{equation}
that include the Berry curvature and quantum metric as
                \begin{align}
                \sigma_{a;bc}^\text{e-iFS} &= - \frac{ie^3}{2} \int \frac{d\bk}{(2\pi)^d} \sum_{p \neq q}  \Omega_{pq}^{bc} \frac{\omega}{ \omega^2 - \left(  \varepsilon_{\bk q}- \varepsilon_{\bk p} \right)^2 } \partial_{k_a}  f (\varepsilon_{\bk p}),
                \label{electric_intrinsic_fermi_surface_photocurrent}\\
                \sigma_{a;bc}^\text{m-iFS} &= - e^3 \int \frac{d\bk}{(2\pi)^d} \sum_{p \neq q}  g_{pq}^{bc} \frac{\varepsilon_{\bk p}- \varepsilon_{\bk q} }{ \omega^2 - \left(  \varepsilon_{\bk q}- \varepsilon_{\bk p} \right)^2 } \partial_{k_a}  f (\varepsilon_{\bk p}),
                \label{magnetic_intrinsic_fermi_surface_photocurrent}
                \end{align}
where $\Omega_{pq}^{bc} = -2 \text{Im}\left( \xi_{pq}^b \xi_{qp}^c \right)$ is the band-resolved Berry curvature which leads to the Berry curvature by summing over one of the band indices as $\Omega_{p}^a = \sum_{q } \epsilon_{abc} \Omega_{pq}^{bc}/2$.
Since the band-resolved Berry curvature determines the helicity-dependent dipole excitations at $\bk$ under the circularly-polarized light.
This term is finite in the system with the electric parity violation because the opposite sign in Fermi-surface deviations at $\pm \bk$ multiplied with the staggered Berry curvature $\Omega_{pq}^{bc} (\bk)  = - \Omega_{pq}^{bc} (-\bk) $ gives nonvanishing contribution.
Thus, it is called the electric intrinsic Fermi-surface (e-iFS) effect.
Contrastingly, the band-resolved quantum metric related to the linearly-polarized-light excitation does show the same sign between $g_{pq}^{bc} (\pm \bk)$ of Eq.~\eqref{magnetic_intrinsic_fermi_surface_photocurrent} leading to the perfect compensation.

One can see the opposite situation in the case of the \PT{}-symmetric system.
The electric intrinsic Fermi-surface effect is forbidden since the retained \PT{} symmetry leads to the zero Berry curvature.
Instead, Eq.~\eqref{magnetic_intrinsic_fermi_surface_photocurrent} including the band-resolved quantum metric offers finite photocurrent response, since the asymmetric electronic band structure of the \PT{}-symmetric system [Fig.~\ref{Fig_real-momentum-structure-oddparitymultipole}(d)] allows for the uncompensated contribution.
Thus, Eq.~\eqref{magnetic_intrinsic_fermi_surface_photocurrent} is the magnetic counterpart of Eq.~\eqref{electric_intrinsic_fermi_surface_photocurrent}, named the magnetic intrinsic Fermi-surface (m-iFS) effect.
As a result, the quantum-geometrical factors lead to the contrasting mechanism for photocurrent generation originating from the characteristic electronic structure purely showing the electric or magnetic parity violation.

The photocurrent conductivity arising from the mechanism based on particle-hole excitations is written by
                \begin{align}
                &\sigma_{a;bc}^\text{e-inj} =  \lim_{\tau \to +\infty} \frac{i\pi e^3 \tau}{2} \int \frac{d\bk}{(2\pi)^d} \sum_{p \neq q} \left(  v_{pp}^a -v_{qq}^a \right) \Omega_{pq}^{bc}I_{pq}(\omega),
                \label{electric_injection}\\
                &\sigma_{a;bc}^\text{m-inj} =  \lim_{\tau \to +\infty} \pi e^3 \tau  \int \frac{d\bk}{(2\pi)^d} \sum_{p \neq q} \left(  v_{pp}^a -v_{qq}^a \right) g_{pq}^{bc}I_{pq}(\omega),
                \label{magnetic_injection}\\
                &\sigma_{a;bc}^\text{shift} =  -\frac{\pi e^3}{2} \int \frac{d\bk}{(2\pi)^d} \sum_{p \neq q} \text{Im} \left( \left[ D_{k_a} \xi^b \right]_{pq} \xi_{qp}^{c} + \left[ D_{k_a} \xi^c \right]_{pq} \xi_{qp}^b   \right) I_{pq}(\omega),
                \label{shift_current}\\
                &\sigma_{a;bc}^\text{gyro} =  \frac{i \pi e^3}{2} \int \frac{d\bk}{(2\pi)^d} \sum_{p \neq q} \text{Re} \left( \left[ D_{k_a} \xi^b \right]_{pq} \xi_{qp}^{c} - \left[ D_{k_a} \xi^c \right]_{pq} \xi_{qp}^b   \right) I_{pq}(\omega).
                \label{gyration_current}
                \end{align}
All the formulas include the factor representing the particle-hole excitations
                \begin{equation}
                        I_{pq} (\omega) = \left\{ f (\varepsilon_{\bk p}) -f (\varepsilon_{\bk q}) \right\} \delta ( \omega + \varepsilon_{\bk q} -\varepsilon_{\bk p}).
                        \label{particle_hole_resonant_factor}
                \end{equation} 
Note that the diverging behavior due to $\tau$ in Eqs.~\eqref{electric_injection} and \eqref{magnetic_injection} is bounded by scattering effects~\cite{de-Juan2017-ph}.
$ D_{k_a} $ denotes the derivative covariant under the gauge transformation by which different energy eigenstates are not admixed with each other.
For instance, in the case of the U(1) gauge associated with the non-degenerate energy spectrum, the covariant derivative is defined by
                \begin{equation}
                \left[ D_{k_a} O \right]_{pq} = \partial_{k_a} O_{pq} - i \left( \xi_{pp}^a - \xi_{qq}^a \right) O_{pq},
                \end{equation}
where the additional terms including the intraband Berry connection $\xi_{pp}^a$ make $\left[ D_{k_a} O \right]_{pq}$ gauge covariant as 
                \begin{equation}
                        \left[ D_{k_a} O \right]_{pq} \to e^{i \phi_p} \left[ D_{k_a} O \right]_{pq} e^{-i \phi_q},
                \end{equation}
under the gauge transformation $\ket{u_{\bk p}} \to \ket{u_{\bk p}} \exp{(-i\phi_p)}$.

Recalling that the absorptive part (`antisymmetric' in Table~\ref{Table_EM_decomposition_response_function}) in the resonant optical conductivity is 
                \begin{equation}
                \text{Re} \,\sigma_{ab}(\omega) \sim \int \frac{d\bk}{(2\pi)^d} \sum_{p \neq q} \xi_{pq}^a \xi_{qp}^b I_{pq} (\omega),
                \end{equation}
the photocurrent response described by $\sigma_{a;bc}^\text{PH}$ can be explained by two steps; irradiating light excites particle-hole pairs similarly to the linear optical response, and the created particle and hole are rectified along opposite directions by `director' determined by the microscopic parity-violating property.
For instance, the director is the group-velocity difference in bands $v_{pp}^a - v_{qq}^a$ for the injection current mechanism.

\begin{figure}[htbp]
                \centering
                \includegraphics[width=0.95\linewidth,clip]{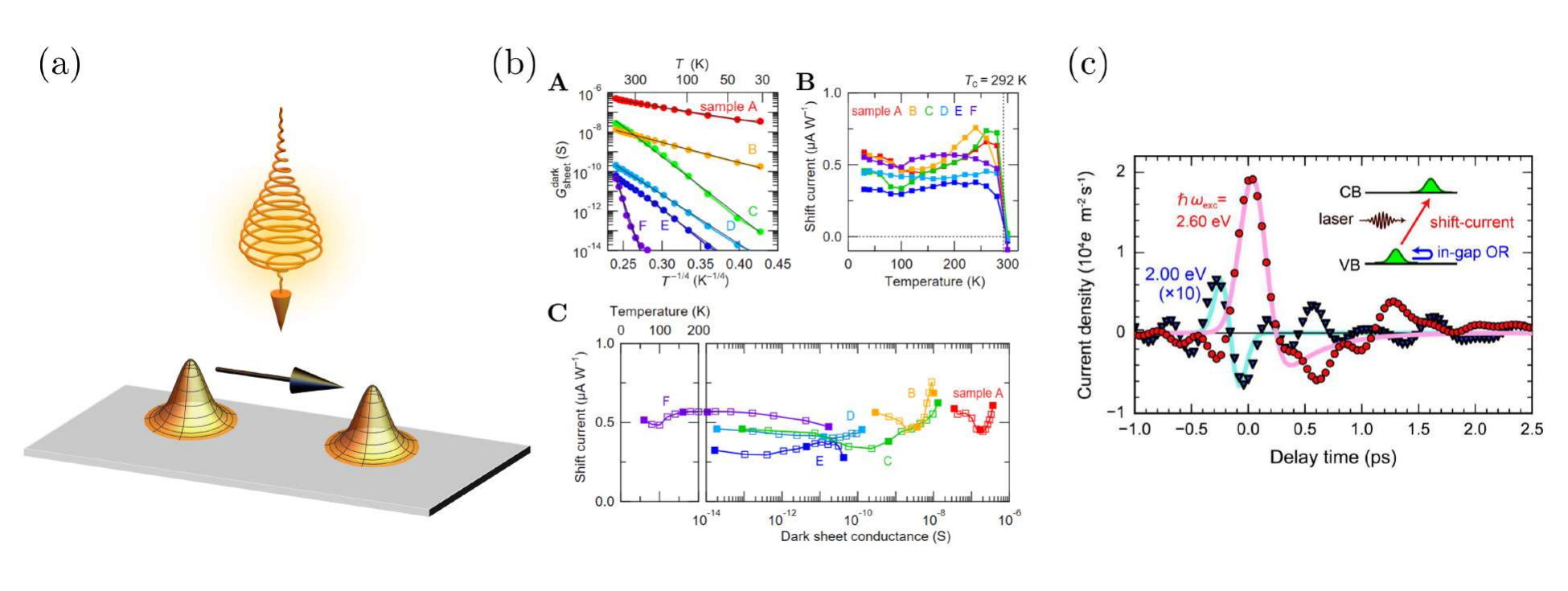}
                \caption{
                        (a) Gyration mechanism for the photocurrent generation.
                        Irradiating circularly-polarized light causes the shift of wave packets (orange-colored) resulting in the photocurrent.
                        (b) Disorder-tolerant property of the shift-current response in a nonmagnetic polar semiconductor SbSI.
                        \textbf{A:} Temperature dependence of the DC electric (dark sheet) conductivity for a series of samples.
                        Plots are $\log{ \sigma_\text{sheet}^\text{dark}}$ with respect to $T^{-1/4}$ and show the linear relation indicating the variable-range hopping transport.
                        \textbf{B:} Temperature dependence of the photocurrent response of the samples.
                        \textbf{C:} Relation between the DC electric conductivity and the shift current with varying samples and temperature.
                        Over a wide range of the DC electric conductivity related to the disorder concentration, the photocurrent attributed to the shift current does show negligible changes in magnitude.
                        (c) Time-resolved measurement of the current density of Sn$_2$P$_2$S$_6$.
                        Compared to the applied pulsive electric field, the electric current profiles differ between that of the shift current (red dot plot) and that comes from the electric polarization excited by in-gap-frequency lights.
                        Panels (b) from \cite{Hatada2020-nu} (\copyright~National Academy of Sciences) and (c) from \cite{Sotome2019-qs} (\copyright~AIP Publishing).
                }
                \label{Fig_gyration_experiment}
                \end{figure}

The shift- and gyration-current mechanisms similarly consist of the director.
When we do not consider the degeneracy of band energy at each momentum $\bk$ for simplicity, the two formulas are recast as
                \begin{align}
                &\sigma_{a;bb}^\text{shift} =  \pi e^3 \int \frac{d\bk}{(2\pi)^d} \sum_{p \neq q} R_{pq;b}^a g_{pq}^{bb} I_{pq}(\omega),
                \label{shift_current_with_shift_vector}\\
                &\sigma_{a;xy}^\text{gyro} =  i \pi e^3 \int \frac{d\bk}{(2\pi)^d} \sum_{p \neq q} \left( R_{pq;+}^a \left| \xi_{pq}^+ \right|^2 -R_{pq;-}^a      \left| \xi_{pq}^- \right|^2  \right) I_{pq}(\omega),
                \label{gyration_current_with_shift_vector}
                \end{align}
where we consider $b=c$ for the shift current and $(b,c) = (x,y)$ for the gyration current without loss of generality because the photocurrents of each effect respond to the linearly-polarized and circularly-polarized lights, respectively.
For the gyration-current mechanism, the circularly polarized light is incident along the $z$ direction by which the photo-electric field is in the $xy$ plane.     
The directors are given by the linear shift vector 
                \begin{equation}
                R_{pq;b}^a = - \partial_{k_a} \text{arg} \, \xi_{pq}^b  + \xi_{pp}^a - \xi_{qq}^a,
                \label{linear_shift_vector}
                \end{equation}
for the shift current mechanism~\cite{von-Baltz1981-yf} and the circular shift vector  
                \begin{equation}
                R_{pq;\pm}^a = - \partial_{k_a} \text{arg} \, \xi_{pq}^{\pm}  + \xi_{pp}^a - \xi_{qq}^a,
                \label{circular_shift_vector}
                \end{equation}
for the gyration current mechanism~\cite{Watanabe2021-bt} with the circular Berry connections $\xi_{pq}^\pm = \left( \xi_{pq}^x \pm i \xi_{pq}^y \right)/\sqrt{2}$.
Since the intraband Berry connection $\xi_{pp}^a$ roughly represents the position of the wave packet with $(\bk,p)$, the shift vectors of Eqs.~\eqref{linear_shift_vector}~\eqref{circular_shift_vector} correspond to the wave-packet positional shift along the $x_a$ direction during the dipole transition $p \leftrightarrow q$~\cite{Morimoto2016-ne,Fregoso2017-pn} [Fig.~\ref{Fig_gyration_experiment}(a)].
As a result, the directors of the shift- and gyration-current mechanisms are interpreted based on the real-space picture, while those of the injection current mechanism are based on the momentum-space picture.

As in the case of the intrinsic Fermi-surface effects, Eqs.~\eqref{electric_injection} and~\eqref{magnetic_injection} are respectively allowed by the electric and magnetic parity violations due to the relevant quantum-geometrical quantities.
The \T{}-even but \PT{}-odd nature of the linear shift vector indicates that the shift current effect is unique to the \T{}-symmetric materials, while the case is opposite for the gyration current characteristic of the \PT{}-symmetric materials.
Note that the shift and gyration current effects can also be explained by the quantum-geometrical quantities defined in the two-band model, that is, the Christoffel symbols which can be halved into the \T{}- and \PT{}-symmetric parts~\cite{Ahn2020-ec}.
Notably, the photocurrent response is similarly induced by the orbital order.
For instance, photocurrent generation due to the orbital-current order has been predicted~\cite{Watanabe2021-yv,Artamonov1992-vr} and expected to give implications for the wide range of exotic quantum phases~\cite{Murayama2021-zg}.

So far, we have explained the photocurrent mechanism derived within the independent particle approximation.
It may be desirable to develop more elaborate treatments for quantitative estimations; \textit{e.g.}, disorder-scattering and electron-correlation effects.
For instance, the photocurrent performance of the electric and magnetic injection effects suffer from hard degradation because the response decreases in inversely proportional to the increasing scattering rate $\gamma \sim \tau^{-1}$ of Eqs.~\eqref{electric_injection} and \eqref{magnetic_injection}. 

The disorder scattering modifies the shift mechanism written by Eqs.~\eqref{shift_current_with_shift_vector} and~\eqref{gyration_current_with_shift_vector}.
However, although the scattering may smear out the resonant condition represented by $\delta ( \hbar \omega + \varepsilon_{\bk q} -\varepsilon_{\bk p})$ in Eq.~\eqref{particle_hole_resonant_factor}, the shift mechanism is expected to show strong tolerance for the disorder effects.
Supporting this argument, in Ref.~\cite{Hatada2020-nu}, the robustness against the disorder concentration has been successfully observed in doped polar semiconductors, in which the shift-current response undergoes negligible degradation despite the high doping level as much as that allowing for sizable electric conductivity. 
Theoretical studies of the shift-current mechanism are in agreement with the experimental results as the response does not significantly change unless the disorders smear out the multiband property relevant to optical excitations~\cite{Morimoto2016-ne,Ishizuka2021-zj} [Fig.~\ref{Fig_gyration_experiment}(b)].

The gyration current similarly leads to disorder-tolerant photocurrent response to the circularly-polarized light because of its similarity to the shift current [Eq.~\eqref{gyration_current_with_shift_vector} and Fig.~\ref{Fig_gyration_experiment}(a)].
Although there exists no experimental demonstration of the disorder-concentration dependence of the gyration current, it may be feasible by the systematic study of doped magnetic semiconductors such as Mn-based \PT{}-symmetric magnets (BaMn$_2$\textit{Pn}$_2$, SrMn$_2$\textit{Pn}$_2$, and so on)~\cite{Watanabe2018-cu}.
It is noteworthy that the ultrafast spectroscopy is informative in identifying the shift mechanism of the gyration-current response as evidenced in Ref.~\cite{Sotome2019-ba,Sotome2019-qs} reporting that the photocurrent generation is distinguished from the polarization current~\cite{Bass1962-lf} by its temporal form [Fig.~\ref{Fig_gyration_experiment}(c)].

Lastly, let us comment on the disorder effect on the Fermi-surface effects [Eq.~\eqref{photocurrent_fermi_surface_total}] which is non-negligible, particularly in the case the low-frequency light irradiation such as terahertz light.
The significant contributions have been observed by detailed studies of frequency dependence of the photocurrent generation~\cite{Olbrich2014-bd,Hild2023-kd} where extrinsic scattering effects play a key role, while the resonant contributions are dominant in the frequency range of visible light~\cite{Plank2018-hv,Matsubara2022-mb}.
The extrinsic mechanism for the photocurrent generation has been formulated in the framework of the semiclassical theory.
The formulas give unified descriptions for the nonreciprocal conductivity and photocurrent generation.
Let us exemplify it by considering the Berry curvature dipole effect given by
                \begin{equation}
                \sigma_{a;bc}(\omega) = -\frac{e^3}{4} \int \frac{d\bk}{(2\pi)^d}  \sum_p  \left( \frac{\tau}{1+i\omega \tau} \epsilon_{acd} \Omega_p^d \partial_{k_b} f_p  + \frac{\tau}{1-i\omega \tau} \epsilon_{abd} \Omega_p^d \partial_{k_c} f_p\right),
                \label{bcd_contribution_unified}
                \end{equation}
in which $\text{Re}\,\sigma_{a;bc}$ converges to Eq.~\eqref{BCD_term} in the dc limit ($\omega \to 0$), and $\text{Im}\,\sigma_{a;bc}$ reproduces Eq.~\eqref{bcd_photocurrent} in the optical limit $\omega \tau \gg 1$~\cite{Sodemann2015-bf,Du2019-sh}.
Because the extrinsic mechanism can contribute to the photocurrent generation~\cite{Belinicher1982-kw,Du2019-sh,Isobe2020-ax,Ma2022-hw}, the skew scattering and side jump effects should be taken into account when we discuss the low-frequency photocurrent response.

The semiclassical treatments may be reasonable for the nonlinear Drude effect [Eq.~\eqref{nonlinear_drude_photocurrent}] and the Berry-curvature-dipole effect of Eq.~\eqref{bcd_contribution_unified}, whereas one has to carefully consider disorder effects on the intrinsic Fermi-surface effects [Eq.~\eqref{intrinsic_fermi_surface_photocurrent}].
Previous theoretical studies have corroborated the possibility of observing the photocurrent response to the light in the in-gap frequency regime, which is related to the intrinsic Fermi-surface effects~\cite{Belinicher1986-tk}.
More rigorous considerations of electric-field correction to the Hamiltonian, namely optical Stark effect, have recently concluded that the in-gap photocurrent is absent in a steady state, while there exists transient photocurrent~\cite{Taguchi2016-pm,Onishi2022-do,Pershoguba2022-mx}. 
The absence of such off-resonant photocurrent generation may be circumvented by including the coupling to the phonon~\cite{Golub2022-yk} and by taking into account the correction to the distribution function~\cite{Shi2023-fr,Matsyshyn2023-ma,Shi2024-pr}.


\section{Availability of \PT{}-symmetric Magnets and Physical Properties}
\label{Sec_control}

\subsection{Switching the compensated magnets}

In general, the antiferromagnetic state is hard to observe because of the zero magnetization, as is evident from the fact that the antiferromagnetism was microscopically identified centuries after the discovery of ferromagnet~\cite{Shull1949-yb}.
The difficulty in manipulation is partly because there is no universal macroscopic field coupled to the antiferromagnetic state in contrast to known controllable order parameters of ferromagnetic and ferroelectric states.
It is, however, not the case for a series of antiferromagnetic where magnetic anisotropy and crystal structure play a key role.

For instance, if the antiferromagnetic order shows the multi-axial magnetic symmetry, the stimuli for the switching ferromagnetic order may control the antiferromagnetic order as well~\cite{Jungwirth2016-gj}.
Let us consider the biaxial antiferromagnets where the N\'eel vector orientation favors either of the $x$ or $y$ direction.
The N\'eel vector directed along the $x$ axis can be switched into the $y$-axis by the external magnetic field along the $x$ axis since it is energetically favorable for the staggered magnetic moments to be perpendicular to the magnetic fields as in the case of the spin-flop transition.
This spin-flop-like mechanism has been applied to the biaxial antiferromagnet IrMn~\cite{Marti2014-bq}.
Similar manipulation of antiferromagnetic order is feasible by making use of the external fields having the same symmetry as the magnetic field such as spin-polarized electric current and pure spin current~\cite{Gomonay2010-ht,Moriyama2018-lc,Reichlova2015-sf}.
The written antiferromagnetic state can be detected in the anisotropic magnetoresistance by which the $x$- or $y$-collinear antiferromagnetic states manifest different electric resistivity typically as much as a few percent~\cite{Shick2010-lh}.
The difference in the magnetoresistance may be improved significantly with a tunneling junction~\cite{Park2011-cx}. 

The spin-flop-like switching of antiferromagnetic order can be applied to various materials showing multi-axial magnetoanisotropy such as Fe$_2$O$_3$ and NiO, but it does not allow us to distinguish the direction of the N\'eel vector, $\bm{L}$ and $-\bm{L}$.
Similarly, the anisotropic magnetoresistance depends on the antiferromagnetic order in the even-order of its order parameter as $\delta \rho (\b{L}) \propto ||\bm{L}||$.

Being in high contrast to partially switchable antiferromagnetic materials, another series of magnets have properties favorable for spintronic applications.
The key is the sublattice degree of freedom with which the antiferromagnetic order can be characterized by the ferroic ordering of even- and odd-parity magnetic multipole moments as discussed in Sec.~\ref{SecSub_multipolar_classification}.
For the case of even-parity magnetic multipole order, one may identify high controllability of antiferromagnetic order by using the external magnetic field, piezomagnetic property, and electric current~\cite{Smejkal2022-rk}.
We do not present detailed discussions of antiferromagnets characterized by the even-parity multipolar order by leaving them to other literature such as Ref.~\cite{Han2023-du,Nakatsuji2022-vr}. 
In the following, we review the availability of the odd-parity magnetic multipolar class including \PT{}-symmetric magnets.

The scheme of switching \PT{}-symmetric magnets differs between the insulators and metals in which we can make use of the magnetoelectric effect and antiferromagnetic Edelstein effect, respectively.
For the former case, the antiferromagnetic domain state may be distinguished by the magnetoelectric free energy
                \begin{equation}
                \mathcal{F}_\text{ME} = -\alpha_{ab} E_a H_b,
                \end{equation}  
where the coefficient $\alpha_{ab}$ depends on the polarization state of the \PT{}-symmetric magnets as it changes the sign ($\alpha_{ab}\to -\alpha_{ab}$) when the magnetic moments are flipped.
By applying both the magnetic and electric fields to be aligned with the symmetry of the allowed magnetoelectric components in $\alpha_{ab}$, the antiferromagnetic state can be manipulated in the deterministic manner distinguishing the sign of the N\'eel vector.
The scheme is applied to the magnetoelectric annealing~\cite{Borisov2005-pm} and has served the successful observation of switching of the antiferromagnetic order~\cite{Van_Aken2007-zf}.
Note that the magnetoelectric energy is typically small and hence manipulation requires the system with good insulating properties to apply the large electric bias.

The different mechanism is available for the switching of the \PT{}-symmetric magnetic metals.
Although the good metallic properties may be unfavorable in the context of magnetoelectric switching, we can utilize the hidden spin polarization coupled to the electrons' momentum instead.
The locally-noncentrosymmetric crystal, a key to \PT{}-symmetric magnetic order, manifests the locking between spin, sublattice, and momentum even in the paramagnetic state as introduced in Sec.~\ref{SecSub_locally_noncentrosymmetric}.
This intriguing locking gives rise to the transport phenomenon written by
                \begin{equation}
                \delta m_{a}^p = \chi_{ab}^p J_b,
                \label{antiferromagnetic_edelstein_effect}
                \end{equation}
with the spin magnetization response $\bm{m}^p$ to the electric current at the $p$-th sublattice~\cite{Yanase2014-mw,Zelezny2014-qr}.
Owing to the locally-noncentrosymmetric symmetry, the response function obeys the relation $\chi_{ab}^p =-\chi_{ab}^q $ where the sites $(p,q)$ are interchanged under the \Pa{} operation.
As a result, the response denotes the correlation between the electric current and antiferromagnetic spin polarization, because the induced spin polarization is compensated in total ($\sum_p \delta \bm{m}^p = \bm{0}$).
By analogy with the Edelstein effect (magnetization response to electric current, inverse magnetogalvanic effect)~\cite{Levitov1985-oy,Edelstein1990-jr}, it is called antiferromagnetic Edelstein effect~\cite{Watanabe2017-qk}.

                \begin{figure}[htbp]
                \centering
                \includegraphics[width=0.80\linewidth,clip]{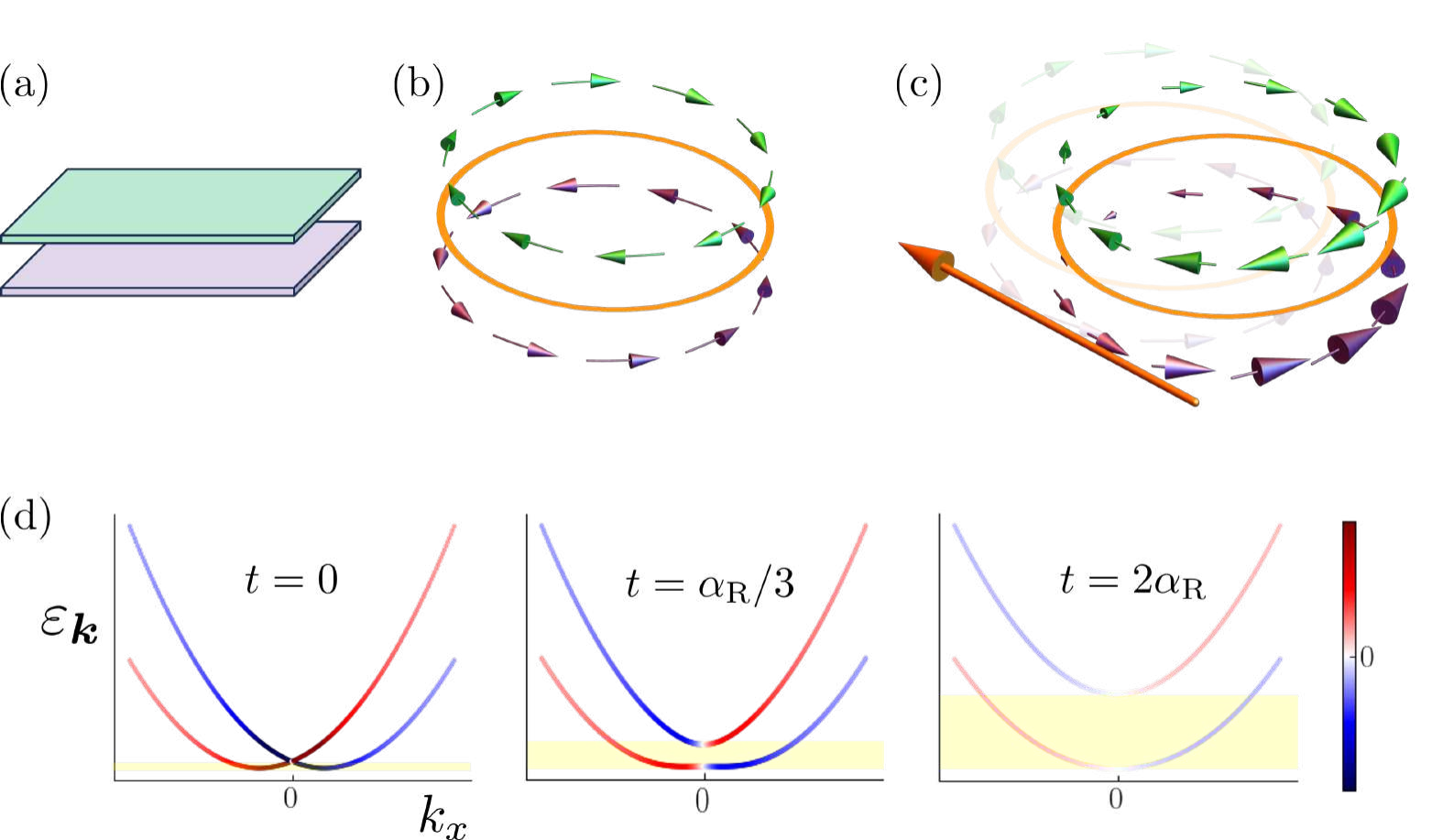}
                \caption{(a) Bilayer system. The green and purple layers are interchanged under the parity operation.
                        (b) Fermi surface (orange-colored circle) and its hidden spin-momentum locking.
                        The momentum-resolved spin polarization is depicted by green and purple arrows for the upper and lower layers of (a).
                        Owing to the locally polar symmetry of the bilayer system, the hidden spin polarization shows the Rashba-type distribution. 
                        (c) Sketch for the antiferromagnetic Edelstein effect.
                        Under the electric-current flow (orange arrow), the imbalanced distribution of the momentum-resolved spin polarization leads to the transverse spin polarization in the opposite manner between the layers.
                        (d) Energy spectrum of the bilayer system colored by the hidden spin polarization with varying the tunneling hopping parameter $t$.
                        The yellow-colored shade indicates the range of chemical potential in which only the lower band is occupied.
                        }
                \label{Fig_AFM-Edelstein}
                \end{figure}

The microscopic grounds for the antiferromagnetic Edelstein effect have been established similarly to the (ferromagnetic) Edelstein effect.
Let us illustrate the mechanism in the framework of the semiclassical theory~\cite{Zelezny2014-qr}.
Note that we here assume the paramagnetic state of a locally-noncentrosymmetric crystal since the mechanism does not require the antiferromagnetic order.
As in Eq.~\eqref{boltzmann_fermi_surface_shift}, the electric current induces the Fermi-surface shift in metals as
                \begin{equation}
                \delta f (\varepsilon_{\bk}) \sim \tau \bm{E} \cdot \nabla_\bk  f (\varepsilon_{\bk}) = \tau \sigma^{-1} \bm{J} \cdot \nabla_\bk f (\varepsilon_{\bk}) = \tau \sigma^{-1} \bm{J} \cdot \nabla_\bk \varepsilon_{\bk} \partial_\varepsilon f (\varepsilon)|_{\varepsilon = \varepsilon_\bk},
                \end{equation}
in which $\tau$ denotes the phenomenological relaxation time and $\sigma$ is the longitudinal conductivity.
In agreement with the form of current-induced response in Eq.~\eqref{antiferromagnetic_edelstein_effect}, the relaxation-time dependence is canceled out as $\tau \sigma^{-1} \propto \tau^0$ by taking the current as the external stimulus.
Owing to the staggered deviation at $\pm \bk$ [$\delta f (\varepsilon_{\bk}) = -\delta f (\varepsilon_{-\bk})$] in the paramagnetic state, the local magnetization is not compensated after the summation over the momentum [Fig.~\ref{Fig_AFM-Edelstein}(c)] as 
                \begin{equation}
                \delta  m_a^p \sim \int d\bk \sigma_a^p (\bk) \delta  f (\varepsilon_{\bk})  \neq 0,
                \end{equation}
different from that in equilibrium where the hidden spin polarization at each crystal momentum $\sigma_a^p (\bk)$ vanishes in total due to the degeneracy between the opposite momentums ($ \bm{m}^p = \bm{0}$ for every sublattice) [Fig.~\ref{Fig_AFM-Edelstein}(b)].
Notably, the antiferroic magnetization occurs only in the presence of the metallic conductivity as is inferred by the Fermi-surface term $\nabla_\bk f (\varepsilon_\bk)$, being contrasting to the magnetoelectric effect.

We note that the structure of hidden spin polarization is determined by the local site symmetry of sites~\cite{Zelezny2017-ov}.
When the locally-noncentrosymmetric crystal structure consists of the two sublattices, the symmetry of $\chi_{ab}^p$ in Eq.~\eqref{antiferromagnetic_edelstein_effect} is identified by the sublattice-dependent antisymmetric spin-orbit coupling of Eq.~\eqref{sublattice_dependent_soc} such as $\chi_{yx}^p = -\chi_{xy}^p$ for the bilayer system [Eq.~\eqref{bilayer_bloch_diagonal}] and $\chi_{xz}^p$ for the zigzag chain [Eq.~\eqref{zigzag_chain_asoc_gvector}].
In the terminology of crystallography, the antiferromagnetic Edelstein effect occurs due to the gyrotropic site symmetry of each sublattice since the gyrotropy ensures the linear coupling between the momentum $\bk$ and spin polarization $\bm{\sigma}$.
Vanishing antiferromagnetic Edelstein effect ($\chi_{ab}^p = 0$) of the honeycomb net follows from the non-gyrotropic site symmetry offering the hidden spin polarization given by Eq.~\eqref{honeycomb_asoc_gvector}.

To obtain a significant response, it is desirable to maximize the hidden spin polarization on the Fermi surface.
For instance, special crystal symmetry such as the nonsymmorphic symmetry suppresses the inter-sublattice hopping which smears out the sublattice-dependent spin-momentum locking (see also Sec.~\ref{SecSub_locally_noncentrosymmetric}).
On the other hand, small inter-sublattice hopping may contribute to enhancing the hidden spin polarization with the fine-tuning of material parameters.
For the Hamiltonian for the bilayer system [Eq.~\eqref{Hamiltonian_bilayer}], the hidden spin polarization is sizable but undergoes the partial compensation between contributions from the inner and outer Fermi surfaces in the absence of the tunneling hopping ($t=0$).
As depicted in the left panel of Fig.~\ref{Fig_AFM-Edelstein}(d), the energy spectrum is singly occupied only in a narrow chemical-potential range.
Such a favorable parameter range can be broadened by including moderate tunneling hopping.
As in the plot with $t= \alpha_\text{R}/3$ of Fig.~\ref{Fig_AFM-Edelstein}(d), one can obtain the single Fermi surface with a non-negligible hidden spin polarization over a wider range of chemical potential.
Note that the hidden spin polarization gets weak if $t$ is much larger than the spin-orbit coupling (right panel of Fig.~\ref{Fig_AFM-Edelstein}(d)).
It has been shown in Ref.~\cite{Yanase2014-mw} that the antiferromagnetic Edelstein effect is significant when the chemical potential is placed in the range where the single Fermi surface appears.
The spin-current generation stemming from the antiferromagnetic Edelstein effect similarly undergoes enhancement~\cite{Suzuki2023-qg}.

When the induced antiferroic magnetization $\left\{ \delta \bm{m}^p \right\}$ has the same symmetry as that of the equilibrium spin configuration, it can directly act on the \PT{}-symmetric magnetic order as the so-called N\'eel spin-orbit torque~\cite{Zelezny2014-qr}.
As the symmetry of the response is determined by the locally-noncentrosymmetric crystal structure, the induced spin torque is field-like; the favorable antiferroic spin polarization depends not on the antiferromagnetic state but on the direction of applied current~\cite{Manchon2019-xj}.
The field-like nature allows for deterministic current-induced switching without any auxiliary fields.

                \begin{figure}[htbp]
                \centering
                \includegraphics[width=0.95\linewidth,clip]{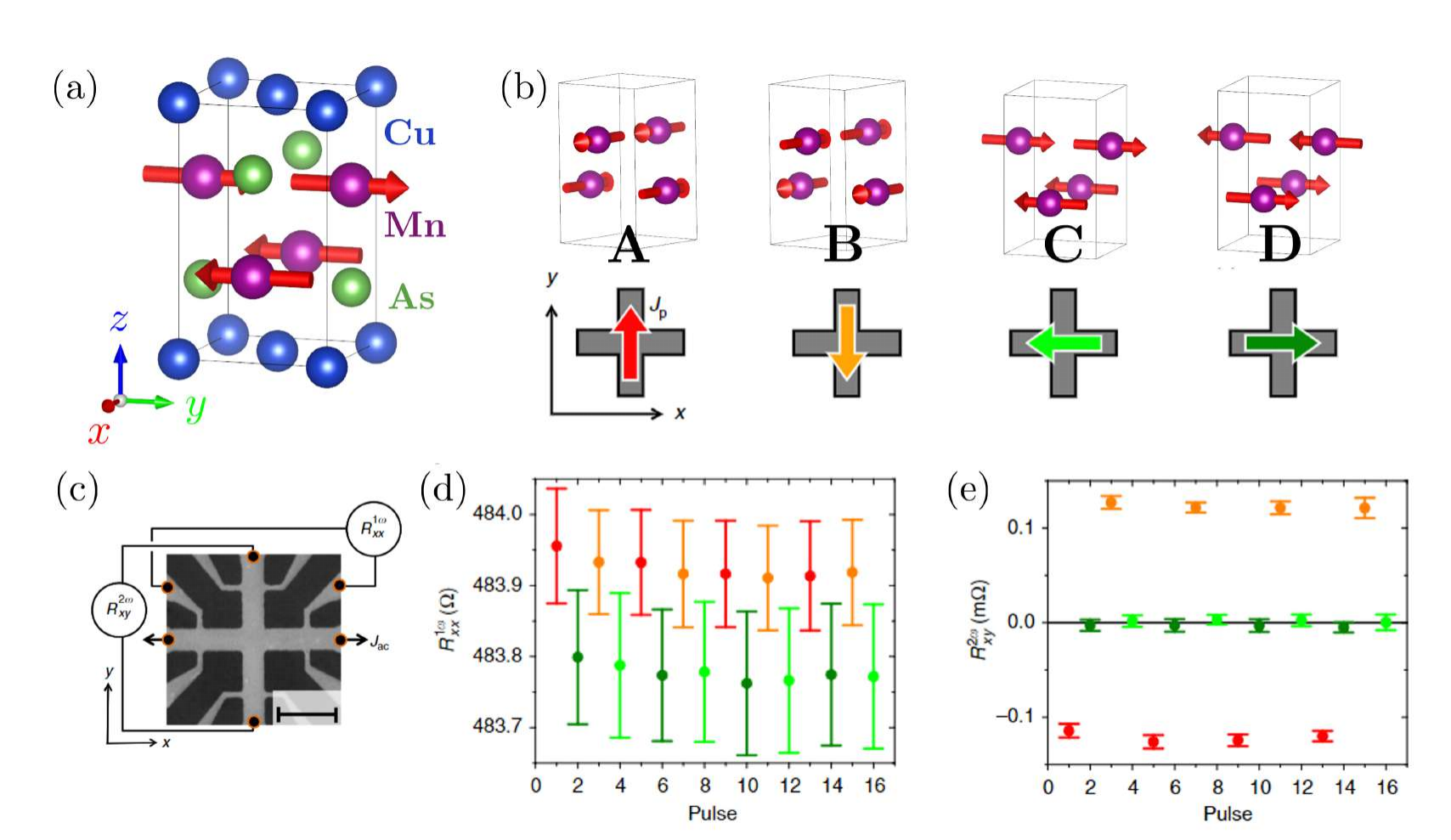}
                \caption{
                        (a) Crystal and magnetic structures of tetragonal CuMnAs~\cite{Wadley2015-ra}.
                        Owing to the biaxial in-plane magnetoanisotropy, the N\'eel vector points to the favorable directions, that is $x$ and $y$ axes.
                        (b) Four domain states (A, B, C, D) of the \PT{}-symmetric magnetic order of CuMnAs (upper panel).
                        In the lower panel, the four colored arrows denote the electric current driving the system to each domain state; \textit{e.g.}, $\bm{J}_\text{p} \parallel +\hat{y}$ for the domain state A.
                        (c) Geometry of measurements for the linear [(d)] and second-order nonlinear conductivity [(e)] of CuMnAs.
                        (d) Anisotropic magnetoresistance of CuMnAs during the current-induced switching of antiferromagnetic domains.
                        The domain state is marked by the colors depicted in the lower panel of (b).
                        The signals distinguish the domain states with different N\'eel-vector orientations such as the A and C domain states but do not distinguish those with the same orientations as A and B. 
                        (e) Nonreciprocal Hall response during the current-induced switching.
                        The domain states (A, B) with the opposite N\'eel vectors are distinguished, while there is no signal for the domain states C and D due to the symmetry constraint.
                        A part of the panel (b) and panels (c,d,e) are taken from \cite{Godinho2018-fn}.
                }
                \label{Fig_cumnas-nonreciprocal-experiment}
                \end{figure}

Following the theoretical predictions, 
the switching of antiferromagnetic states has been experimentally demonstrated by \PT{}-symmetric magnetic metals such as CuMnAs~\cite{Wadley2016-lo,Olejnik2017-lc,Wadley2018-dx} [Fig.~\ref{Fig_cumnas-nonreciprocal-experiment}(a)] and Mn$_2$Au~\cite{Bodnar2018-ud}.
The domain state written by the large pulsive electric current has been monitored by the anisotropic magnetoresistance related to the biaxial magnetoanisotropy.
Both of CuMnAs and Mn$_2$Au show the biaxial in-plane magnetoanisotropy by which the N\'eel vector is favorably directed along the $x$ and $y$ axes.
It follows that there are four possible domain states written by the N\'eel vector $\bm{L}$ as $\bm{L} \parallel \pm \hat{x},~\pm \hat{y}$ [Fig.~\ref{Fig_cumnas-nonreciprocal-experiment}(b)].
These magnetic materials show the locally polar symmetry as in the case of the bilayer system [Eq.~\eqref{bilayer_bloch_diagonal}]~\cite{Barthem2013-lf,Wadley2015-ra}, and the large electric current drives the N\'eel vector to be perpendicular to the applied direction as the current $\bm{J} \parallel \hat{x}$ leads to $\bm{L} \parallel \hat{y}$.

The domain state of biaxial antiferromagnets can be detected if there is appreciable anisotropic magnetoresistance in antiferromagnetic metals whose electronic structure may undergo significant modification by the magnetic order~\cite{Elmers2020-rp}.
The signal is quantified by the N\'eel-vector dependent part of the electric resistance $\delta \rho_{ab} (\bm{L})$.
In the case of CuMnAs, one can find the difference in $ \delta \rho_{ab} (\bm{L})$ as 
                \begin{equation}
                        \delta \rho_{xx} (\bm{L}\parallel \hat{x}) \neq \delta \rho_{xx} (\bm{L}\parallel \hat{y}),
                        \label{AMR_symmetry_CuMnAs}
                \end{equation}
by which the antiferromagnetic state is monitored [Fig.~\ref{Fig_cumnas-nonreciprocal-experiment}(d)].
The order-induced anisotropy in the electronic property has been similarly probed via optical measurements~\cite{Saidl2017-gk,Grzybowski2017-he,Wadley2018-dx,Sapozhnik2018-af,Grigorev2021-wm}.

The readout based on anisotropic magnetoresistance may not be direct evidence for the perfect switching because the anisotropic magnetoresistance does not distinguish all the possible domain states completely; \textit{i.e.}, the magnetoresistance is equivalent between the domain states interchanged by the \Pa{} operation, for example, $\delta \rho_{xx} (\bm{L}\parallel +\hat{x}) = \delta \rho_{ab} (\bm{L}\parallel -\hat{x})$.
In addition, one may take account of thermal effects implying that observed signals are possibly attributed to the nonequilibrium state not to the domain state in equilibrium~\cite{Zhang2019-nd,Chiang2019-ob,Cheng2020-pb,Churikova2020-xr,Surynek2020-lt}.
In contrast to the readout by the anisotropic magnetoresistance, the polarity can be identified by the nonreciprocal conductivity which is sensitive to the magnetic parity violation (see also Sec.~\ref{SecSubSub_nonreciprocal_conductivity}).
The observed change in the odd-parity nonreciprocal response indicates the perfect switching of the \PT{}-symmetric magnetic order~\cite{Godinho2018-fn} [Fig.~\ref{Fig_cumnas-nonreciprocal-experiment}(e)], where the domain states with the opposite N\'eel vectors have been separately detected by the nonlinear Hall conductivity with the antisymmetry
                \begin{equation}
                \sigma_{x;yy} (\bm{L}\parallel +\hat{y}) = - \sigma_{x;yy} (\bm{L}\parallel -\hat{y}).
                \end{equation}
Here, the N\'eel-vector dependence of the nonreciprocal conductivity is explicitly denoted as in Eq.~\eqref{AMR_symmetry_CuMnAs}.
Note that $\sigma_{x;yy} (\bm{L} \parallel \pm \hat{x})= 0$ while $\sigma_{y;xx} (\bm{L} \parallel \pm \hat{x}) \neq  0$.
Similarly, the \PT{}-symmetric magnetic state can be distinguished unambiguously by odd-parity physical phenomena introduced in preceding sections such as the magnetopiezoelectric effect, photocurrent response, and so on~\cite{Song2021-qd,Zhang2019-uy}.

The scheme for perfect switching based on the antiferromagnetic Edelstein effect has been proposed in several candidate materials and has been subsequently generalized to identify the switchability of \PT{}-symmetric magnets in a group-theoretical framework~\cite{Watanabe2018-xp}.
Among a lot of candidate materials for \PT{}-symmetric magnetic states, a class of the magnetic states, that is the magnetic toroidal state (Fig.~\ref{Fig_uniform_toroidal}), enables the current-induced manipulation.
Similarly to the nonmagnetic and noncentrosymmetric materials possessing the electric parity violation, \PT{}-symmetric magnets are classified into the polar and nonpolar classes concerning the unitary symmetry including no \T{} operation.
For instance, the \PT{}-symmetric magnetic states of CuMnAs and Mn$_2$Au show the polar symmetry in the unitary part of their magnetic point group ($2mm$), while that of BaMn$_2$As$_2$ is noncentrosymmetric but nonpolar ($\bar{4}2m$) [Eq.~\eqref{BaMn2As2_magnetic_point_group}].
As a result, the former can be manipulated by the electric current, whereas the latter cannot be done.
Note that the criterion based on the toroidal degree of freedom applies to general cases including \PT{}-violating and noncentrosymmetric magnets.
The presence of the magnetic toroidal moment ensures the feasibility of current-induced switching (see also Appendix~\ref{SecApp-magnetic_materials}).
This criterion is also generalized to cover the switchability of nonpolar magnets with the help of strain~\cite{Watanabe2018-xp} such as demonstrated in ferromagnetic semiconductor~\cite{Chernyshov2009-dy}.

\subsection{Control of electronic property with \PT{}-symmetric magnetic order}

The proof-of-concept experiment of electrical control of antiferromagnets paved the way for intensive investigations of its reading-writing scheme.
It is anticipated that one can realize unprecedented designing of physical properties if the \PT{}-symmetric magnetic order is under our control.
For instance, since the topologically-nontrivial electrons are hosted on the antiferromagnetic conductors~\cite{Tang2016-au}, strong modulation of the electronic bands may be feasible through current-aided manipulation~\cite{Smejkal2017-qb} and by the spin-flop transition~\cite{Masuda2016-ml}.

Furthermore, owing to the \PT{}-symmetric but \Pa{}-violating symmetry, the symmetry and its characteristic responses can be tuned in a nontrivial manner.
As tabulated in Fig.~\ref{Fig_2_space-time-classification}, the \PT{} symmetry forbids uniform electric and magnetic polarizations and thereby keeps the spin degeneracy intact at each crystal momentum.
Since the degenerate states show the intimate coupling between the spin and sublattice degree of freedom, the spin-momentum-sublattice locking can exert unique transport phenomena like the layer Hall effect~\cite{Gao2021-ka} where the anomalous Hall response occurs in the staggered way between the twined layers.
The degeneracy is easily lifted by the external fields violating the \PT{} symmetry such as the external magnetic and electric fields.
The available elimination of \PT{}-ensured degeneracy implies that one can manipulate the physical phenomena arising from the ferroelectric and ferromagnetic order.
For instance, it has been shown that the electric-field effect gives rise to the uniform Berry curvature and associated magnetooptical and magnetogalvanic effects in layered materials such as CrI$_3$~\cite{Jiang2018-pk,Jiang2018-bd,Huang2018-zp} and even-layer MnBi$_2$Te$_4$~\cite{Du2020-tg}.

Similarly, the external magnetic field leads to the physical phenomena originating from the \PT{}-symmetry violation.
For instance, induced small canting of antiferromagnetic moments leads to sizable anomalous Hall response which is called chiral Hall effect~\cite{Lux2020-oo}.
This sizable response is in contrast to the typical case of canted antiferromagnets where the anomalous Hall response may not be significant when the induced magnetization is small.
Intriguingly, the spin-momentum texture experiences the magnetic-field-induced correction that is entirely different from the typical case.
To corroborate the induced spin-momentum structure, let us consider the bilayer system undergoing the layer-dependent spin polarization.
The obtained \PT{}-symmetric antiferromagnetic state can be found in layered magnetic materials.
When the magnetic field is in the out-of-plane direction, the combination of the \PT{}-symmetric magnetic order and external field results in lifting the spin-layer-coupled degeneracy at every crystal momentum and thereby in the Rashba-type spin-momentum locking.

                \begin{figure}[htbp]
                \centering
                \includegraphics[width=0.90\linewidth,clip]{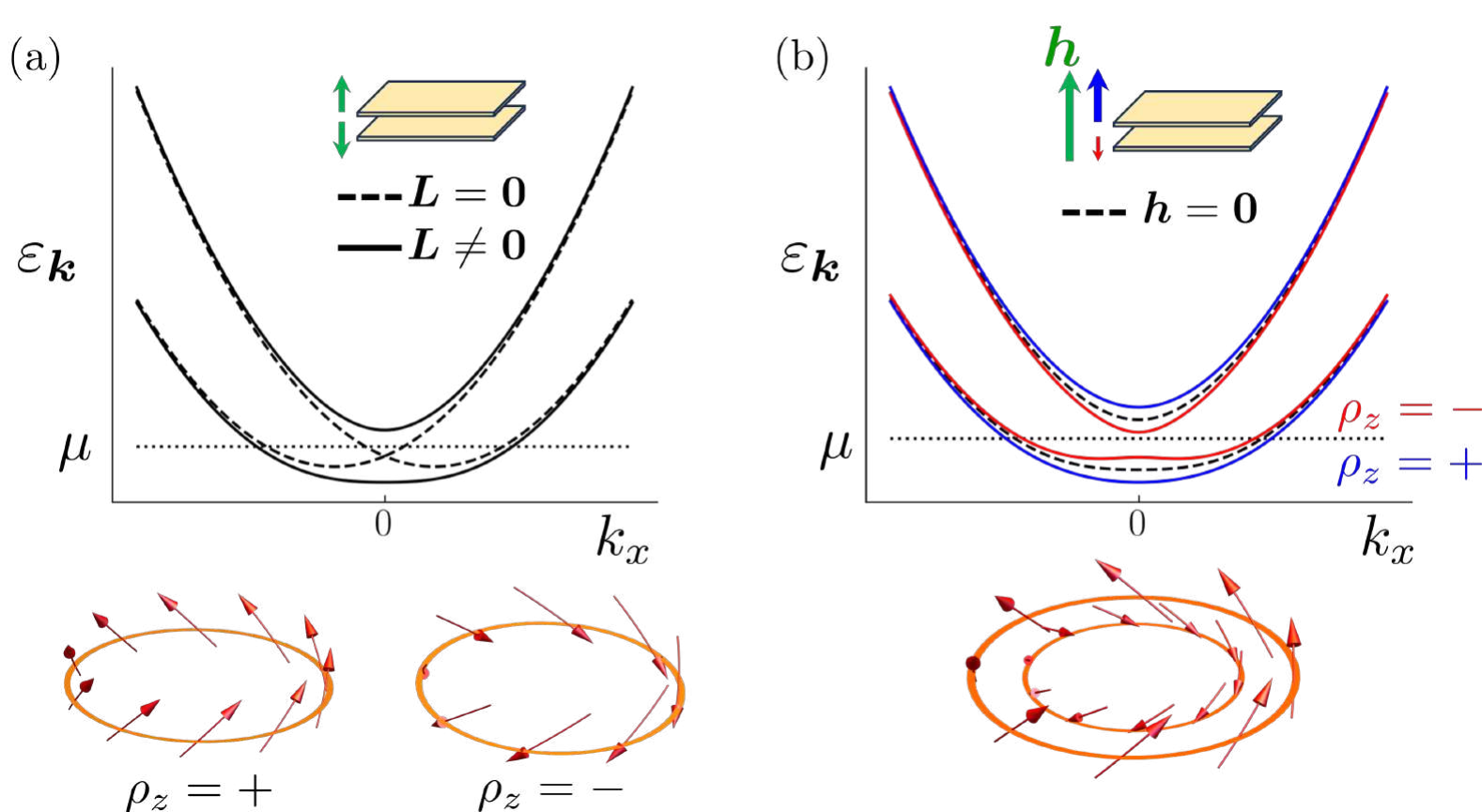}
                \caption{
                        Sketch for the mechanism of magnetic ASOC in the bilayer antiferromagnet.
                        (a) Energy spectrum $\varepsilon_\bk$ with the zero interlayer tunneling $t=0$. 
                        Energy bands in the antiferromagnetic phase with a finite molecular field are shown by solid lines while those in the nonmagnetic phase are by dashed lines.
                        The magnetic structure is in-plane ferromagnetic and interlayer antiferromagnetic as depicted in the inset.
                        Lower panels indicate the spin-momentum texture on the Fermi surface for $\bm{L} \neq \bm{0}$ parametrized by the chemical potential $\mu$ (red arrows denote the spin polarization at each momentum).
                        The Fermi surface, being doubly degenerate due to the \PT{} symmetry, is decomposed into those in the upper ($\rho_z = +1$) and lower layers ($\rho_z = -1$).
                        (b) Energy spectrum 
                        obtained by adding the out-of-plane Zeeman field $\bm{h}$.
                        Solid lines in blue (red) show the energy of electrons on the upper (lower) layers.
                        Dashed lines are equivalent to the solid lines in (a) with $\bm{h}=0$.
                        The degenerate Fermi surface 
                        is split into two showing the opposite helical spin texture with different out-of-plane polarization as depicted in the lower panel, where the spin-momentum structure on each Fermi surface is illustrated.   
                }
                \label{Fig_magnetic-ASOC}
                \end{figure}

The resultant Rashba spin-orbit coupling stems from the purely magnetic modification of electronic structure and does not require the noncentrosymmetric crystal structure, being in sharp contrast to the known cases of spin-momentum locking such as those found in nonmagnetic polar crystals.
Thus, such magnetically-induced antisymmetric spin-orbit coupling is dubbed with \textit{magnetic antisymmetric spin-orbit coupling} (magnetic ASOC)~\cite{Watanabe2020-oe}.
Let us exemplify the magnetic ASOC by taking the model Hamiltonian for the bilayer system [Eq.~\eqref{Hamiltonian_bilayer}].
We further add the antiferromagnetic molecular field and Zeeman field written by
                \begin{equation}
                H_\text{mag} = \sum_{\rho_z = \pm}-\left( \bm{h}  + \bm{L} \rho_z \right)\cdot \bm{\sigma},
                \end{equation}
with $\bm{L} \parallel \hat{z}$.
Note that $\bm{\sigma}$ and $\bm{\rho}$ are Pauli matrices for the spin and layer degrees of freedom, respectively.
In the case of zero tunneling hopping ($t=0$), the momentum-resolved spin polarization quantifying the magnetic ASOC is analytically obtained as
                \begin{equation}
                \Braket{\bm{\sigma}} (\bk,\rho_z) = \frac{\bm{G}}{||\bm{G} ||}, ~\bm{G} = -\alpha_\text{R} \hat{\bm{g}}_\bk \rho_z+ \bm{h} +\bm{L} \rho_z,
                \end{equation} 
for the lower-energy eigenstates of the upper ($\rho_z = +1$) and lower ($\rho_z = -1$) layers. 
The hidden Rashba spin-orbit coupling is given with the vector $\hat{\bm{g}}_\bk = \left( k_y, -k_x, 0 \right)$.
The two lower energy spectrums are originally degenerate due to the \PT{} symmetry, and their energies are slightly separated by the Zeeman field as much as $\sim |\bm{h}|$.
The resultant difference in Fermi-surface volumes and Fermi wavelengths ($k_\text{F}^\pm $) between the layers gives rise to the uncompensated Rashba-like spin-momentum texture (Fig.~\ref{Fig_magnetic-ASOC}).

Furthermore, the magnetic ASOC may take a form different from that of the hidden ASOC determined by the crystal structure in the nonmagnetic phase.
For instance, under the in-plane magnetic field such as $\bm{h} \parallel \hat{x}$, the induced magnetic ASOC has the form of $k_x \sigma_z$ being irrelevant to the existing Rashba spin-orbit coupling hidden by the layer degree of freedom.  
The interlayer tunneling hopping $t$ plays an essential role in producing this type of magnetic ASOC.
This is in sharp contrast to the fact that the large inter-subsector hopping weakens the hidden spin polarization governed by the paramagnetic Hamiltonian.

The magnetic ASOC may offer tunable spin-momentum locking, while the conventional ASOC is usually determined by the noncentrosymmetric crystal structure whose parity violation is hard to control externally.
Considering vast \PT{}-symmetric magnetic materials~\cite{Watanabe2018-cu}, we can find various types of magnetic ASOC different from the Rashba ASOC, such as the Dresselhaus-type~\cite{Watanabe2020-oe} and the chiral ones.
The concept of the magnetic ASOC can be comprehensively generalized to magnetically-induced electric parity violations.
Then, combining the \PT{}-symmetric magnet with the external magnetic field, one can obtain the various physical phenomena unique to the nonmagnetic and noncentrosymmetric crystals; \textit{e.g.,} piezoelectricity, nonlinear Hall effect arising from the Berry curvature dipole~\cite{Watanabe2020-oe}, photogalvanic effects for charge and spin current~\cite{Merte2023-vt}, and so on.
The emergence of the Berry curvature dipole is identified by the discussions parallel to those of the magnetic ASOC because the spin and momentum-space Berry curvature have the same space-time symmetry (\Pa{}-even and \T{}-odd).
The magnetically-tunable responses discussed above can be systematically identified with the help of magnetic symmetry analysis~\cite{Erb2020-jb,Watanabe2020-oe,Yatsushiro2022-ui} and are expected to be a key to rich spintronic phenomena based on antiferromagnetic materials.


\section{Parity-violating Superconducting Responses}
\label{Sec_SC_and_MPV}

In the preceding sections, we overviewed the physical phenomena induced by the magnetic parity violation.
Various emergent responses and itinerant properties originate from the electronic structure unique to the parity violation.
Similarly, such intriguing electronic structures have a significant influence on the quantum phases emerging in metals such as superconductivity.
Noncentrosymmetric superconductors, superconductors with the lack of the \Pa{} symmetry, have been intensively studied in light of the strong spin-orbit coupling as found in heavy fermion and van der Waals materials~\cite{NCSC_book,Ideue2021-wi}.
Those studies mainly worked on modifications of the basic superconducting properties such as critical magnetic fields and pairing states.
This section is devoted to discussions of physical phenomena unique to the parity violation in superconductors.

Let us consider candidates for the noncentrosymmetric superconductors other than known materials crystalizing in the noncentrosymmetric structure.
The parity violation can be invoked in the following three ways; (1) parity-violating order concurrently existing with the superconductivity, (2) the supercurrent injection, and (3) multiple superconducting transitions where the \Pa{} parity is different between those superconducting pairing potentials~\cite{Watanabe2022-hk}.

Firstly, the parity violation due to the spontaneous order is ascribed to the nonmagnetic and magnetic origins.
Specifically, for the former the ferroelectric- or piezoelectric-like order has been identified to coexist with the superconductivity in doped SrTiO$_3$~\cite{Rischau2017-aw} and heavy-fermion systems~\cite{Hu2017-fx}, while for the latter the odd-parity magnetic multipolar order may be involved in the superconductivity of a locally-noncentrosymmetric superconductor CeRh$_2$As$_2$~\cite{Kibune2022-vn,Kitagawa2022-si}.
The order-induced modifications of the electronic structures may assist the emergence of exotic superconductivity such as the finite-momentum pairing state in the presence of the magnetic parity violation~\cite{Sumita2016-nb,Sumita2017-cr}. 
Although similar interplay between the superconductivity and magnetic parity violation can be realized in the noncentrosymmetric superconductors under the external magnetic field~\cite{NCSC_book,Smidman2017-jl,Kaur2005-jl,Wakatsuki2017-ft,Dimitrova2003-rl}, the odd-parity magnetic order may offer more significant modification of the electronic property with the energy scale of the Hund's coupling, which can be much larger than the Zeeman coupling, and cause a prominent parity-violating effect on the superconductivity.

Secondly, the parity violation can be built into the superconductor by the supercurrent injection.
The superfluidity allows the current to flow without the Joule heating and thereby realizes the parity-violating phase even in prototypical $s$-wave superconductors.
Owing to its space-time symmetry, the biased electric current gives rise to the magnetic parity violation.
The resulting finite-momentum superconductivity has been verified in experiments detecting the optical signals unique to the parity violation~\cite{Yang2019-cw,Nakamura2020-wt,Vaswani2020-ij}.

Lastly, the superconductivity itself can break the \Pa{} symmetry.
Since the \Pa{} parity is definite in the conventional pairing states such as the even-parity $s$-wave and odd-parity for $p$-wave superconductivity in centrosymmetric materials, the parity violation does not happen at the single superconducting transition.
Note that the \Pa{} operation is effectively retained in the odd-parity superconducting states because the \Pa{} symmetry is preserved in the form of its combination with the U(1) gauge transformation.
This is a significant difference from the odd-parity multipolar order.
On the other hand, spontaneous symmetry breaking occurs if the system exhibits multiple superconducting transitions.
Evidencing such possibility, Ginzburg-Landau analysis of the multiple superconducting transitions showed that the $s+ip$-wave pairing state, an exotic pairing state showing the purely magnetic parity violation, is stable when both of $s$- and $p$-wave superconductivity emerge~\cite{Wang2017-co}.
Interestingly, the multiband nature in the normal phase allows for richer properties of magnetically parity-violating superconductivity in the Bogoliubov-quasiparticle spectrum~\cite{Kanasugi2022-pb,Kitamura2023-ep}.
Following the parallel discussion of the electrically-switchable \PT{}-symmetric magnets, one can identify the possibility of supercurrent-induced switching of the magnetically parity-violating superconductivity, such as between the $s \pm i p$-wave states.
When the superconducting order parameters have the same symmetry as the magnetic toroidal order, it is called anapole superconductivity~\cite{Kanasugi2022-pb} and allows supercurrent-induced switching.
The spontaneously parity-violating superconductivity demands the stringent condition of sizable fluctuations leading to the multiple Cooper instability and thus material realization requires further studies, while such fluctuations have been implied in a heavy-fermion superconductor~\cite{Ishizuka2021-ry,Aoki2022-uk}.

The superconductivity-induced symmetry breaking is similarly found in the chiral superconductivity manifesting the \T{}-symmetry breaking.
The chiral superconductivity, however, differs from the parity-violating superconductivity as it can occur in the single transition labeled by a multi-dimensional irreducible representation~\cite{Sigrist1991-bd,Sigrist_Sr2RuO4}.

Considering the playgrounds for the parity-violating superconductivity, one may be interested in their characteristic physical responses.
Recently, active experimental and theoretical research has identified various nonreciprocal responses related to superconductivity.
For instance, tremendous interest has been drawn by the recent discoveries of the nonreciprocal charge transport of superconductors such as nonreciprocal conductivity, superconducting diode effect, and nonreciprocal Josephson effect~\cite{Ideue2021-wi,Nadeem2023-fd,Nagaosa2024-fr}.
Furthermore, the superconducting property shows up in various responses including cross-correlated and optical responses as follows.

Here, we review the cross-correlated responses in superconductors.
Let us consider the free energy in the superconducting state dependent on external fields (magnetic field $\bm{H}$, stress $\hat{s}$, and vector potential $\bm{A}$)
        \begin{equation}
        F = F (\bm{H},\hat{s},\bm{A}).
        \end{equation}
Importantly, the vector potential characterizes the supercurrent injection through the London equation.
Taking the derivative with respect to each field, we obtain the conjugate quantities as 
                \begin{equation}
                \bm{M} = -\frac{\partial F}{\partial \bm{H}},~ \hat{\varepsilon} = -\frac{\partial F}{\partial \hat{s}},~ \bm{J}^\text{sc} = - \frac{\partial F}{\partial \bm{A}}, 
                \end{equation}
that is, magnetization, strain, and supercurrent.
The free energy is transformed as
                \begin{equation}
                        F (\bm{H},\hat{s},\bm{A}) \to  F (\bm{H},\hat{s},-\bm{A}),
                \end{equation}
under the \Pa{} operation and 
                \begin{equation}
                        F (\bm{H},\hat{s},\bm{A}) \to  F (-\bm{H},\hat{s},-\bm{A}),
                \end{equation}
under the \T{} operation.
Then, corrections to the free energy arising from the parity violation is given by
                \begin{equation}
                \delta F_\text{cc} = - \alpha_{ab} H_a A_b - \beta_{ab} s_a A_b,
                \label{superconducting_cross_correlation} 
                \end{equation} 
up to the bilinear form.
The coefficient $\alpha_{ab}$ is \T{}-even but \PT{}-odd, while the parity of $\beta_{ab}$ is opposite for each operation.
It follows that the cross-correlation $\alpha_{ab}$ ($\beta_{ab}$) is unique to the superconductors with the electric (magnetic) parity violation.
One can obtain the cross-correlated response from the former coupling as
                \begin{equation}
                M_a = \alpha_{ab} J^\text{sc}_b,
                \label{superconducting_Edelstein_effect}
                \end{equation}
that is the superconducting Edelstein effect~\cite{Edelstein1995-cp} being in stark contrast to the normal Edelstein (inverse magnetogalvanic) effect in which the ohmic current leads to the magnetization~\cite{Levitov1985-oy,Edelstein1990-jr}~\footnote{
        Note that the inverse response written by $J^\text{sc}_b = \alpha_{ab} H_a $ does not occur in the DC regime due to the Bloch's no-go theorem~\cite{Ohashi1996-ko}.}.
Notably, the magnetization response of Eq.~\eqref{superconducting_Edelstein_effect} can be significant due to nontrivial contribution from the topological edge states of nodal superconductors~\cite{Ikeda2020-kh}.

For the latter coupling in Eq.~\eqref{superconducting_cross_correlation}, one can identify the strain response to the supercurrent
                \begin{equation}
                \varepsilon_a = \beta_{ab} J^\text{sc}_b,
                \end{equation} 
called superconducting piezoelectric effect~\cite{Chazono2022-pd,Chazono2023-cy}.
As in the case of the magnetization-supercurrent correlation, the superconducting piezoelectric effect is clearly distinguished from the conventional (inverse) piezoelectric and magnetopiezoelectric effects in terms of the fields stimulating the strain, since the latter two effects are involved in the electric field and ohmic current, respectively.
One can refer to Eqs.~\eqref{piezoelectric_repsonse} and \eqref{current_induced_strain} for comparison.
Unlike them, superconducting Edelstein and superconducting piezoelectric effects are the responses to the supercurrent.

\begin{figure}[htbp]
                \centering
                \includegraphics[width=0.90\linewidth,clip]{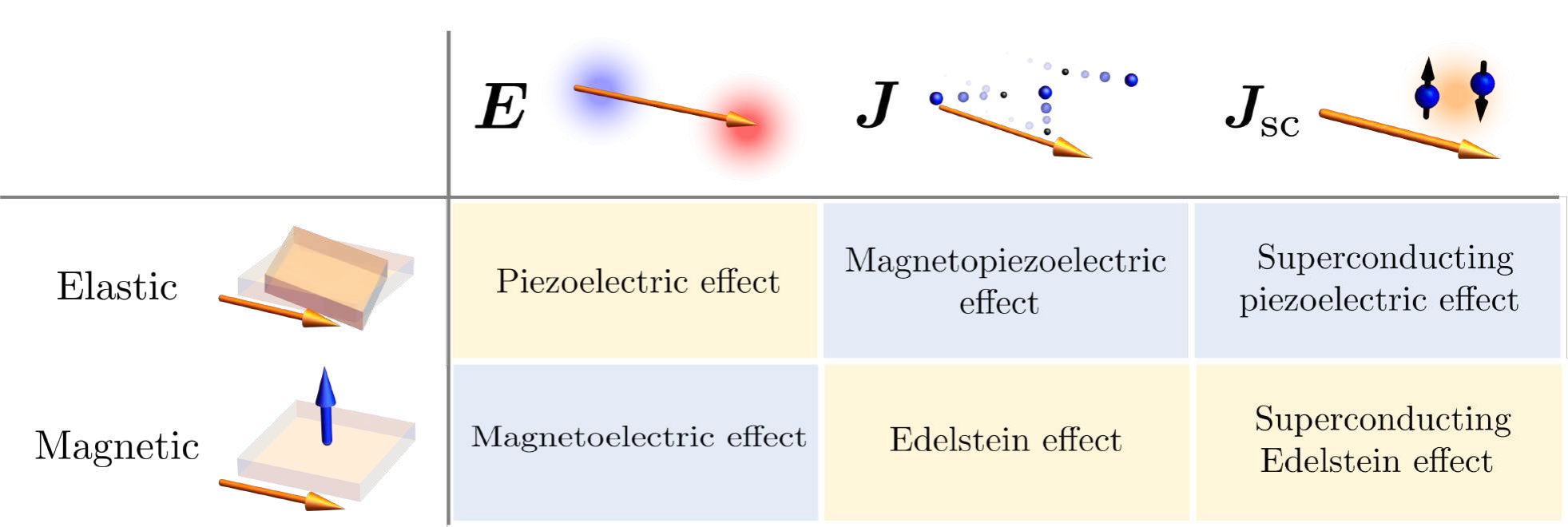}
                \caption{
                        Classification of the cross-correlated responses.
                        The cross correlation is tabulated for the strain (elastic) and magnetization (magnetic) responses to the electrical stimuli, that is electric field $\bm{E}$, ohmic electric current $\bm{J}$, and supercurrent $\bm{J}_\text{sc}$.
                        The responses are colored in blue and yellow depending on the type of parity breaking; the responses shaded in yellow are induced by the electric parity violation, while those in blue are induced by the magnetic parity violation.  
                        }
                \label{Fig_superconducting_cross_correlation}
                \end{figure}

We summarize the cross-correlated responses in Fig.~\ref{Fig_superconducting_cross_correlation}. 
To be of interest, it has been shown that the superconducting piezoelectric effect displays an abrupt change during the crossover between the two superconducting phases in the Rashba superconductor~\cite{Chazono2022-pd}, that is, helical and Fulde-Ferrell states~\cite{Dimitrova2003-rl,Kaur2005-jl}.
This implies that the cross-correlated response plays a key to identifying the exotic phase induced by the coupling between superconductivity and magnetic parity violation.
Notably, the supercurrent injection in bulk can be replaced with the biased Josephson junction~\cite{Kapustin2022-lf} because the vector potential plays a role equivalent to the gradient of the superconducting phase due to the minimal coupling.
For instance, the giant superconducting Edelstein effect is possibly realized in the biased Josephson junction, which could be essential for superconductor-based spintronics.
Although the research of superconducting nonreciprocal responses has advanced significantly as reviewed below and elsewhere~\cite{Nagaosa2024-fr,Nadeem2023-fd}, cross-correlated responses in superconductors have not been explored experimentally, pointing to the next issue in superconducting science.

Next, we discuss the nonreciprocal responses in superconductors.
The free energy can be expanded to nonlinear order with respect to the external fields as well.
Considering only the effects of vector-potentials, we obtain the correction up to the cubic components as
                \begin{equation}
                \delta F_{\bm{A}} = - \frac{1}{2} \rho_{ab} A_a A_b - \frac{1}{3} f_{abc} A_a A_b A_c.
                \label{freeenergy_with_NRSF}
                \end{equation} 
The quadratic correction results in the well-known Meissner response, while the cubic coefficient $f_{abc}$ gives \Pa{}-odd, \T{}-odd, and \PT{}-even corrections.
Then, one arrives at the nonreciprocal response induced by the magnetic parity violation
                \begin{equation}
                \frac{1}{2}  \left[ J_a^\text{sc} (\bm{A}) - J_a^\text{sc} (-\bm{A}) \right] =  f_{abc} A_b A_c.
                \label{nonreciprocal_Meissner_effect}
                \end{equation}
This formula represents the nonreciprocal Meissner effect, by which the supercurrent response differs depending on the direction of the supercurrent shielding the external magnetic field~\cite{Watanabe2022-gh}.
The nonreciprocal response is determined by the nonreciprocal correction to the superfluid weight $f_{abc}$, namely, nonreciprocal superfluid weight.
Its relevance to the superfluid weight follows from the relation
                \begin{equation}
                f_{abc}= \lim_{\bm{A} \to \bm{0}} \partial_{A_a} \rho_{bc}.
                \end{equation}
The nonreciprocal superfluid weight plays a key role in various nonreciprocal responses of superconductors, not only the nonreciprocal Meissner effect [Eq.~\eqref{nonreciprocal_Meissner_effect}], but also the nonreciprocal conductivity~\cite{Hoshino2018-ub} and the nonreciprocal optical responses~\cite{Watanabe2022-hk}.

For the nonreciprocal current generation of Eq.~\eqref{nonreciprocal_current_generation_formula}, let us again consider the photocurrent response as in Sec.~\ref{SecSubSub_photocurrent}.
The perturbative calculations are straightforwardly performed if one works on the molecular-field (BCS) approximation for the superconductivity~\cite{Xu2019-xk,Watanabe2022-hk} where the electronic excitation is attributed to the Bogoliubov quasiparticles.
The nonreciprocal optics of superconductors have unique features that are not found in the normal conducting phase.
One is the optical excitation associated with the van Hove singularity in the Bogoliubov spectrum, which gives the peak of the DOS at the gap edge, and another is the anomalous nonreciprocal optical responses.

To introduce these features, we consider the expression for the photocurrent conductivity in the clean limit.
The formula consists of two contributions 
                \begin{equation}
                \sigma_{a;bc} ( 0;\omega,-\omega)  = \sigma_{a;bc}^\text{reg} + \sigma_{a;bc}^\text{ano}.
                \label{total_photocurrent_superconducting_state}
                \end{equation}
The first and second terms are conventional and anomalous contributions, respectively. 
The conventional contribution ($\sigma^\text{reg}$) is given by the expressions similar to those in the normal state [Eq.~\eqref{total_photocurrent_response_normal}]; \textit{e.g.}, it originates from the Fermi-surface excitation and the resonantly-created Bogoliubov quasiparticles~\cite{Xu2019-xk,Watanabe2022-hk}.
The Fermi-surface contribution may be significant in nodal-gap superconductors such as those manifesting the Bogoliubov Fermi surface~\cite{Agterberg2017-rl}, although it disappears in the full gap superconductors at zero temperatures.
Note that the Berry connection contributing to various nonreciprocal optical responses is defined in the parameter space described by the vector potential, not by the momentum.
This is because the electrons and holes are treated on equal footing in the framework of the Bogoliubov-de Gennes Hamiltonian and thereby the minimal coupling does not imply the equivalence between the derivatives with respect to the vector potential ($\partial_{\bm{A}}$) and to the crystal momentum ($\partial_\bk$).

\begin{figure}[htbp]
                \centering
                \includegraphics[width=0.95\linewidth,clip]{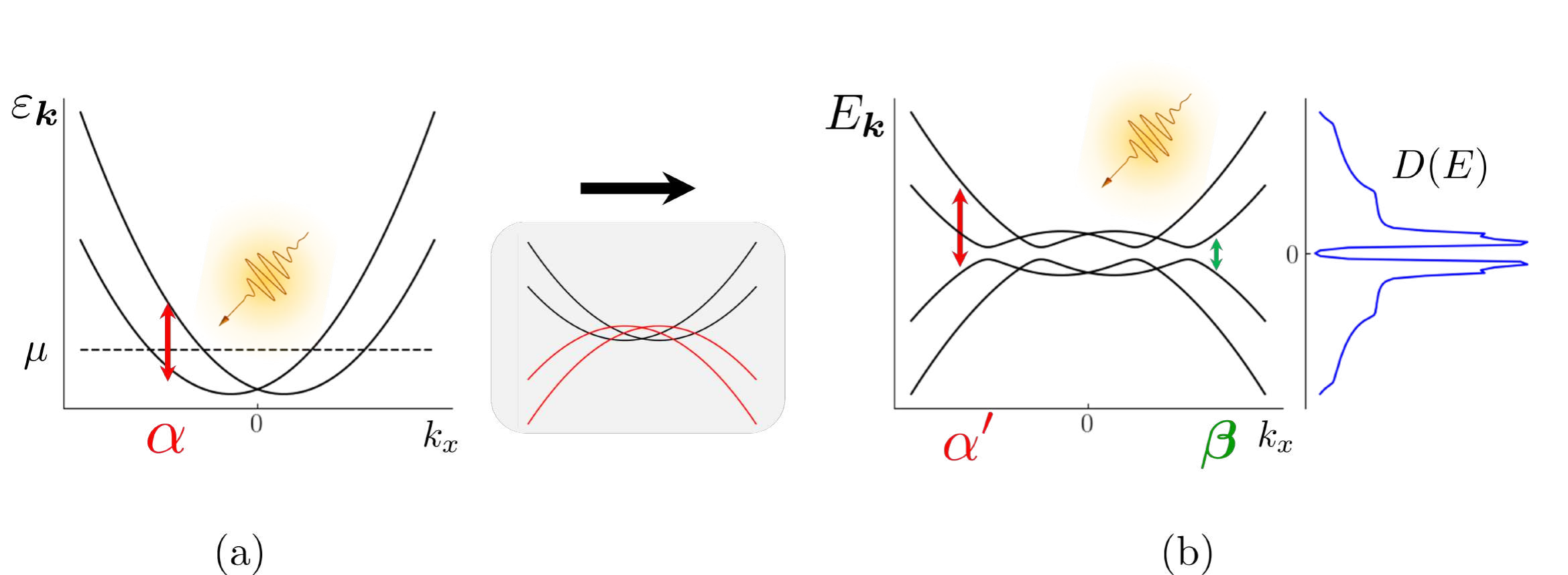}
                \caption{
                        Schematics of resonant optical excitations in the normal and superconducting states.
                        (a) Energy spectrum $\varepsilon_\bk$ in the normal state of the Rashba model and its optical excitation across the chemical potential $\mu$ denoted by $\alpha$.
                        (b) Energy spectrum $E_\bk$ of Bogoliubov quasiparticles in the superconducting state, optical paths ($\alpha',\beta$), and density of states $D (E)$.
                        The energy spectrum is obtained by introducing the pair potential to the Rashba Hamiltonian in (a).
                        The high-frequency optical excitation $\alpha'$ is almost the same as that in the normal state ($\alpha$), while the optical path $\beta$ bridging the coherence peaks is unique to the superconducting state.    
                        }
                \label{Fig_sc-optical-path}
                \end{figure}

Let us consider the resonant optical response of superconductors included in the conventional contribution $\sigma_{a;bc}^\text{reg}$.
An intriguing property of the resonant superconducting optical responses is the optical excitation available after the formation of the superconducting gap.
When the system undergoes the superconducting transition, the superconducting gap gives rise to the van Hove singularity, which is related to the characteristic behaviors of physical properties.
In terms of the optical response, the energy spectrum of Bogoliubov quasiparticles indicates the optical excitation $\beta$ bridging the van Hove singularities as $\omega \sim 2 |\Delta|$ ($\Delta$: pair potential) in addition to the high-frequency optical path $\alpha'$ [Fig.~\ref{Fig_sc-optical-path}(b)].
We note that the path $\alpha'$ gives the optical response similar to that given by the path $\alpha$ in the normal state [Fig.~\ref{Fig_sc-optical-path}(a)] if the frequency of light is much higher than the superconducting gap energy.
Thus, the optical path $\beta$ is unique to the superconducting state, however, it was considered to make no contribution to the optical response due to the selection rule related to the \T{} symmetry~\cite{SchriefferBook}.

The absence of optical excitations relevant to the optical path $\beta$ has been recently revisited in the context of the superconducting fitness~\cite{Ahn2021-jb} quantifying the multiband nature of the pair potential~\cite{Fischer2013-dh,Ramires2016-xe}.
The newly-contributing optical path $\beta$ does not work in prototypical superconductors such as the conventional $s$-wave superconductors, whereas it is not the case if one takes into account the strong-coupling effect, multi-band nature in the normal state, and supercurrent flow~\cite{Bickers1990-mm,Ahn2021-jb,Jujo2022-qw}. 
Consistent with this argument, the peak structure stemming from the van Hove singularity has been identified in the optical spectrum for the linear~\cite{Ahn2021-jb} and second-order optical responses~\cite{Tanaka2023-nv}.

In contrast to the conventional contributions discussed above, the anomalous contribution to the photocurrent conductivity Eq.~\eqref{total_photocurrent_superconducting_state} is characteristic of the parity-violating superconductors and has no counterpart in the normal phase.
The formula is explicitly given by the two terms both of which are written by the total derivative with respect to the vector potential as 
                \begin{equation}
                \sigma_{a;bc}^\text{ano} = - \frac{1}{2\omega^2} f_{abc} + \frac{1}{4\omega^2} \sum_{p\neq q} \lim_{\bm{A} \to \bm{0}} \partial_{A_a} \left[ J_{pq}^b J_{qp}^c \left( \frac{1}{\omega - \varepsilon_{p}+ \varepsilon_{q}} + \frac{1}{\varepsilon_{p} - \varepsilon_{q}} \right) \left\{ f (\varepsilon_p) - f (\varepsilon_q) \right\} \right],
                \label{anomalous_photocurrent_total}
                \end{equation}
with the paramagnetic current operator $\hat{J}^a$ and indices ($p,q$) for the energy eigenstates.
We note that the gapful superconductivity is assumed in Eq.~\eqref{anomalous_photocurrent_total}.
It should be noticed that the chain rule between $\partial_{\bm{A}}$ and $\partial_\bk$ holds in the normal state, and thereby the anomalous contribution vanishes due to the periodicity in the momentum space~\cite{Michishita2021-rb}.
The nonreciprocal superfluid weight covers various superconducting properties such as the Meissner and photocurrent responses, similarly to the (linear) superfluid weight~\cite{Ferrell1958-uz}.

The anomalous contribution is comprised of the nonreciprocal superfluid density and the second term of Eq.~\eqref{anomalous_photocurrent_total}.
The latter part is proportional to the vector-potential derivative of the reactive part in the linear optical conductivity
                \begin{equation}
                        \sigma_{ab}^\text{reg} (\omega) =\frac{1}{i\omega }  \sum_{p\neq q} J_{pq}^b J_{qp}^c \left( \frac{1}{\omega - \varepsilon_{p}+ \varepsilon_{q}} + \frac{1}{\varepsilon_{p} - \varepsilon_{q}} \right) \left\{ f (\varepsilon_p) - f (\varepsilon_q)  \right\}.
                \end{equation}
Given this expression, the second term in Eq.~\eqref{anomalous_photocurrent_total} is called the conductivity derivative.
By taking the DC limit, the conductivity derivative becomes the derivative of the Berry curvature as
                \begin{align}
                        \frac{i}{4\omega}\lim_{\bm{A} \to \bm{0}} \partial_{A_a} \sigma_{bc}^\text{reg} (\omega)
                                &=\frac{1}{4\omega^2}\lim_{\bm{A} \to \bm{0}} \partial_{A_a}  \sum_{p\neq q} \left[ J_{pq}^b J_{qp}^c \left( \frac{1}{\omega - \varepsilon_{p}+ \varepsilon_{q}} + \frac{1}{\varepsilon_{p} - \varepsilon_{q}} \right) \left\{ f (\varepsilon_p) - f (\varepsilon_q) \right\}\right],\\ 
                                &=  \frac{i}{4\omega} \lim_{\bm{A} \to \bm{0}} \epsilon_{bcd} \, \partial_{A_a} \left( \sum_p \Omega_{p}^d f_p \right). 
                \end{align}
Note that the derivative of the Berry curvature is totally different from the Berry curvature dipole defined in Eq.~\eqref{BCD_term} where the derivative acts only on the Berry curvature.
Being concerned with the anomalous photocurrent response at the frequency below the superconducting gap, we can attribute the response to the derivative of the superfluid weight $f_{abc}$ and the total Berry curvature $\partial_{A_a} \left( \sum_p \Omega_p^d f_p \right)$.
These terms give rise to the photocurrent responses to the linearly and circularly polarized light, respectively.

An intriguing property of the anomalous contribution is the diverging behavior in the low-frequency limit; \textit{i.e.}, the nonreciprocal superfluid weight leads to the diverging photocurrent conductivity $\sigma_{a;bc} \propto \omega^{-2}$ with the frequency of irradiating light $\omega$, and the derivative of the Berry curvature does $\sigma_{a;bc} \propto \omega^{-1}$.
The diverging behavior does not result from the assumption of the clean limit in Eq.~\eqref{total_photocurrent_superconducting_state}, while scattering effects restrict the normal contributions to a finite value in the DC limit~\cite{Du2019-sh,Michishita2021-rb}.
The robustness of the low-frequency divergence has been confirmed in the numerical studies by varying the scattering rate~\cite{Watanabe2022-hk}.
The low-frequency divergence and disorder-tolerant nature may be advantageous for applications to optoelectronic devices based on superconductors.

Furthermore, the two anomalous contributions show the contrasting space-time symmetry and are therefore conveniently classified by the electric and magnetic parity violations.
The nonreciprocal superfluid weight is \T{}-odd and \PT{}-even as mentioned around Eq.~\eqref{freeenergy_with_NRSF}, while the derivative of the Berry curvature has the same space-time symmetry as the Berry curvature dipole, that is \T{}-even and \PT{}-odd.
It implies that the anomalous nonreciprocal optical response is a convenient tool for identifying the symmetry of parity violation.
Under the irradiation of the low-frequency light, the \PT{}-symmetric superconductors show significant photocurrent response to the linearly-polarized light and the \T{}-symmetric superconductors do that to the circularly-polarized light.

The anomalous mechanism works not only in the photocurrent response but also in other nonreciprocal optical responses in superconductors.
In the DC limit, the anomalous terms are determined by several indicators irrespective of the frequency $(\omega_1, \omega_2)$ in the response function $\sigma_{a;bc} (\omega_1,\omega_2)$~\cite{Watanabe2022-hk}.
For the gapful superconductors, the formula is given by
                \begin{equation}
                \sigma_{a;bc}^\text{ano} (\omega_1,\omega_2) = \frac{1}{2\omega_1 \omega_2} f_{abc} - \frac{i }{8} \lim_{\bm{A} \to \bm{0}} \left[  \frac{1}{\omega_1} \epsilon_{acd} \, \partial_{A_b} \left( \sum_{p}\Omega_{p}^d f_p \right) + \frac{1}{\omega_2} \epsilon_{abd} \, \partial_{A_c} \left( \sum_{p}\Omega_{p}^d f_p \right) \right],
                \end{equation}
to which the nonreciprocity in the superfluid weight and the Berry curvature universally contribute.
Thus, the low-frequency divergence is similarly predicted in various responses including the second-harmonic generation ($\omega_1 = \omega_2$).
It may be relevant to recent experiments where the supercurrent-induced parity breaking leads to the second-harmonic generation~\cite{Nakamura2020-wt}.

Finally, it should be noted that the nonreciprocal and superconducting optical responses may be informative for quantifying the parity mixing in the superconducting pairing.
The superconducting gap symmetry does not show the definite parity under the \Pa{} violation in noncentrosymmetric systems, and the resultant parity mixing is expected to give rise to intriguing superconducting properties~\cite{NCSC_book}.
It has been a longstanding issue whether such parity mixing can be evaluated in the experiments.     
In this regard, the nonreciprocal optical response was found to be enhanced in the presence of moderate parity mixing~\cite{Watanabe2022-hk,Tanaka2023-nv}, similar to the nonreciprocal paraconductivity~\cite{Wakatsuki2018-bd}.
The quantitative estimates of the parity mixing can be obtained by future optical measurements of tunable materials, such as Li$_2$(Pt$_{3-x}$Pd$_x$)B~\cite{Badica2005-he} where the parity mixing presumably changes by chemical substitution.


\section{Summary and Outlook}
\label{Sec_summary}

The role of magnetic parity violation in solid state electron systems has been reviewed with the comparison to that of the electric parity violation, a known type of parity violation found in the materials crystalizing in a noncentrosymmetric structure.
Since the magnetic parity violation is accompanied by the time-reversal-symmetry breaking, it may be considered less fundamental than the electric parity violation.
The preserved space-time symmetry, however, allows us to make clear distinctions between physical phenomena induced by either the electric or magnetic parity violation; \textit{i.e.}, the \T{}-symmetric system manifests the physical properties arising from the electric parity violation, while the 
\PT{}-symmetric system shows phenomena unique to the magnetic parity violation.

The physical responses originating purely from the magnetic parity violation have been intensively discussed in the field of multiferroics in terms of the magnetoelectric phenomena.
In addition, recent studies have been devoted to the investigation of magnetically parity-violating physical phenomena, which have advanced the earlier studies of the electric parity violation, such as for the photocurrent response.
The identified responses are complementary to those induced by the electric parity violation.
The complementary roles of magnetic and electric parity violation have been highlighted and grasped by their contrasting electronic structures.

The magnetic parity violation may lead to nontrivial physical phenomena due to its combination with other quantum nature of materials, for example, giant optoelectronic responses enhanced by the quantum-geometrical effect of topological electrons, nonreciprocity in the electromagnetic responses in superconductors, and so on.
Despite the rapid growth in theoretical understandings of magnetic parity violation and the ubiquity of \PT{}-symmetric magnetic materials~\cite{Watanabe2018-cu} (see also Appendix~\ref{SecApp-magnetic_materials}), the material realization and experimental observations of emergent responses remain elusive.
Promising materials include those having the Dirac electrons such as quasi-two-dimensional Mn compounds (\textit{e.g.}, even-layer MnBi$_2$Te$_4$ and EuMnBi$_2$)~\cite{Zhang2019-oi,Sakai2022-dq} and massless Dirac systems (\textit{e.g.}, CuMnAs)~\cite{Tang2016-au,Smejkal2017-qb,Linn2023-ix}.

The material realization may be feasible in playgrounds other than the bulk materials.
The magnetic parity violation has been realized by metamaterials where the magnetic islands are arrayed~\cite{Lehmann2019-xp} and the ferromagnetic thin films are patterned~\cite{Matsubara2022-mb,Hild2023-kd}.
The high tunability of physical responses is advantageous, and it has been demonstrated, for example, for the photocurrent generation in the magnetic metamaterials~\cite{Matsubara2022-mb}.
Furthermore, the magnetic parity violation may be found in the fictitious fields of the topological solitons.
Recent studies clarified that the emergent magneto-multipolar fields of the magnetic hopfion give rise to various spin-charge-coupled phenomena~\cite{Pershoguba2021-mr,Liu2022-se}.

The systematics based on the preserved space-time symmetry is expected to be applied to a broad range of physical responses other than those explained in this review. 
For instance, the \T{}-even/\T{}-odd decomposition is similarly found in various physical phenomena such as second-harmonic generation~\cite{Bhalla2022-ff}, parametric conversion~\cite{Wang2010-zc,Werake2010-eb} and photo-induced and current-induced spin current responses~\cite{Young2013-fx,Hamamoto2017-ul,Kim2017-zp,Xu2021-jt,Xiao2021-if,Matsubara2022-mb,Hayami2022-lb,Adamantopoulos2022-rt,Merte2023-vt}.
The mechanism for each response has been mainly investigated in the framework of band electrons' Hamiltonian, whereas further precise treatments including disorders and interactions may allow for quantitative estimates and imply the potential of the responses in probing the quantum nature of matter.
For instance, parity-violating responses may show enhancement due to the electron correlation effect~\cite{Peters2018-mz} and collective modes such as phonons~\cite{Okamura2022-px}, spin fluctuation~\cite{Yokouchi2017-iy,Morimoto2019-ad,Ishizuka2020-qo,Iguchi2024}, plasmon~\cite{Sano2021-jh,Toshio2022-ig}, and exciton~\cite{Morimoto2020-nx,Sotome2021-es,Kaneko2021-dw}.


\section*{Acknowledgement}

H.W. expresses a lot of thanks to Naoto Nagaosa, Riki Toshio, and Yoshihiro Michishita for valuable discussions and comments.
The authors thank Inti Sodemann Villadiego for his fruitful comments.
This review partly features our works done with collaborators.
We here express our courtesy to them, particularly to
Akito Daido,
Atsuo Shitade,
Hinako Murayama,
Hiroto Tanaka,
Junta Iguchi.
Kohei Shinohara,
Kousuke Ishida,
Liu Yizhou,
Masakazu Matsubara,
Michiya Chazono,
Yugo Onishi,
Yuki Shiomi.
The authors were supported by
Grant-in-Aid  for  Scientific  Research  on 
Innovative Areas “J-Physics” (Grant No. JP15H05884), SPIPITS 2020 of Kyoto University,
and JSPS KAKENHI (Grants 
No.~JP15K05164,
No.~JP15H05745,
No.~JP16H00991,
No.~JP18H01178,
No.~JP18H04225,
No.~JP18H05227,
No.~JP18J23115,
No.~JP20H05159,
No.~JP21K18145,
No.~JP21J00453,
No.~JP22H04933, 
No.~JP22H01181,
No.~JP23K17353,
No.~JP23K13058,
No.~JP24H00007).

We made use of the useful software \texttt{vesta} for visualization of crystal and magnetic structures~\cite{Momma2011-jl}.

\appendix

\section{Table of magnetic multipolar magnets}
\label{SecApp-magnetic_materials}

Many magnetic materials are characterized by the zero propagation vector $\bm{q}=\bm{0}$ of order parameter, by which the unit cell does not change at the magnetic phase transition.
Zero propagation vector implies that the seemingly antiferroic magnetic order does induce uniform fields through the coupling to the crystal sublattice degree of freedom.
From the viewpoint of symmetry, such a uniform field is classified into the even-parity and odd-parity magnetic multipolar fields in terms of parity under the space-inversion (\Pa{}) operation.

Let us consider the point group symmetry of those magnetic multipolar materials. 
The even-parity magnetic multipolar order breaks the \T{} symmetry as well as the \PT{} symmetry while preserving the \Pa{} symmetry.
On the contrary, the odd-parity magnetic multipolar order, though it similarly breaks the \T{} symmetry, is \PT{}-even and \Pa{}-odd.
Then, when we assume the centrosymmetric symmetry of the paramagnetic state, the point group of the magnetic state is comprised of the space-inversion operation ($g=I$) but lacks parity-time-reversal operation ($g= I \theta$) in the case of even-parity magnetic multipolar systems, while it is comprised of $g=I\theta$ but lacks $g=I$ for the odd-parity magnetic multipolar systems.
Owing to the preserved \Pa{} or \PT{} symmetry, either even-parity or odd-parity magnetic multipolar field is allowed.

Letting $\bm{G}$ be the magnetic point group, the coset decomposition is obtained as 
        \begin{equation}
            \bm{G} = \bm{H}_\text{even} \cup I \cdot \bm{H}_\text{even},
            \label{even_parity_Mag-coset_decomposition}
        \end{equation}
for the even-parity magnetic multipolar systems and 
        \begin{equation}
            \bm{G} = \bm{H}_\text{odd} \cup I\theta \cdot \bm{H}_\text{odd},
            \label{odd_parity_Mag-coset_decomposition}
        \end{equation}
for the odd-parity magnetic multipolar systems.
The subgroup $\bm{H}$ of $\bm{G}$ is convenient to identify the emergent responses induced by each magnetic multipolar field; \textit{e.g.} a response unique to the odd-parity magnetic multipolar systems is subject to symmetry constraints of $\bm{H}_\text{odd}$ and thereby may be zero even when it is not forbidden by the \PT{} symmetry.

In Tables~\ref{Table-even_parity_multipolar},~\ref{Table-odd_parity_multipolar}, we tabulate the pairs of $\bm{G}$ and $\bm{H}_\text{even/odd}$ with some characteristics and candidate materials.
Notably, since we consider the \Pa{}-symmetric or \PT{}-symmetric magnetic point groups, the even- and odd-parity magnetic multipolar fields show up without being admixed with each other.
The even-parity magnetic multipolar symmetry is characterized by whether the magnetic dipolar field (M) and piezomagnetic effect (PM) are allowed.
For instance, if the magnetic dipole field is active (M: \cm), the magnetic materials denoted by the symmetry show the anomalous Hall effect.

Similarly, the odd-parity magnetic multipolar symmetry is characterized by the magnetic toroidal moment (T), magnetic quadrupole moment (MQ), and magnetopiezoelectric effect (MPE).
If `T: \cm', the \PT{}-symmetric magnetic order may be switchable with the electric current~\cite{Watanabe2018-xp} (see also Sec.~\ref{Sec_control}).
Magnetic materials with `MQ: \cm' allow for various odd-parity responses formulated by the rank-2 and \PT{}-symmetric axial tensor such as the magnetoelectric effect, photocurrent response under the circularly-polarized lights (see Sec.~\ref{SecSubSub_photocurrent}), and so on.
Magnetic materials with `MPE: \cm' similarly host odd-parity responses described by the \PT{}-symmetric tensor sharing the same symmetry as that of the magnetopiezoelectric effect [\textit{e.g,} nonreciprocal electric conductivity (Sec.~\ref{SecSubSub_nonreciprocal_conductivity}) and photocurrent response under the linearly-polarized and unpolarized lights (Sec.~\ref{SecSubSub_photocurrent})].
Note that `T' is always active if `MQ' is active because the magnetic quadrupole moments include the magnetic toroidal moment.

\begin{longtable}{LLcccc}
\caption{
Magnetic point groups (MPG) $\bm{G}$ hosting pure even-parity magnetic multipolar fields.
Each group is decomposed by its subgroup $\bm{H}_\text{even}$ by Eq.~\eqref{even_parity_Mag-coset_decomposition}.
Each item is characterized by the magnetic dipole moment (M), piezomagnetic activity (PM), and candidate material.
In some cases, candidate materials are missing (N/A) to our knowledge.
}
\label{Table-even_parity_multipolar}\\

\hline
\text{MPG} 			&\bm{H}_\text{even}	 	& M & PM  &\multicolumn{2}{c}{Candidate materials}  \\ 
\hline
\endfirsthead

\multicolumn{6}{c}{\tablename\ \thetable\ (\textit{cont.})} \\
\hline
\text{MPG} 			&\bm{H}_\text{even}	 	& M & PM  &\multicolumn{2}{c}{Candidate materials}  \\ 
\hline
\endhead

\endfoot

\endlastfoot

                m\bar{3}m  & 432  	    &               &              &\multicolumn{2}{c}{N/A}                    \\
                m\bar{3}m'  & 4'32'  	    &               &\cm              &\matfig{Cd$_2$Os$_2$O$_7$}{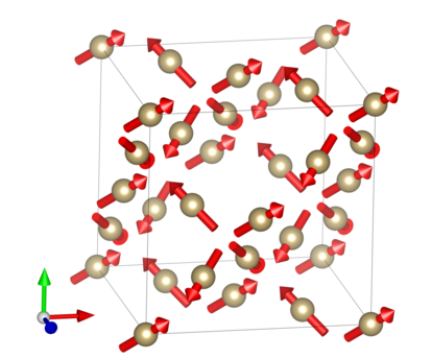}                    \\ 
                m\bar{3}  & 23  	    &               &\cm              &\matfig{NiS$_2$}{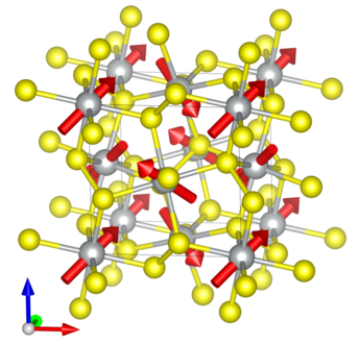}                    \\
                6/mmm  &  622 	&              &\cm              &  \multicolumn{2}{c}{N/A}                  \\
                6/mm'm'  &  62'2' 	&\cm               &\cm              &\matfig{Mn$_5$Ge$_3$}{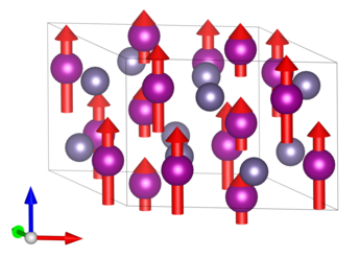}                    \\ 
                6'/m'mm'  &  6'22' 	&               &\cm              &\matfig{CrNb$_4$S$_8$}{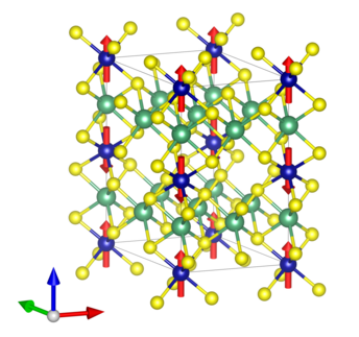}                    \\ 
                6/m &  6 	    &\cm               &\cm              &     \multicolumn{2}{c}{N/A}               \\
                6'/m' &  6' 	    &               &\cm              &      \multicolumn{2}{c}{N/A}              \\
                \bar{3}m & 32  	&               &\cm              & \matfig{CoF$_3$}{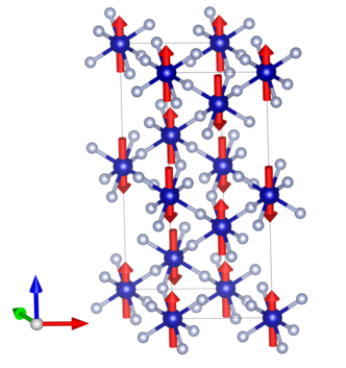}                   \\ 
                \bar{3}m' & 32'  	&\cm               &\cm              & \matfig{Mn$_3$Ir}{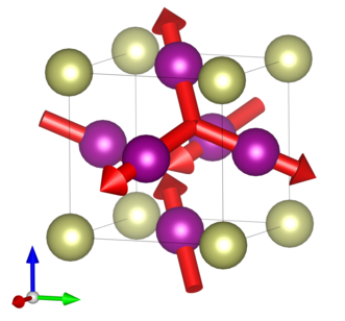}                   \\ 
                \bar{3} & 3  	&\cm               &\cm              &   \matfig{Mn$_3$NiN}{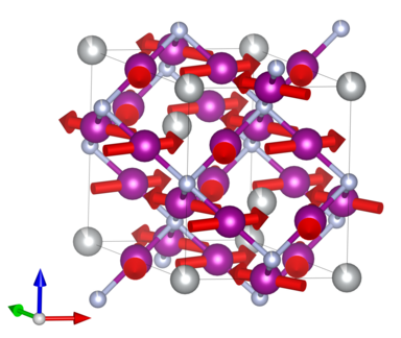}                 \\ 
                4/mmm  &  422 	&               &\cm              &\matfig{KMnF$_3$}{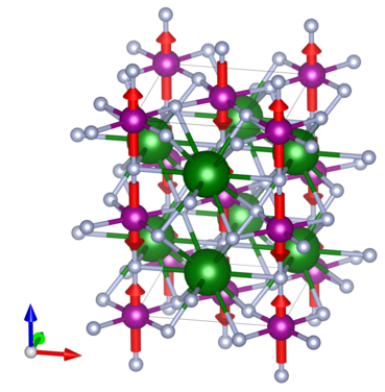}                    \\
                4/mm'm'  &  42'2' 	&\cm               &\cm              & \matfig{Nd$_2$NiO$_4$}{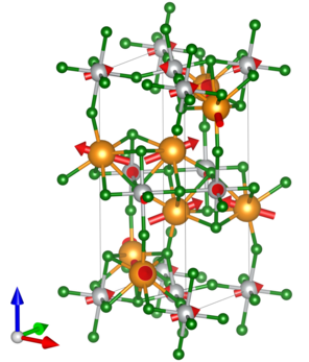}                   \\ 
                4'/mm'm  &  4'22' 	&               &\cm              &\matfig{MnF$_2$}{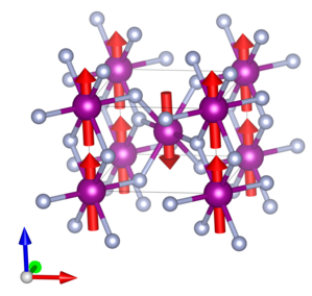}                    \\ 
                4/m  &  4 	    &\cm               &\cm              & \matfig{MnV$_2$O$_4$}{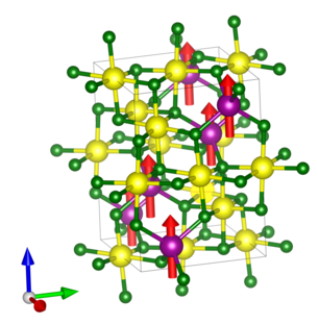}                   \\
                4'/m  &  4' 	    &               &\cm              & &                   \\
                mmm  & 222  	&               &\cm              &\matfig{MnTe}{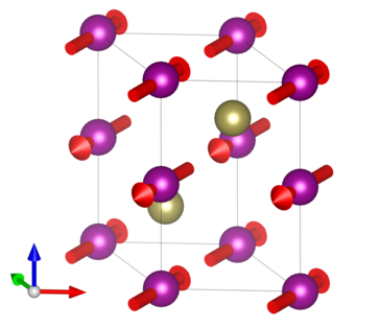}  \\
                m'm'm  & 2'2'2  	&\cm               &\cm              &\matfig{Mn$_3$Sn}{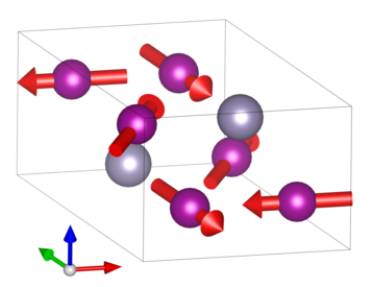}  \\ 
                2/m  &  2 	    &\cm               &\cm              &\matfig{MnCO$_3$}{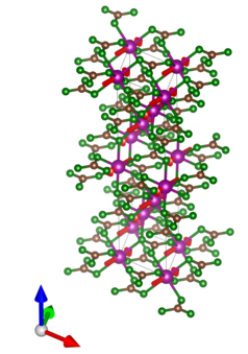}                    \\ 
                2'/m'  &  2' 	    &\cm               &\cm              &\matfig{KMnF$_4$}{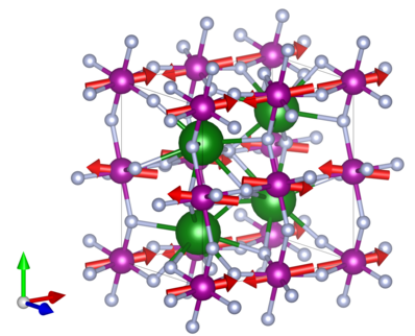}                    \\ 
                \bar{1}  &  1 	    &\cm               &\cm              &\matfig{RbMnF$_4$}{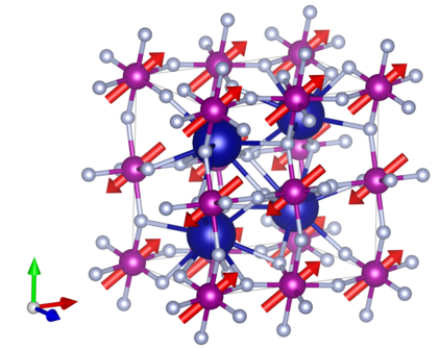}                    \\ 
\hline
\end{longtable}

\begin{longtable}{LLccccc}
\caption{
Magnetic point groups $\bm{G}$ hosting pure odd-parity magnetic multipolar fields.
Each group is decomposed by its subgroup $\bm{H}_\text{odd}$ by Eq.~\eqref{odd_parity_Mag-coset_decomposition}.
Each item is characterized by the magnetic toroidal moment (T), magnetic quadrupole moment (MQ), magnetopiezoelectric activity (MPE), and candidate material.
In some cases, candidate materials are missing to our knowledge.
}
\label{Table-odd_parity_multipolar}\\

\hline
\text{MPG} 			&\bm{H}_\text{odd}	 	& T & MQ & MPE  &\multicolumn{2}{c}{Candidate materials}  \\ 
\hline
\endfirsthead

\multicolumn{7}{c}{\tablename\ \thetable\ (\textit{cont.})} \\
\hline
\text{MPG} 			&\bm{H}_\text{odd}	 	& T & MQ & MPE  &\multicolumn{2}{c}{Candidate materials}  \\ 
\hline
\endhead

\endfoot

\endlastfoot

                m'\bar{3}'m'  & 432  	    &&\cm               &              &\multicolumn{2}{c}{N/A}                    \\
                m'\bar{3}'m  & \bar{4}3m  	   & &               &\cm              &\multicolumn{2}{c}{N/A}                    \\ 
                m'\bar{3}'  & 23  	    &  &\cm             &\cm              &\multicolumn{2}{c}{N/A}                    \\
                6'/mmm'  &  \bar{6}m2 	& &              &\cm              &\multicolumn{2}{c}{N/A}                    \\
                6/m'mm  &  6mm 	&\cm &\cm              &\cm              &\multicolumn{2}{c}{N/A}                    \\
                6/m'm'm'  &  622 	& &\cm              &\cm              &\multicolumn{2}{c}{N/A}                    \\
                6'/m  &  \bar{6} 	& &              &\cm              &\matfig{U$_{14}$Au$_{51}$}{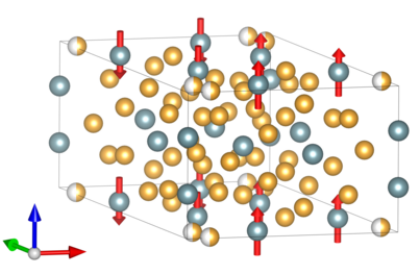}                    \\
                6/m'  &  6 	&\cm &\cm              &\cm              &    \matfig{PbMn$_2$Ni$_6$Te$_3$O$_{18}$}{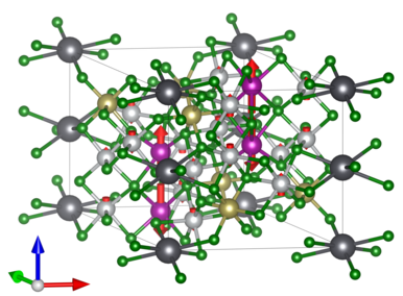}                \\
                \bar{3}'m' & 32  	& &\cm              &\cm              & \matfig{Cr$_2$O$_3$}{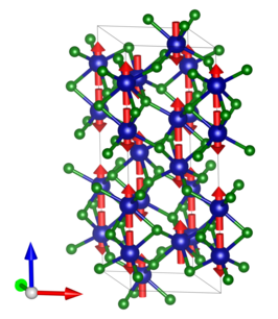}                   \\
                \bar{3}'m & 3m  	&\cm &\cm              &\cm              & \matfig{Ca$_2$YZr$_2$Fe$_3$O$_{12}$}{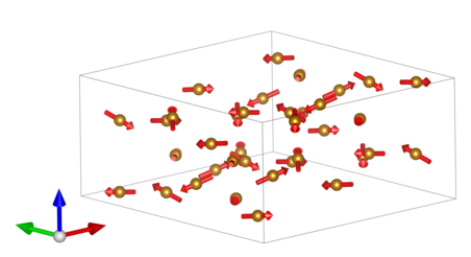}                   \\
                \bar{3}' & 3  	&\cm &\cm              &\cm              & \matfig{MnGeO$_3$}{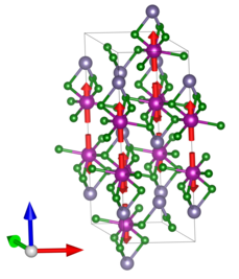}                   \\
                4/m'm'm'  &  422 	&&\cm               &\cm              & \matfig{GdB$_4$}{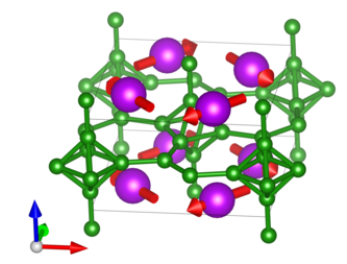}                   \\ 
                4/m'mm  &  4mm 	&\cm&\cm               &\cm              & \matfig{Co$_3$Al$_2$Si$_3$O$_{12}$}{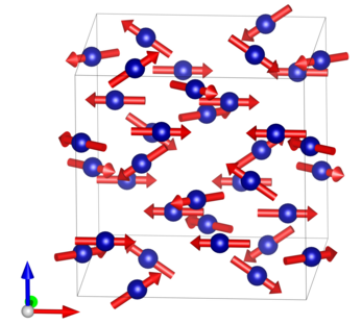}                   \\ 
                4'/m'm'm  &  \bar{4}2m 	&&\cm               &\cm              &\matfig{BaMn$_2$As$_2$}{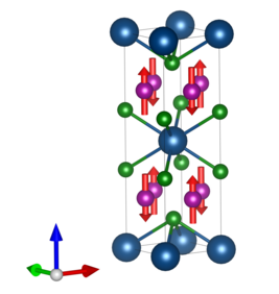}                    \\ 
                4/m'  &  4 	&\cm&\cm               &\cm              &  \matfig{TlFe$_{1.6}$Se$_2$}{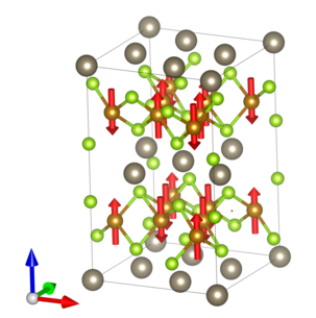}                   \\ 
                4'/m'  &  \bar{4} 	&&\cm               &\cm              &\matfig{KOsO$_4$}{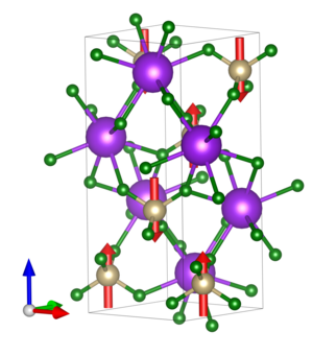}                    \\ 
                m'm'm'  &  222 	&&\cm               &\cm              & \matfig{LiMnPO$_4$}{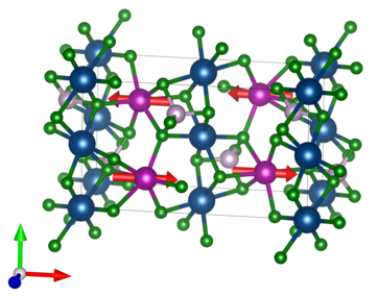}                   \\ 
                mmm'  &  mm2 	&\cm&\cm               &\cm              &\matfig{CuMnAs}{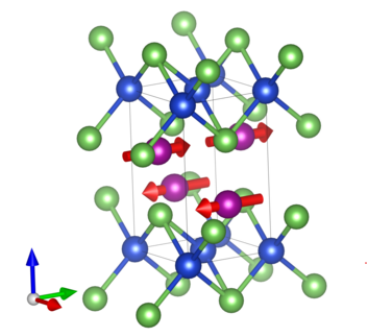}                    \\ 
                2/m'  &  2 	&\cm&\cm               &\cm              &\matfig{Na$_2$RuO$_4$}{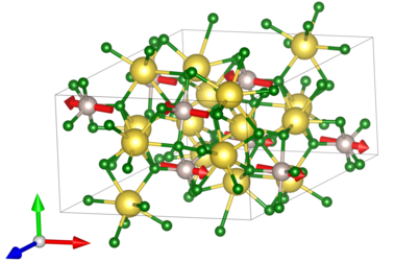}                    \\ 
                2'/m  &  m 	&\cm&\cm               &\cm              &\matfig{SrMn$_2$As$_2$}{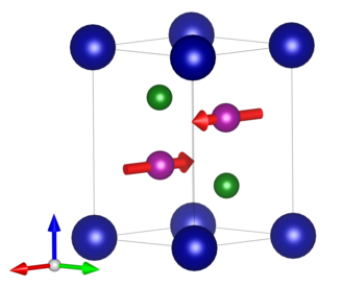}                    \\ 
                \bar{1}'  &  1 	&\cm&\cm               &\cm              &\matfig{YbMn$_2$Sb$_2$}{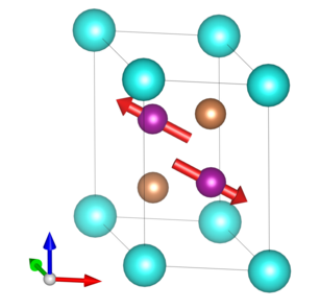}                    \\ 
\hline
\end{longtable}

\end{document}